\documentclass[twocolumn,aps,prl,showpacs,amsmath,superscriptaddress,longbibliography,notitlepage]{revtex4-2}

\usepackage{amssymb}
\usepackage{dcolumn}
\usepackage{bm}
\usepackage{graphicx}
\usepackage{amsmath}
\usepackage{xcolor}
\graphicspath{{pict/}{}}
\usepackage[pdfstartview=FitH, CJKbookmarks=true, bookmarksnumbered=true, bookmarksopen=true, colorlinks=true, pdfborder=001, citecolor=blue, linkcolor=blue, linktocpage=true] {hyperref}

\def\ket#1{\left|#1\right\rangle}
\def\bra#1{\left\langle#1\right\rangle}

\begin{document}

\preprint{APS/123-QED}
\title{Occupation-selective topological pumping from Floquet gauge fields}

\author{Wenjie Liu}\email{liuwenjie@dlut.edu.cn} 
\affiliation{School of General Education, Dalian University of Technology, Panjin 124221, China}

\author{Ching Hua Lee}
\email{phylch@nus.edu.sg}
\affiliation{Department of Physics, National University of Singapore, Singapore 117551, Republic of Singapore}

\author{Zhoutao Lei}\email{leizht3@nus.edu.sg} 
\affiliation{Department of Physics, National University of Singapore, Singapore 117551, Republic of Singapore}

\date{\today}

\begin{abstract}
Topological pumping is conventionally governed by single-particle band topology. 
Here we show that promoting tunneling to a dynamical, occupation-conditioned variable fundamentally reshapes this paradigm, leading to \emph{occupation-selective topological pumping}. 
In a periodically driven one-dimensional superlattice with density-dependent hopping, two-body bound states (doublons) acquire Chern numbers distinct from those of single particles and exhibit quantized transport even when the single-particle pump is trivial, including counter-propagating responses. 
We identify a dynamical-gauge-field mechanism that induces topological phase transitions in the bound-state sector absent from the single-particle spectrum. 
Furthermore, the gauge field concentrates Berry curvature into sharply localized resonant regions without compromising adiabatic quantization. 
A Floquet realization with ultracold atoms is proposed to realize such occupation-selective pumping.
Our results reveal a mechanism for occupation-selective topological responses that can persist across higher-occupancy bound states.
\end{abstract}

\maketitle

\noindent\emph{{Introduction}.---}
Topological phases exhibit robust quantized responses governed by global invariants~\cite{HasanKane2010,QiZhang2011,Cooper2019}, and in driven settings, \emph{topological pumping} provides quantized transport under adiabatic cyclic modulation~\cite{Thouless1983,2023TPandTopo,LaserPhotonReview2025}.
Topological pumping has been realized in ultracold atoms~\cite{Nakajima2016,Lohse2016,Walter2023HubbardPump,Viebahn2024InteractionsPump} and explored across diverse platforms~\cite{ArguelloLuengo2024HubbardPump,2022PhysRevLett.129.140502,Sridhar2024SyntheticPump,Liu2025SCProcessorPump,xh3v-tky4}, revealing interaction-induced breakdown~\cite{Walter2023HubbardPump,Tuloup2023Breakdown} or interaction-enabled pumping~\cite{2017PhysRevA.95.063630,2020PhysRevA101023620,2023PhysRevResearch.5.013020,2024PhysRevLett133140202,Viebahn2024InteractionsPump,ArguelloLuengo2024HubbardPump}, disorder-modified responses~\cite{PhysRevA.101.052323,Cerjan2020DisorderedPhotonic,Liu2025SCProcessorPump}, as well as non-integer~\cite{2014PhysRevA90053623,2016PhysRevB94235139,2017PhysRevLett118230402,2018PhysRevB.98.094434,2024PhysRevA.109.053311,Juergensen2025FractionalPump} and nonlinear~\cite{Jurgensen2021QuantizedNonlinear,Jurgensen2022ChernNumber,Fu2022NonlinearThouless,Mostaan2022QuantizedTopological,Jurgensen2023QuantizedFractional,Xiao2025Nonlinear,2025PhysRevA111033306,Ravets2025KerrPump,202596f5-qszj} regimes.
Yet, even in interacting settings, pumping is still almost always formulated on top of a fixed transport structure: correlations can reshape the dynamics, but the tunneling rule itself remains \emph{occupation independent}. Topology therefore continues to be defined within a common single-particle framework, albeit dressed by interactions. This raises a fundamental question: \emph{can topological pumping itself become occupation selective, with different occupancy sectors exhibiting distinct quantized transport under the same driving cycle?}

A natural route toward such occupation-selective pumping is to promote tunneling into a \emph{dynamical, occupation-conditioned} variable through density-dependent hopping, which acts as a microscopic dynamical gauge field~\cite{2009PhysRevLett103133002,Keilmann2011DDPeierls,Greschner2014DDGauge,2016PhysRevLett116205301,2017PhysRevLett.118.146403,2018PhysRevLett.121.237401,2022CommunPhys5238,2022PhysRevLett.129.180401,2024PhysRevLett132023401,2024NatCommun155807,2025lw8k-7h6p,arXivLei2025}.
More broadly, lattice gauge theories provide a microscopic framework for matter coupled to dynamical gauge fields and underpin phenomena such as confinement and string breaking~\cite{Kogut1979,1983RevModPhys55775,Wiese2013,Schweizer2019Z2,Banuls2020QT,Cheng2024EmergentU1,2022PRXQuantum3040316,2022PRXQuantum3040317,Zhang2025MicroConfinement,PhysRevLett.109.175302,2013PhysRevLett111110504,2016PhysRevX6011023,2020PhysRevB102014308,2020PhysRevX10021041,De2024StringBreakingDynamics,GonzalezCuadra2025StringBreakingRydberg,Cochran2025VisualizingStrings,2025mwy1-v9hk}.
Recent progress has enabled controlled studies of correlated matter--gauge dynamics and constraint-induced effects~\cite{2017PhysRevLett118266601,2021PhysRevLett126130401,2022PRXQuantum3020345,2023PhysRevLett131220402,2014PhysRevLett112201601,2019PhysRevLett122250401,2022PhysRevB105125123,2023PhysRevLett131050401,2024PhysRevLett133216601,2025CommunPhys}, with dynamical lattice gauge fields realized in platforms ranging from ultracold atoms~\cite{PhysRevLett.109.175302,Zohar_2016,2019NP151168,2019NP151161,2023PhysRevLett.131.080403,2025PhysRevA111013319} to superconducting circuits~\cite{2013PhysRevLett111110504}.
At the same time, controlled dissipation offers additional handles on nonequilibrium gauge physics~\cite{2025wztw-l8wg}.
These developments suggest that dynamical gauge fields are not only a platform for correlated gauge phenomena, but also a possible route to engineer topological transport.

In this Letter, we demonstrate that a dynamical, occupation-conditioned tunneling mechanism gives rise to a fundamentally new form of topological transport. 
Focusing on a periodically driven one-dimensional superlattice with density-dependent hopping, we show that two-body bound states form isolated bands with well-defined Chern numbers that can differ from those of single particles. 
We establish that the dynamical gauge field reshapes the effective hopping structure of bound states, driving topological phase transitions that are absent in the single-particle description. 
As a consequence, quantized pumping becomes occupation selective: doublons can undergo quantized transport in parameter regimes where the single-particle pump is trivial, and the pumping direction can be independently tuned across occupancy sectors. 
We further reveal that the dynamical gauge field localizes Berry curvature into narrow regions in parameter space, compressing the geometric response without spoiling adiabatic quantization. 
Finally, we outline a feasible Floquet scheme with ultracold atoms to realize and probe these effects.

\begin{figure}[t!]
\center
\includegraphics[width=0.45\textwidth]{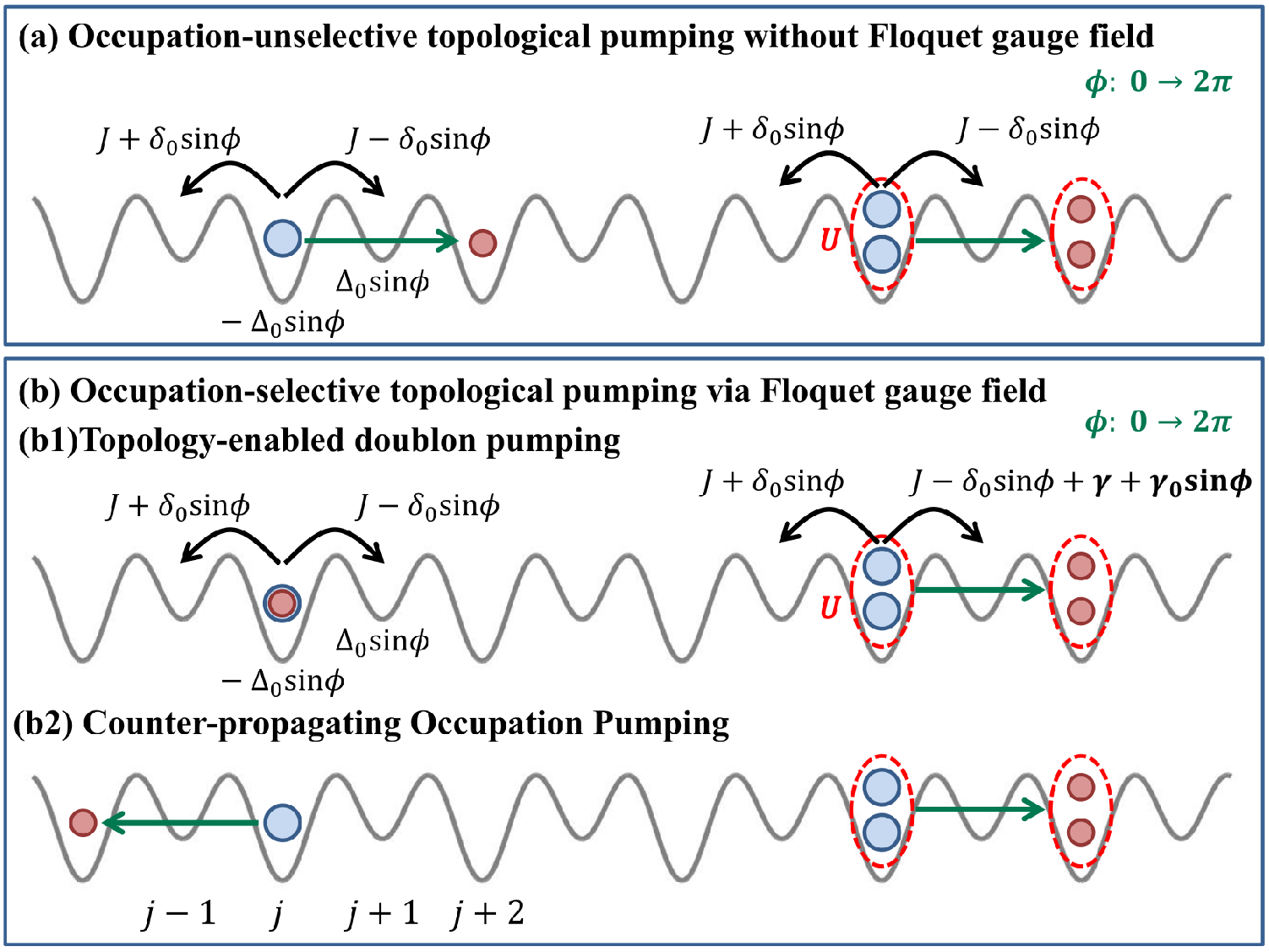}
\caption{\textbf{Schematic illustration of occupation-selective topological pumping beyond the single-particle paradigm.}
(a) Occupation-unselective pumping without a dynamical gauge field: in the conventional Thouless pump, single particle moves one unit cell over a cycle [$\phi = 0$ (blue circle) $\to$ $\phi = 2\pi$ (red circle)], and doublon follows the same path.
(b) Occupation-selective pumping via a dynamical gauge field: density-dependent tunneling renders the hopping amplitude dynamical, allowing the pumping response to depend on particle occupation.
(b1) Topology-enabled doublon pumping: doublon exhibits quantized transport even when the single-particle pump is topologically trivial, whose dynamical realization is shown in Fig.~\ref{fig:DensityEvolution}(a).
(b2) Counter-propagating pumping: the pumping directions experienced by single and double-boson state sectors can be opposite, with the corresponding dynamics illustrated in Fig.~\ref{fig:DensityEvolution}(b).
\label{fig:Diagram}}
\end{figure}

\noindent\emph{{A driven Superlattice with Dynamical Gauge Field}.---}
We consider interacting bosons in a one-dimensional periodically driven superlattice with \emph{density-dependent tunneling}, as illustrated in Fig.~\ref{fig:Diagram}. 
The system is described by
\begin{equation}\label{Bosonmodel}
\begin{aligned}
& \hat{H}(t)=\sum_j^N\left[J+\delta_0 \sin (\pi j+\phi(t))\right] \hat{a}_j^{\dagger} \hat{a}_{j+1}+\text { H.c. } \\
& +\sum_j^{N / 2} \hat{a}_{2 j-1}^{\dagger}\left[\gamma+\gamma_0 \sin \phi(t) \left(\hat{n}_{2 j-1}+ \hat{n}_{2 j}\right)\right] \hat{a}_{2 j}+\text { H.c. } \\
& +\sum_j^N \Delta_0 \cos (\pi j+\phi(t)) \hat{n}_j+\frac{U}{2} \sum_j^N \hat{n}_j\left(\hat{n}_j-1\right)
\end{aligned}
\end{equation}
where $J$ and $U$ denote the hopping and on-site interaction, $\delta_0$ and $\Delta_0$ the modulated hopping and staggered potential, and $\gamma$ ($\gamma_0$) the occupation-independent (dependent) intra-cell tunneling. 
The density-dependent term acts as a dynamical gauge field, making the tunneling amplitude occupation-conditioned~\cite{2022PhysRevLett.129.180401,2024PhysRevLett132023401}, and thereby enabling occupation-selective topology in which multi-particle states acquire independent Chern numbers and quantized pumping. 
Such a Hamiltonian can be implemented via Floquet engineering in ultracold atoms through fast modulation of tunneling and interactions  (see details in the Supplemental Materials~\cite{2026supplemental}).

\begin{figure}[htp]
\center
\includegraphics[width=0.5\textwidth]{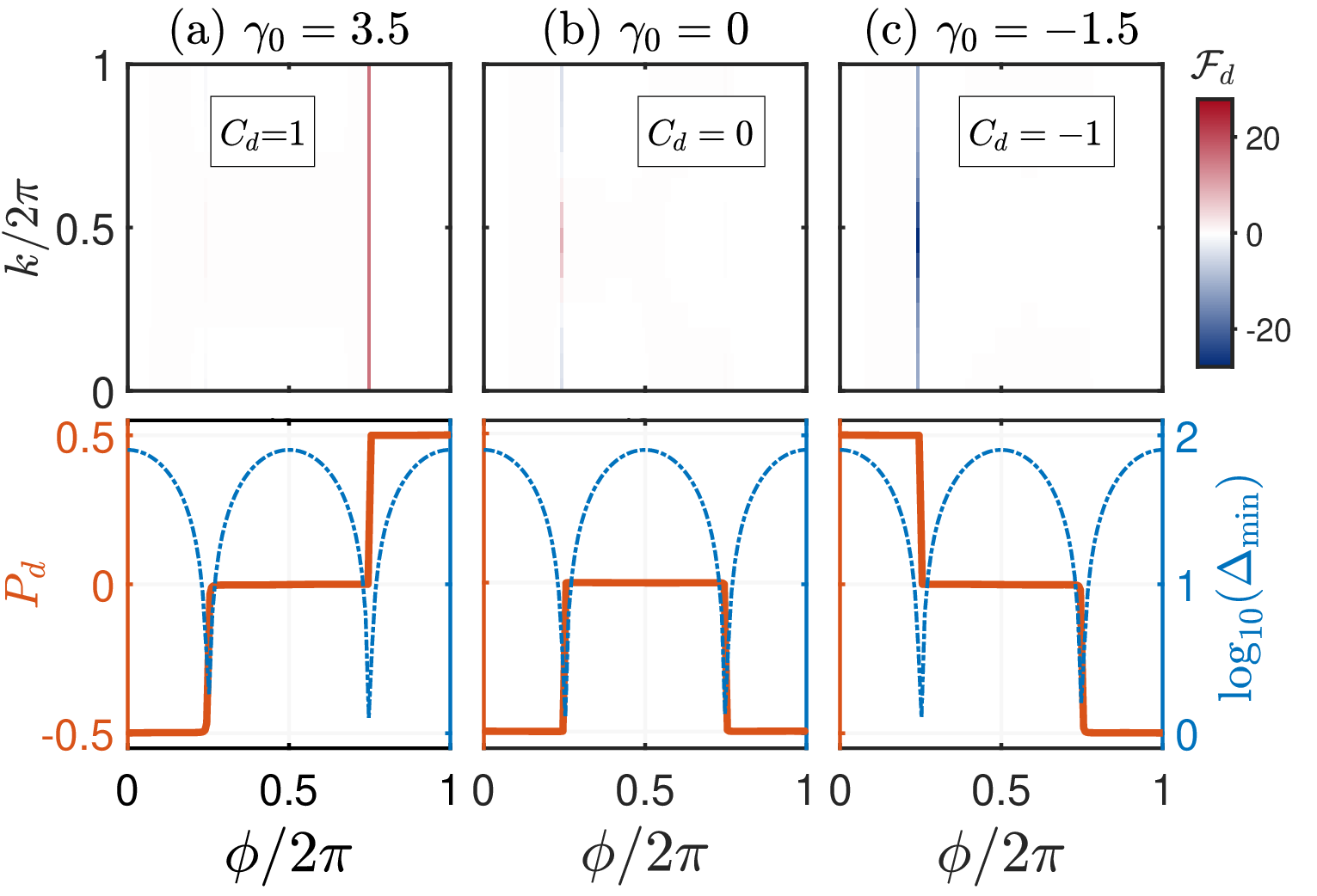}
\caption{\textbf{Topological characterization and response of the doublon bound-state band.}
Dynamical-gauge-field–induced Berry curvature and quantized polarization of the lower two-body bound-state band for different density-dependent hopping amplitudes $\gamma_0$.
Columns (a)--(c) correspond to $\gamma_0=3.5$, $0$, and $-1.5$, respectively.
Upper row: Berry curvature $\mathcal{F}_{d}$ in the $(\phi,k)$ parameter space. 
The curvature exhibits extremely sharp, localized peaks and valleys, leading to a highly discontinuous appearance in the linear color scale. These sharp local features are a consequence of the dynamical gauge field and do not affect the global topological properties.
Lower row: Variation of the polarization $P_{d}$ over one driving period (left axis) and the minimum energy gap $\Delta_{\min}$ between the two bound-state bands after scanning the Brillouin zone (right axis).
The sharp transitions in $P_{d}$ reflects a dynamical-gauge-field–enhanced resonant switching of the Wannier center.
A quantized winding of $P_{d}$ is observed for $\gamma_0=3.5$ and $-1.5$, corresponding to $C_d=+1$ and $-1$, respectively, while no winding occurs for $\gamma_0=0$, indicating a topologically trivial phase.
\label{fig:boundspectrum}}
\end{figure}

The Hamiltonian conserves total particle number, and under periodic boundary conditions, a global cotranslational symmetry preserves the center-of-mass quasimomentum 
$k$, allowing well-defined multi-particle Bloch bands~\cite{2026supplemental}.
We focus on the two-particle sector in the strong-interaction regime $U=100$ (with $J=1$, $\Delta_0=20$, $\delta_0=0.5$), where two isolated doublon bands remain separated from the scattering continuum throughout the cycle. 
We analyze the lower bound band and denote its Chern number by $C_d$,
\begin{equation}
C_d=\frac{1}{2\pi}\int_{0}^{\pi}\! dk \int_{0}^{2\pi}\! d\phi \, \mathcal{F}_d(k,\phi)
\end{equation}
with
\begin{equation}
\mathcal{F}_d
= i \left(
\langle \partial_{\phi} \psi_d | \partial_{k} \psi_d \rangle
-
\langle \partial_{k} \psi_d | \partial_{\phi} \psi_d \rangle
\right).
\end{equation}
As shown in Fig.~\ref{fig:boundspectrum}, localized Berry-curvature structures yield $C_d=\pm1$ for $\gamma_0=3.5$ and $-1.5$, while $\gamma_0=0$ gives a trivial phase $C_d=0$.

We further compute the many-body polarization via the projected position operator $\hat X_P=\hat P_d \hat X \hat P_d$. 
If $\{\lambda_n\}$ are its eigenvalues, the polarization reads
\begin{equation}
P_d(\phi)
=
\frac{1}{2\pi}
\,\mathrm{Im}\,
\ln
\left[
\prod_{n=1}^{N/2} \lambda_n(\phi)
\right]
\end{equation}
whose winding over one cycle reproduces $C_d$. This is verified in the bottom row of  Fig.~\ref{fig:boundspectrum}, which showcases quantized windings of $\mp 1$ for $\gamma_0=-1.5,3.5$, and vanishing winding for $\gamma_0=0$. 
Although the polarization $P_d(\phi)$ exhibits sharp transitions near $\phi=\pi/2,3\pi/2$, the evolved state can be adiabatically locked to the instantaneous band. This is ensured by the finite minimum gap $\Delta_\text{min}$
between the two doublon bound-state bands and the gap-adapted driving protocol [Eq.~\eqref{DrivenFrequency}], which slows the evolution near minimal gaps. Consequently, nonadiabatic transitions are suppressed, and the state accurately tracks the sharply varying Wannier center. These sharp features therefore reflect a geometric compression of the topological response rather than a breakdown of adiabaticity.

Notably, the dynamical gauge field strongly localizes Berry curvature near high-symmetry points, producing sharp kink-like polarization profiles. 
This reflects a geometric compression of the topological response into narrow resonant windows without inducing nonadiabaticity, indicating that the dynamical gauge field both induces new topological phases and sharpens the pumping response. 

\noindent\emph{{Doublon Band Topology and Gauge-Field–Induced Phase Transitions}.---}
In the strong-interaction regime $U \gg (J,\delta_0,\gamma,\gamma_0,\Delta_0)$, the low-energy sector is spanned by on-site doublons. 
Treating the remaining terms perturbatively, we derive an effective single-particle Hamiltonian via degenerate perturbation theory~\cite{2026supplemental},
\begin{equation}\label{EffectiveModel}
\begin{aligned}
\hat{H}_{\text {eff }} & =\sum_j^{N}\left[\mathcal{J}_1+\mathcal{J}_2 \frac{1-\cos (\pi j)}{2}\right] \hat{b}_j^{\dagger} \hat{b}_{j+1}+\text { H.c. } \\
& +\sum_{j=1}^N 2 \Delta_0 \cos (\pi j+\phi) \hat{b}_j^{\dagger} \hat{b}_j+\mu \sum_{j=1}^N \hat{b}_j^{\dagger} \hat{b}_j
\end{aligned}
\end{equation}
where $\hat b_j^\dagger|\mathbf0\rangle=|2\rangle_j$ creates a doublon. 
The effective couplings are $\mathcal{J}_1=\tfrac{2}{U}[J+\delta_0 \sin (\pi j+\phi)]^2$, $\mathcal{J}_2=\tfrac{4}{U}(\gamma+\gamma_0 \sin \phi(t))(J-\delta_0 \sin \phi)+\tfrac{2}{U}(\gamma+\gamma_0 \sin \phi)^2$, and $\mu=U+\tfrac{2}{U}(J-\delta_0 \sin \phi)^2+\tfrac{2}{U}(J+\delta_0 \sin \phi)^2+\mathcal J_2$, all inheriting the driving dynamics. 
$\mu$ is a global shift and does not affect topology. 

Fourier transformation yields two doublon bands whose polarizations agree quantitatively with the full many-body results~\cite{2026supplemental}, validating the effective description above. 
At $\phi=\pi/2,\,3\pi/2$, Eq.~\eqref{EffectiveModel} reduces to the SSH model~\cite{1980PhysRevB222099,1988RevModPhys60781},while generic values of $\phi$ generate more complicated though conceptually analogous unit cells. 
Because density-dependent hopping renormalizes the effective intra-cell tunneling, the doublon SSH configuration can differ from that experienced by a single-particle, whose hoppings are directly given in Eq.~\eqref{Bosonmodel}, and thus host distinct topology.

\begin{figure}[htp]
\center
\includegraphics[width=0.45\textwidth]{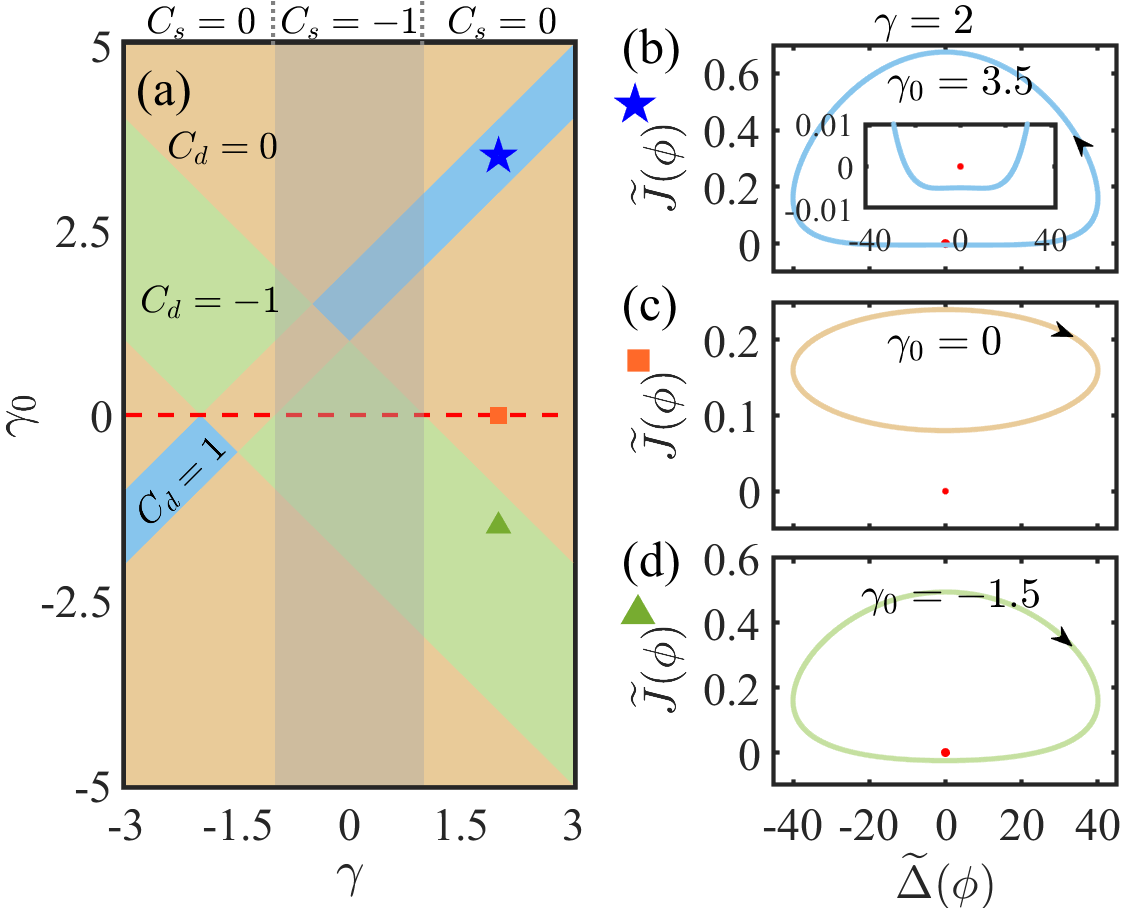}
\caption{\textbf{Dynamical-gauge-field–induced doublon topology beyond the single-particle phase diagram.}
(a) Chern number $C_d$ of the lower doublon band in the $(\gamma,\gamma_0)$ plane, with $\gamma$ (static) and $\gamma_0$ (dynamical gauge-field) intra-cell hopping.
The gray shaded region indicate the corresponding single-particle topology $C_{s}=-1$; along the density-independent limit $\gamma_0=0$ (red dashed line), the doublon topology follows that of the single-particle one. 
For $\gamma_0\neq0$, the dynamical gauge field generates topology unique to the doublon sector, including $C_{d}=\pm1$ even when $C_{s}=0$, establishing occupation-selective topology. 
(b)–(d) Pumping trajectories in the effective parameter space $(\widetilde{\Delta}(\phi),\widetilde{J}(\phi))$ at fixed $\gamma=2$ for $\gamma_0=3.5,0,-1.5$. 
The red dot marks the band-touching point.  As $\gamma_0$ is tuned away from zero, the trajectory evolves from not enclosing to encircling this degeneracy, directly revealing a dynamical-gauge-field–driven topological phase transition from trivial to nontrivial winding, with opposite circulation corresponding to opposite signs of $C_{d}$.
\label{fig:trajectory}}
\end{figure}

Topological transitions occur when the doublon gap closes. 
From the spectrum analytically obtained from Eq.~\eqref{EffectiveModel}, the phase boundaries in the space of density independent/dependent hopping amplitudes $(\gamma,\gamma_0)$ follow from band touching and are given by $\gamma_0=\pm\gamma+1$ and $\gamma_0=\pm2\pm\gamma$. 
Away from these lines, the Chern number $C_d$ is computed numerically, yielding the phase diagram in Fig.~\ref{fig:trajectory}(a), where the single-particle topology $C_s$ is shown for comparison. 
Along $\gamma_0=0$, the doublon topology follows $C_s$. 
For $\gamma_0\neq0$, however, the dynamical gauge field independently controls the doublon sector, producing regions with $C_d=\pm1$ even when $C_s=0$, and allowing opposite Chern numbers between sectors.

To expose the geometric origin, Eq.~\eqref{EffectiveModel} can be cast into a two-level form characterized by an effective staggered potential $\widetilde{\Delta}(\phi)=2\Delta_0\cos\phi$ and a tunneling imbalance $\widetilde{J}(\phi)=\big|\tfrac{2}{U}(J-\delta_0 \sin \phi)^2+\mathcal{J}_2\big|-\big|\tfrac{2}{U}(J+\delta_0 \sin \phi)^2\big|$. 
The pumping cycle traces a closed trajectory in the $(\widetilde{\Delta}(\phi),\widetilde{J}(\phi))$ plane. 
The red point in Fig.~\ref{fig:trajectory}(b)–(d) marks the degeneracy where the gap vanishes. 
For $\gamma_0=0$ the trajectory does not enclose it ($C_d=0$), whereas finite $\gamma_0$ deforms the path to encircle the degeneracy, producing a gauge-field–induced topological transition. 
Opposite circulation for $\gamma_0=3.5$ and $-1.5$ yields $C_d=\pm1$, and at the critical lines the trajectory passes through the degeneracy, confirming bulk gap closing.

\noindent\emph{{Quantized Doublon Pumping Beyond Single-Particle Topology}.---}
We now demonstrate the dynamical consequence of occupation-selective topology through adiabatic pumping. 
As established above, doublon bands acquire Chern numbers controlled by the density-dependent hopping $\gamma_0$, even when the single-particle sector is topologically trivial. 
Such a bound-state topology directly gives rise to quantized transport as a dynamical consequence.

We initialize the system in the lower doublon band using a two-body Wannier state, well approximated in the nearly flat-band regime by a localized doublon $|\boldsymbol{\psi}_0\rangle = |2\rangle_{2j-1}$. 
Adiabaticity is challenged by transient gap constrictions near $\phi=\pi/2$ and $3\pi/2$, where Landau--Zener transitions are most likely. 
To suppress excitations, we adopt a gap-adapted driving protocol,
\begin{equation}\label{DrivenFrequency}
\phi(t)=\omega t , \qquad \omega=\eta(\Delta_{\min})^{2},
\end{equation}
which slows the evolution near minimal gaps while remaining efficient elsewhere~\cite{2026supplemental}. 
The state evolves as $|\boldsymbol{\psi}(t)\rangle=\mathcal{T}\exp[-i\int_{t_0}^t \hat H(t')dt']|\boldsymbol{\psi}_0\rangle$. 
We monitor the density $n_j(t)=\langle\hat n_j\rangle$ and the center-of-mass displacement $\Delta X(t)=X(t)-X(0)$ with $X=\langle\hat X\rangle$. 

\begin{figure}[b!]
\center
\includegraphics[width=0.45\textwidth]{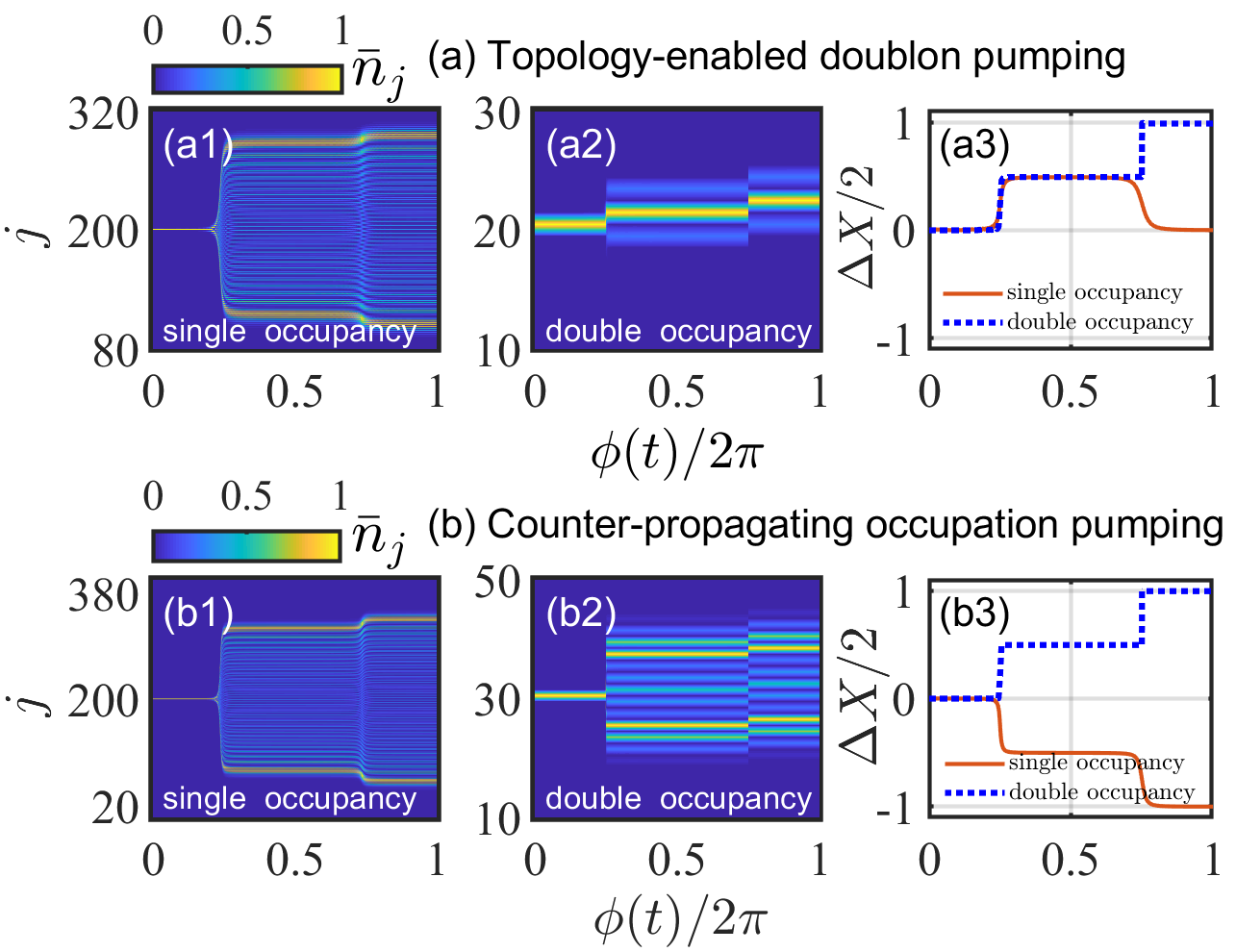}
\caption{\textbf{Dynamics of dynamical-gauge-field-induced occupation-selective topological pumping.}
The phase $\phi(t)$ is driven according to Eq.~(\ref{DrivenFrequency}), where the driving frequency is chosen to ensure an adiabatic evolution over one pumping cycle.
(a) Topology-enabled doublon pumping $(\gamma = 2,\ \gamma_0 = 3.5)$.
Time evolution of the site-resolved density for a single particle (a1) and for a two-particle bound state (doublon) (a2) over one pumping cycle, with color indicating the renormalized occupation $\bar{n}_j$.
Although the single-particle pump is topologically trivial, the doublon undergoes quantized transport.
The resulting center-of-mass displacement $\Delta X/2$ is shown in (a3), where solid red and dashed blue curves denote the single- and double-occupancy sectors, respectively.
(b) Counter-propagating occupation pumping $(\gamma = 0.5,\ \gamma_0 = 2)$.
The same quantities are shown in (b1)--(b3), demonstrating that the single- and double-occupancy sectors are pumped in opposite directions under the same adiabatic driving protocol. 
\label{fig:DensityEvolution}}
\end{figure}

Figure~\ref{fig:DensityEvolution} shows the real-time dynamics of the normalized density $\bar n_j=n_j/n_j^{\max}$ for two representative regimes (parameters fixed at $J=1$, $\Delta_0=20$, $\delta_0=0.5$, $U=100$, $\eta=0.01$). 
For $(\gamma,\gamma_0)=(2,3.5)$ [Fig.~\ref{fig:DensityEvolution}(a)], the single-particle pump is trivial ($\Delta X/2=0$), whereas the doublon shifts rightward by one unit cell ($\Delta X/2=+1$), consistent with $C_d=+1$. 
This realizes topology-enabled doublon pumping in a single-particle trivial regime.
For $(\gamma,\gamma_0)=(0.5,2)$ [Fig.~\ref{fig:DensityEvolution}(b)], the single-particle sector is topological and moves leftward ($\Delta X/2=-1$), while the doublon is pumped rightward ($\Delta X/2=+1$). 
The opposite chiral responses under identical driving demonstrate that transport is determined by occupation-selective topology rather than inherited single-particle invariants.

These results show that quantized doublon pumping is directly controlled by the dynamical gauge field parameter $\gamma_0$.
Nontrivial transport persists even when the single-particle pump is trivial, and the pumping direction can be independently tuned— or reversed—between occupancy sectors, providing a direct and observable dynamical signature of interaction-induced dynamical gauge fields.

\noindent\emph{{Floquet Engineering of Occupation-Selective Pumping with Ultracold Atoms}.---}
The Hamiltonian in Eq.~\eqref{Bosonmodel} can be realized with ultracold bosons in a driven optical superlattice, but crucially in a form tailored to our mechanism: a two-timescale Floquet protocol that embeds an occupation-dependent tunneling directly into a Thouless pumping cycle. 
We consider a one-dimensional superlattice with two sites per unit cell~\cite{Nakajima2016,Lohse2016}. 
A slow modulation $\phi(t)=\omega t$ generates the pump through $J+\delta_0\sin(\pi j+\phi)$ and $\Delta_0\cos(\pi j+\phi)$.
The key ingredient is a high-frequency drive $\Omega\gg\omega$ that generates a dynamical gauge field rather than merely a static correlated hopping. 
Raman-assisted tunneling produces a complex intra-cell hopping $(\alpha+i\beta\sin\Omega t)$~\cite{PhysRevLett.114.125301,PhysRevLett.109.145301}, while the on-site interaction is modulated as $[U+(-1)^jU_0\cos(\Omega t)]/2$ via a Feshbach resonance~\cite{PhysRevLett.90.230401,WinklerK2006,RevModPhys.82.1225,ClarkLoganW2017}. 
As shown in the Supplemental Material~\cite{2026supplemental}, the leading-order Floquet expansion yields an effective intra-cell tunneling
$\gamma + \frac{\beta U_0}{2\Omega}(\hat n_{2j-1}+\hat n_{2j})$
with $\gamma=\alpha$, arising from the commutator between the fast tunneling and interaction drives. 
To realize the form $\gamma_0\sin\phi(t)$ in Eq.~\eqref{Bosonmodel}, the Floquet-induced amplitude $\beta U_0/(2\Omega)$ can be synchronized with the slow pumping cycle by modulating either $\beta$ or $U_0$ at frequency $\omega$, such that $\beta U_0/(2\Omega)=\gamma_0\sin\phi(t)$.

Unlike previous realizations of density-dependent hopping~\cite{2009PhysRevLett.103.133002,2021PhysRevLett.109.203005,2014PhysRevA.89.013624,2014PhysRevLett.113.215303,2016PhysRevLett.116.205301}, the occupation-dependent link here is synchronized with the pumping cycle. 
The resulting dynamical gauge field reshapes the doublon trajectory, inducing topology unique to the doublon sector and enabling topology-enabled doublon pumping and counter-propagating responses, observable via site-resolved center-of-mass measurements.

\noindent\emph{{Conclusion and Discussion}.---}
We have shown that promoting tunneling to a dynamical, occupation-conditioned variable enables a qualitatively new regime of topological transport beyond the single-particle paradigm. 
In this setting, density-dependent hopping acts as a dynamical gauge field that reshapes the effective band structure of bound states, allowing them to acquire independent topological invariants and exhibit quantized pumping responses distinct from those of single particles. 
This mechanism establishes that topological transport in driven systems need not be universally shared across occupancy sectors, but can instead be intrinsically occupation resolved. 
More broadly, our results highlight dynamical gauge fields as a powerful tool for engineering many-body topology and controlling geometric responses in periodically driven systems. 

While the main text has focused on the minimal two-particle setting, the underlying mechanism is more general: as shown in the Supplemental Material~\cite{2026supplemental}, higher-occupancy bound states such as triolons can likewise acquire distinct topological phase structures and quantized pumping responses under the same driving cycle. 
This suggests a systematic route toward a hierarchy of occupation-selective topological transport across different particle-number sectors beyond the single-particle paradigm.

\begin{acknowledgments}
W.L. is supported by the Fundamental Research Funds for the Central Universities, Dalian University of Technology [DUT25RC(3)084]. C.H.L. and Z.L. acknowledge support from the Singapore Ministry of Education Tier II. grants MOE-T2EP50224-0007 and MOE-T2EP50224-0021 (WBS nos. A-8003505-01-00 and A-8003910-00-00).
\end{acknowledgments}


%


\clearpage

\onecolumngrid

\section*{Supplemental Materials: Occupation-selective topological pumping from Floquet gauge fields}

\pagestyle{plain}

\setcounter{page}{1}   

\setcounter{secnumdepth}{3}
\setcounter{tocdepth}{2}

\setcounter{section}{0}
\setcounter{equation}{0}
\setcounter{figure}{0}
\setcounter{table}{0}

\renewcommand{\thesection}{S\arabic{section}}
\renewcommand{\theequation}{S\arabic{equation}}
\renewcommand{\thefigure}{S\arabic{figure}}
\renewcommand{\thetable}{S\arabic{table}}


\section{Two-Body Spectrum and Band Isolation for well-defined topology} \label{originalspectrum}

\begin{figure}[htp]
\center
\includegraphics[width=0.5\textwidth]{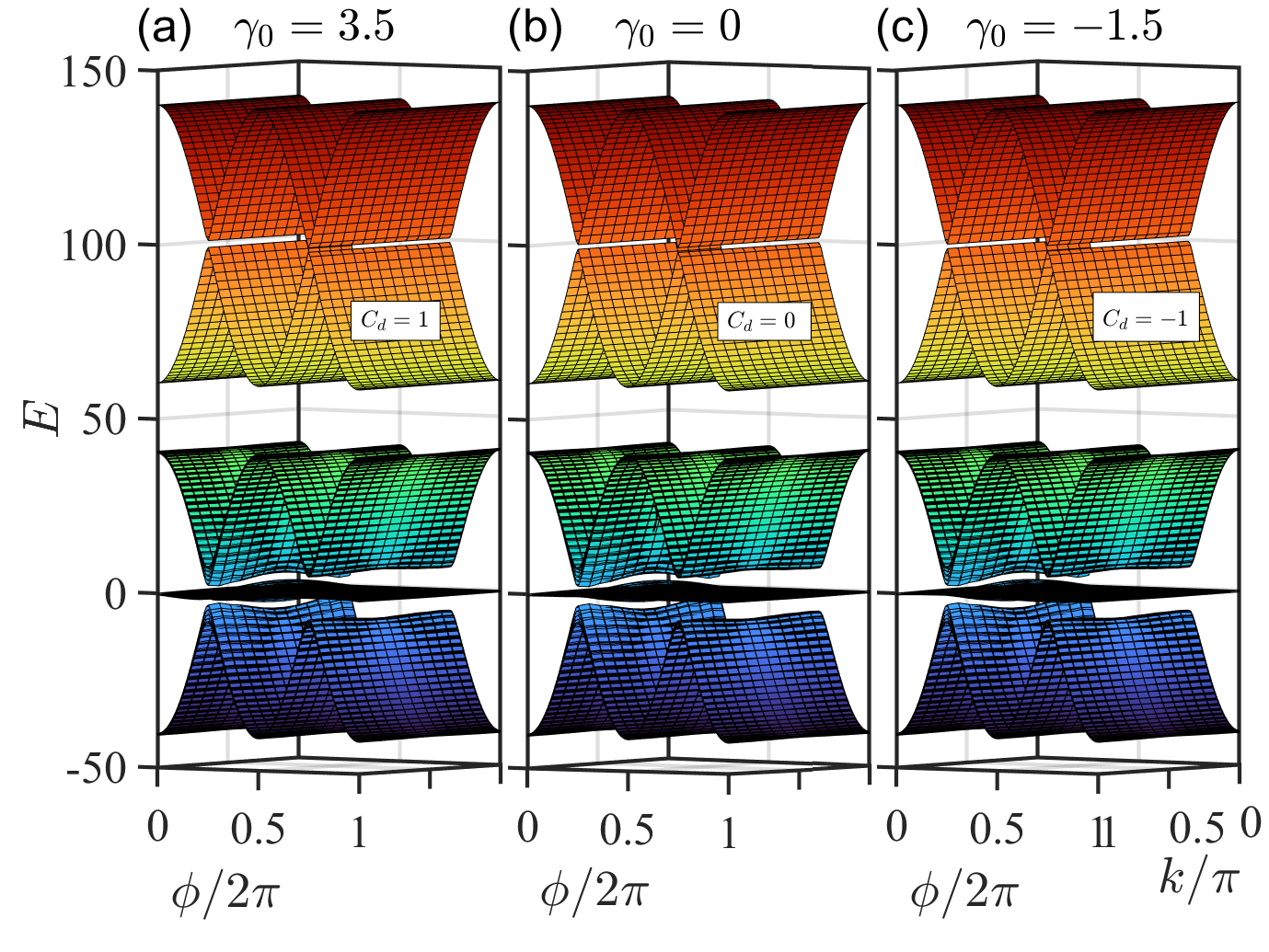}
\caption{\textbf{Full two-particle energy spectrum underlying the topological analysis in the main text.}
Panels (a)–(c) correspond to different values of the density-dependent hopping amplitude $\gamma_0$, which acts as the strength of the dynamical gauge-field–mediated tunneling. 
Throughout the driving cycle, the scattering continuum remains clearly separated from two isolated bound-state bands. 
We focus on the lower bound band and denote its Chern number by $C_d$, as indicated in each panel. 
Tuning $\gamma_0$ modifies the effective intra-cell tunneling of the doublon and drives a topological transition, with $C_d$ changing from $+1$ to $-1$ as $\gamma_0$ varies from $3.5$ to $-1.5$.
\label{fig:originalspectrum}}
\end{figure}

In the main text, the topology of the doublon bands is characterized by the Berry curvature, polarization, and adiabatic energy gaps as shown in Fig.2. 
A prerequisite for such a band-topological description is the existence of spectrally isolated two-body bound-state bands. 
In this section, we present the full two-particle energy spectrum in Fig.~\ref{fig:originalspectrum} and demonstrate the clear separation of the doublon bands from the scattering continuum. 
This spectral isolation ensures that the Chern numbers and polarization defined in the main text are well posed within the bound-state manifold and justifies the band-topological treatment.

With periodic boundary conditions, the cotranslational operator $\hat{T}_{q}$ is described by
\begin{equation}
\begin{aligned}
&\hat{T}_{q}\left|n_{1}, n_{2}, \ldots, n_{N}\right\rangle \\
&=\left|n_{N-q+1}, \ldots, n_{N}, n_{1}, \ldots, n_{N-q)}\right\rangle
\end{aligned}
\end{equation}
where $\left|\left\{n_{j}\right\}\right\rangle=\left|n_{1}, n_{2}, \ldots, n_{N}\right\rangle$ is a Fock state and
$n_{j}$ is the atom number at site $j$.
Given $\hat{T}_{q}^{-1} \hat{H}\hat{T}_{q}=\hat{H}$, the quasimomentum $k$ of the center of mass is a good quantum number.
The single and multiparticle shift operator~\cite{2003PhysRevE.68.056213} guarantee an orthogonal basis
\begin{equation}
|k, \mathbf{n}\rangle=\frac{1}{\sqrt{M}} \sum_{j=0}^{M-1} \exp (ik q j) \hat{T}_{q}^{j}|\mathbf{n}\rangle
\end{equation}
where $\hat{T}_{q}^{j}$ denotes an operator for shifting $j$ cells.
$|\mathbf{n}\rangle$ is the seed state, $M(\leqslant N/q)$ is the obtained state number after repeatedly applying the shift operator $\hat{T}_{q}$ to the seed state.
A series of Fock states $\left\{|\mathbf{n}\rangle, \hat{T}_{q}|\mathbf{n}\rangle, \ldots, \hat{T}_{q}^{M-1}|\mathbf{n}\rangle\right\}$ construct a translational period for the corresponding quasimomentum $k=2 \pi l /(q M)$ with $l=0,1, \ldots, M-1$.
The matrix element becomes
\begin{equation}
\begin{aligned}
\left\langle k^{\prime}, \mathbf{n}^{\prime}|\hat{H}| k, \mathbf{n}\right\rangle &=\left\langle k^{\prime}, \mathbf{n}^{\prime}\left|\hat{T}_{q}^{-1} \hat{H}_{B} \hat{T}_{q}\right| k, \mathbf{n}\right\rangle \\
&=e^{i\left(k^{\prime}-k\right) q}\left\langle k^{\prime}, \mathbf{n}^{\prime}|\hat{H}| k, \mathbf{n}\right\rangle
\end{aligned}
\end{equation}
in the basis ${|k, \mathbf{n}\rangle}$.
As we can see, for $k \neq k^{\prime} $ the matrix element is equal to zero which means the subspaces with different quasimomenta decouple with each other.
As such, the Hamiltonian can be written in the form $\hat{H}=\oplus_{j=1}^{N/q} \hat{H}\left(k_{j}\right)$, where
$\hat{H}\left(k_{j}\right)=\left\langle k_{j}, \mathbf{n}^{\prime}|\hat{H}| k_{j}, \mathbf{n}\right\rangle$.
%

The Hilbert space dimension of model (1) in the main text is
\begin{equation}
\mathcal{D}=\frac{(L+N-1)!}{L!(N-1)!}
\end{equation}
where $L$ indistinguishable bosons are distributed over $N$ lattice sites with a unit cell of size $q$.
When $L$ and $N/q$ are coprime, each seed state returns to itself after $N/q$ applications of the shift operator. Consequently, for each quasimomentum $k$, the Hamiltonian block $\hat{H}(k_j)$ is reduced to a uniform dimension $\mathcal{D}q/N$. Without loss of generality, we assume $L$ and $N/q$ to be coprime so that all multiparticle Bloch bands have identical dimension.
In the main text, we focus on a two-site unit cell $q=2$ with two bosons $L=2$.

By solving the eigenvalue equation
\begin{equation}\label{EigenFuction}
\hat{H}(k)\ket{\psi_{m}(k)}
=
E_{m}(k)\ket{\psi_{m}(k)},
\end{equation}
we obtain the multiparticle Bloch bands. The corresponding eigenstates can be written as
\begin{equation}
\begin{aligned}
\ket{\psi_{m}(k)}
&=\sum_{\mathbf{n}} \psi_{m}(k, \mathbf{n})\ket{k, \mathbf{n}} \\
&=\frac{1}{\sqrt{N/q}}
\sum_{j, \mathbf{n}}
\psi_{m}(k, \mathbf{n})
\exp (i k q j)\,
T_{q}^{j}\ket{\mathbf{n}}
\end{aligned}
\end{equation}
which represents the multiparticle Bloch state in the $m$th band with quasimomentum $k$ and eigenvalue $E_m(k)$.
The topological properties of the $m$th band are characterized by the Chern number defined over the two-dimensional parameter space,
\begin{equation}\label{ChernNumber}
C
=
\frac{1}{2 \pi}
\int_{0}^{2 \pi / q} \mathrm{d}k
\int_{0}^{2\pi} \mathrm{d}\phi \,
\mathcal{F}_{m}(\phi,k)
\end{equation}
where the Berry curvature is
\begin{equation}
\mathcal{F}_m
=
i\Big(
\langle \partial_{\phi} \psi_{m} | \partial_{k} \psi_{m}\rangle
-
\langle \partial_{k} \psi_{m} | \partial_{\phi} \psi_{m}\rangle
\Big).
\end{equation}

\medskip

The multiparticle Bloch states of the $m$th band span the occupied subspace, which is described by the projector
\begin{equation}
\hat{P}
=
\sum_{k}
\ket{ \psi_m(k) }
\bra{ \psi_m(k) }.
\end{equation}
To construct localized Wannier states in a numerically stable manner, we introduce the unitary position operator
\begin{equation}
\hat{X}
=
e^{i \delta_k \hat{x}}
\end{equation}
where the many-body position operator is
\begin{equation}
\hat{x}
=
\sum_{j=1}^{N}
\frac{j \hat{n}_j}{L},
\qquad
\delta_k = \frac{2\pi}{N}.
\end{equation}
Projecting $\hat{X}$ onto the occupied subspace yields
\begin{equation}
\hat{X}_P
=
\hat{P}\,\hat{X}\,\hat{P}.
\end{equation}
The eigenstates of the projected operator $\hat{X}_P$ define the many-body Wannier states~\cite{1982PhysRevB.26.4269,1997PhysRevB.56.12847,2015PhysRevB.91.085119}. Since $\hat{X}_P$ acts within the $m$th band subspace,
it possesses $N/q$ eigenvalues $\{\lambda_n\}$, whose phases encode the spatial centers of the Wannier states.
Importantly, the many-body polarization can be directly extracted from these eigenvalues. Specifically, it is given by
\begin{equation}
P = \frac{1}{2\pi}\,\mathrm{Im}\,
\ln \left( \prod_{n=1}^{N/q} \lambda_n \right).
\end{equation}
This expression identifies the bulk polarization with the 
gauge-invariant Wilson-loop phase encoded in the spectrum of 
the projected position operator.


Figure~\ref{fig:originalspectrum} displays the full two-particle spectrum as a function of the driving phase $\phi$ and the center-of-mass quasimomentum $k$ for representative values of the density-dependent hopping amplitude $\gamma_0$. 
In the strong-interaction regime ($U=100$), two well-defined bound-state bands emerge above the scattering continuum and remain separated from it by a sizable energy gap throughout the pumping cycle. 
All other parameters are fixed at $J=1$, $\Delta_0=20$, $\delta_0=0.5$, and $\gamma=2$. 
Without loss of generality, we focus on the lower-energy bound-state band and denote its topological invariant by $C_d$. 
This clear spectral isolation ensures that the bound-state manifold forms a closed and gapped subspace in which the Bloch states, projectors, Berry curvature, and polarization defined above are well posed. 
Consequently, the Chern numbers and polarization windings computed in the main text can be unambiguously attributed to the isolated doublon band. 
While the energy spectrum establishes the existence of the bound states, the topological properties themselves are encoded in the geometric quantities—Berry curvature and Wilson-loop polarization—introduced above, which provide the direct diagnosis of occupation-selective topology.

\section{Derivation and Validation of the Effective Doublon Hamiltonian} \label{Effectiveness}

\begin{figure}[htp]
\center
\includegraphics[width=0.7\textwidth]{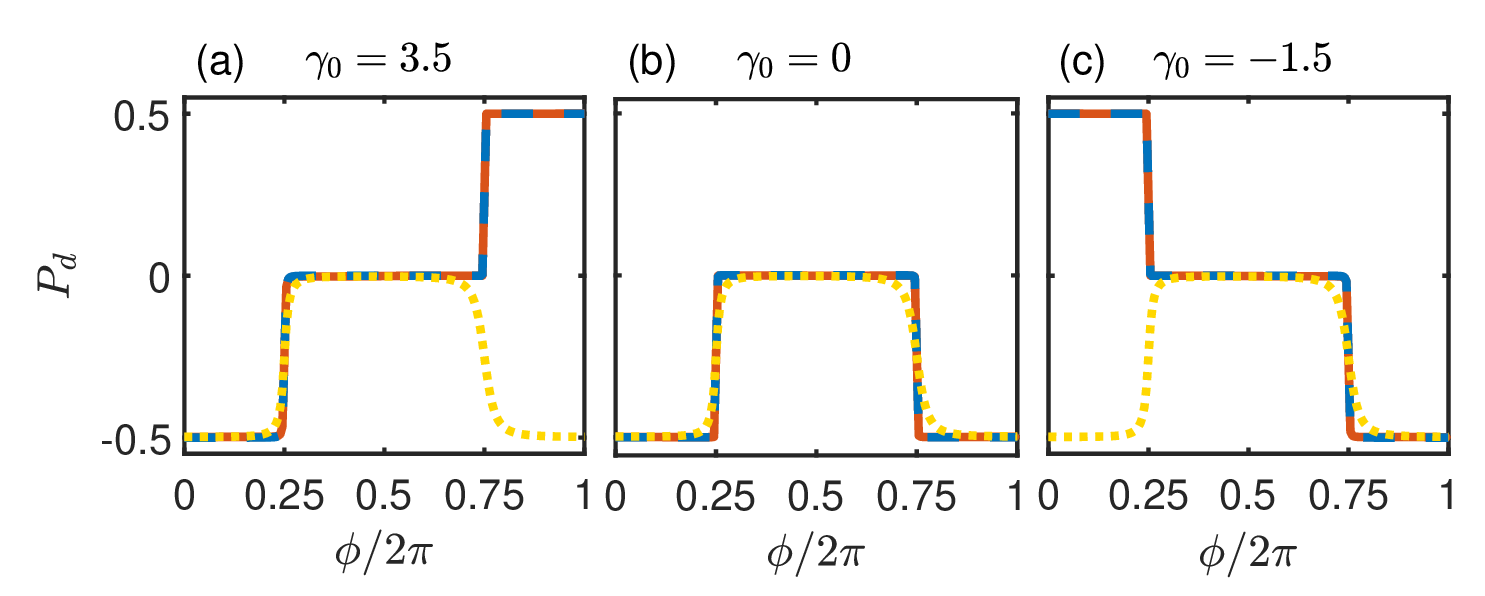}
\caption{\textbf{Validation of the effective doublon Hamiltonian via polarization dynamics.}
Evolution of the polarization $P_d$ of the lower doublon band for different values of the density-dependent hopping amplitude $\gamma_0$. 
Blue dashed lines show the results obtained from the effective single-particle doublon Hamiltonian [Eq.~(5) in the main text], while red solid lines correspond to the full two-particle many-body model [Eq.~(1) in the main text]. 
Yellow dotted lines denote the single-particle polarization computed from the original Hamiltonian [Eq.~(1) in the main text] for comparison. 
Panels (a)–(c) correspond to $\gamma_0 = 3.5$, $0$, and $-1.5$, respectively.
The excellent agreement between the effective and full two-particle results confirms that the derived Hamiltonian accurately captures the doublon band topology in the strong-interaction regime, including the occupation-selective topological transitions absent in the single-particle sector. 
\label{fig:comparisonspectrum}}
\end{figure}

In the strong-interaction regime $U \gg (J,\delta_0,\gamma,\gamma_0,\Delta_0)$, we focus on the two-particle dynamics dominated by on-site doublon states. 
In the main text, we introduced an effective single-particle Hamiltonian $\hat H_{\mathrm{eff}}$ [Eq.~(5)] to explain the dynamical-gauge-field–induced topological phase transitions shown in Fig.3. 
In this section, we provide the derivation of $\hat H_{\mathrm{eff}}$ using degenerate perturbation theory and explicitly obtain the effective tunneling amplitudes and energy shifts generated by virtual processes. 
We further validate the effective description by comparing its polarization with that of the original two-particle model (Fig.~\ref{fig:comparisonspectrum}), thereby establishing its reliability for analyzing doublon band topology.

We separate the Hamiltonian~[Eq.(1)] in the main text into two parts, with the dominant part
\begin{equation} \label{Dominantpart}
\hat{H}_{\text {d}}=\frac{U}{2} \sum_{j} \hat{n}_{j}\left(\hat{n}_{j}-1\right)
\end{equation}
and the perturbation part
\begin{equation}\label{Perturpart}
\begin{aligned}
& \hat{H}_{\mathrm{p}}=\sum_j^N\left[J+\delta_0 \sin (\pi j+\phi(t))\right] \hat{a}_j^{\dagger} \hat{a}_{j+1}+\text { H.c. } \\
& +\sum_j^{N / 2} \hat{a}_{2 j-1}^{\dagger}\left(\gamma+\gamma_0 \sin \phi(t) \hat{n}_{2 j-1}+\gamma_0 \sin \phi(t) \hat{n}_{2 j}\right) \hat{a}_{2 j}+\text { H.c. } \\
& +\sum_j^N \Delta_0 \cos (\pi j+\phi(t)) \hat{n}_j.
\end{aligned}
\end{equation}
The dominant $\hat{H}_{\text {d}}$ can be divided into two degenerate subspaces $\mathcal{U}$ and $\mathcal{V}$.
The subspace $\mathcal{U}\equiv\left\{|2\rangle_{j}\right\}$ contains bound states which two particles populate on the same site, and its degenerate energy is $E_{j}=U$.
The subspace $\mathcal{V}\equiv\left\{|1\rangle_{j}|1\rangle_{k}\right\} $ contains scattering states which two particles populate on different sites, and its degenerate energy is $E_{j k}=0$ with $j\neq k$.
The projection operators onto subspaces $\mathcal{U}$ and $\mathcal{V}$ are written as
\begin{equation}
\hat{P}=\sum_{j}|2\rangle_{j}\langle 2|_{j}
\end{equation}
and
\begin{equation}
\hat{S}=\sum_{j \neq k} \frac{1}{E_{j}-E_{jk}}|1\rangle_{j}|1\rangle_{k}\langle 1|_{k}\langle 1|_{j}.
\end{equation}
Perturbation theory~\cite{1977JPC101289,2011AOP3262793} gives the effective single-particle model in the subspace $\mathcal{U}$ as
\begin{equation}
\hat{H}_{\mathrm{eff}}=\hat{h}_{0}+\hat{h}_{1}+\hat{h}_{2}=E_{j} \hat{P}+\hat{P} \hat{H}_{\mathrm{p}} \hat{P}+\hat{P} \hat{H}_{\mathrm{p}} \hat{S} \hat{H}_{\mathrm{p}} \hat{P}.
\end{equation}
At zeroth-order perturbation, it yields
\begin{equation}\label{ZeroTerm}
\hat{h}_{0}=E_{j} \hat{P}=U\sum_{j}|2\rangle_{j}\langle 2|_{j}.
\end{equation}
At the first-order perturbation level, it is
\begin{equation} \label{FirstTerm}
\begin{aligned}
& \hat{h}_1=\hat{P} \hat{H}_{\mathrm{p}} \hat{P}=\sum_{j=1}^N 2 \Delta_0 \cos (\pi j+\phi(t))|2\rangle_j\langle 2|_j.
\end{aligned}
\end{equation} %
And the second-order perturbation expression is
\begin{equation} \label{SecondTerm}
\begin{aligned}
\hat{h}_2 & =\hat{P} \hat{H}_{\mathrm{p}} \hat{S} \hat{H}_{\mathrm{p}} \hat{P} \\
& =\sum_j^{N}\left[\mathcal{J}_1+\mathcal{J}_2 \frac{1-\cos (\pi j)}{2}\right]|2\rangle_j\langle 2|_{j+1}+\text { H.c. } \\
& +\sum_{j=1}^N(\mu-U)|2\rangle_j\langle 2|_j.
\end{aligned}
\end{equation}
Summing Eqs.~\eqref{ZeroTerm}, ~\eqref{FirstTerm} and~\eqref{SecondTerm}, we obtain the effective single-particle model as
\begin{equation}
\begin{aligned}
\hat{H}_{\text {eff }} & =\sum_j^{N}\left[\mathcal{J}_1+\mathcal{J}_2 \frac{1-\cos (\pi j)}{2}\right] \hat{b}_j^{\dagger} \hat{b}_{j+1}+\text { H.c. } \\
& +\sum_{j=1}^N 2 \Delta_0 \cos (\pi j+\phi(t)) \hat{b}_j^{\dagger} \hat{b}_j+\mu \sum_{j=1}^N \hat{b}_j^{\dagger} \hat{b}_j.
\end{aligned}
\end{equation}
Here, the parameters are defined as $\mathcal{J}_1=\frac{2}{U}\left[J+\delta_0 \sin (\pi j+\phi(t))\right]^2$, $\mathcal{J}_2=\frac{4}{U}\left(\gamma+\gamma_0 \sin \phi(t)\right)\left(J-\delta_0 \sin \phi(t)\right)+\frac{2}{U}\left(\gamma+\gamma_0 \sin \phi(t)\right)^2$ and $\mu=U+\frac{2}{U}\left(J-\delta_0 \sin \phi(t)\right)^2+\frac{2}{U}\left(J+\delta_0 \sin \phi(t)\right)^2+\mathcal{J}_2$.
$\hat{b}_{j}^{\dagger}$ creates two particles at the $j$th site simultaneously, that is, $\hat{b}_{j}^{\dagger}|\mathbf{0}\rangle=|2\rangle_{j}$.
$\mu$ is a constant and makes no contribution to the topological properties.

We use Fourier transformation into momentum space via
\begin{equation}\label{FFT}
\begin{array}{l}
\hat{b}_{2 j}^{\dagger}=\frac{1}{\sqrt{N/q}} \sum_{k} e^{i k 2 j} \hat{b}_{k, \mathrm{e}}^{\dagger}, \\
\hat{b}_{2 j-1}^{\dagger}=\frac{1}{\sqrt{N/q}} \sum_{k} e^{i k(2 j-1)} \hat{b}_{k, \mathrm{o}}^{\dagger},
\end{array}
\end{equation}
to obtain the eigenvalues and eigenstates of the bound-state bands. The subscript
$\mathrm{o}$ ($\mathrm{e}$) indicates odd (even) sites.
$k$ is the quasimomentum and
thereby the effective Hamiltonian [Eq.(5) in the main text] can be decomposed as $\hat{H}_{\text {eff }}(t)=\sum_{k} \hat{H}_{\text {eff }}(k,t)$ in the quasimomentum space.
Each $\hat{H}_{\text {eff }}(k,t)$ turns to a two-level quantum system
\begin{equation}\label{TwoBandHamiltonian}
\hat{H}_{\text {eff }}(k, t)=h_{x} \hat{\sigma}_{x}+h_{y} \hat{\sigma}_{y}+h_{z} \hat{\sigma}_{z}+\mu
\end{equation}
where
\begin{equation}
\left(\begin{array}{l}
h_x \\
h_y \\
h_z
\end{array}\right)=\left(\begin{array}{l}
\left\{\frac{4}{U}\left[J^2+\left(\delta_0 \sin \phi(t)\right)^2\right]+\mathcal{J}_2\right\} \cos k \\
\left(\frac{8}{U} J \delta_0 \sin \phi(t)-\mathcal{J}_2\right) \sin k \\
2 \Delta_0 \cos \phi(t)
\end{array}\right).
\end{equation}
The eigenequation $\hat{H}_{\mathrm{eff}}(k, t)|u(k, t)\rangle=\varepsilon(k, t)|u(k, t)\rangle$ is solved with the eigenvalue
\begin{equation}\label{TwoBoundBand}
\varepsilon_{\pm} =\pm \sqrt{h_{x}^{2}+h_{y}^{2}+h_{z}^{2}}+\mu
\end{equation}
and eigenstate
\begin{equation}
\left|u_{\pm}(k, t)\right\rangle=\left(\begin{array}{c}
\frac{h_{x}-i h_{y}}{\varepsilon_{\pm}-h_{z}} \\
1
\end{array}\right).
\end{equation}

To evaluate the validity of the effective Hamiltonian~[Eq.(5) in the main text], Fig.~\ref{fig:comparisonspectrum} exhibits its polarization which shows excellent agreement with the original Hamiltonian [Eq.(1) in the main text] in the strong-interaction regime, and the parameters are consistent with Fig.2 in the main text. 
The effective single-particle model [Eq.(5) in the main text] proves to be highly accurate in describing the two-body bound bands.
The polarization of the effective model is shown with blue dashed lines in Fig.~\ref{fig:comparisonspectrum}.
The red solid line and yellow dotted line correspond to the polarization results of the original model in the two-particle case and single-particle case, respectively.
The remarkable agreement between the results of the original model and those of the effective single-particle model confirms that the effective model accurately captures the essential physics of the two-particle bound states in the strong-interaction limit.
Furthermore, the polarization analysis reveals that the effective model preserves the key feature of occupation-selective topology: when the single-particle polarization change is zero, the two-particle system can still exhibit a non-zero polarization change, demonstrating that the topological properties arise from the particle occupation mechanism rather than being inherited from the single-particle bands. This comprehensive validation establishes the reliability of our effective single-particle approach for studying the topological behavior of interacting bound states.

\section{Adiabaticity and Gap-Adapted Driving Protocol} \label{Adiabaticcriteria}

\begin{figure}[htp]
\center
\includegraphics[width=0.6\textwidth]{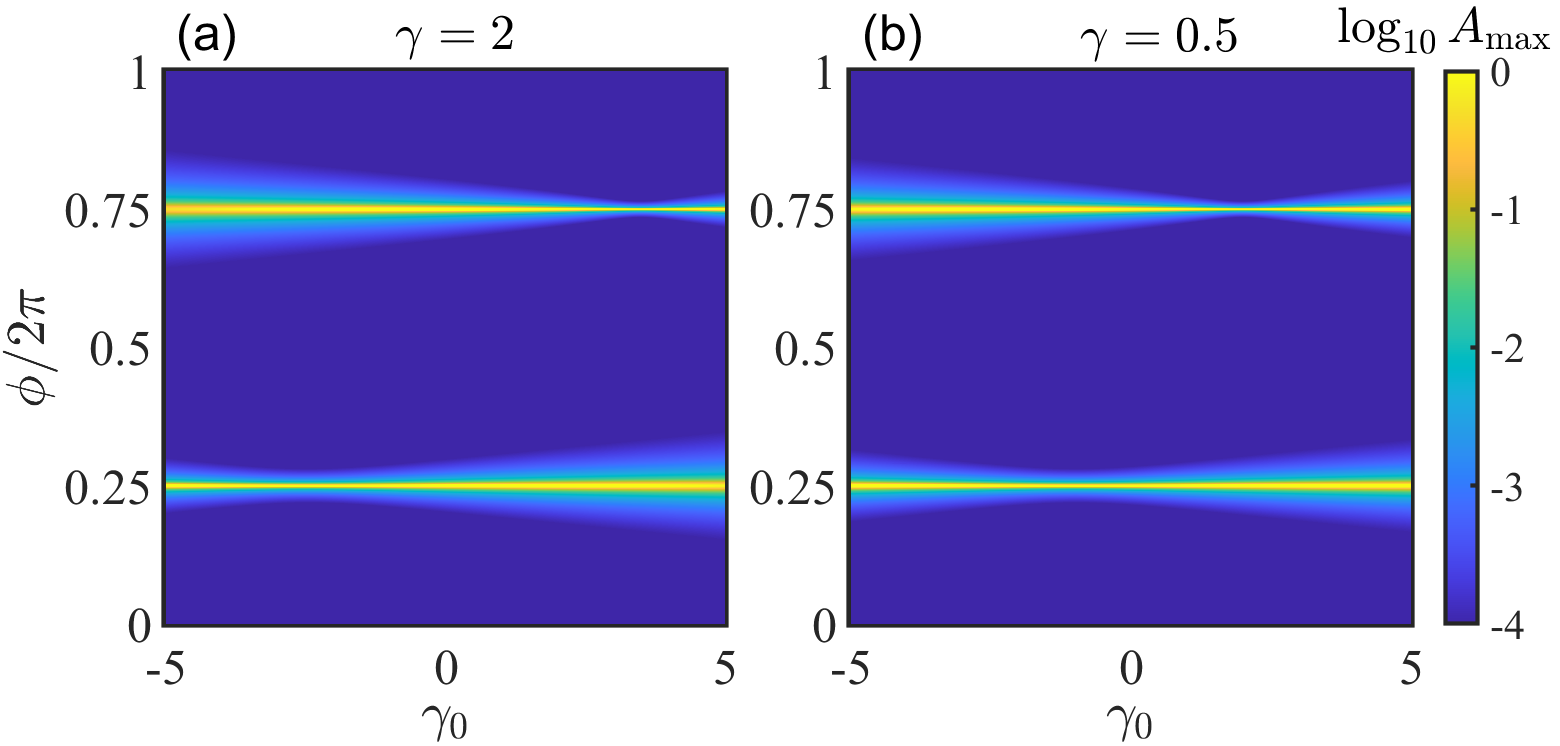}
\caption{\textbf{Adiabaticity landscape of the doublon bound bands.}
Maximum adiabatic criterion 
$A_{\max} = \max_{k} \left[ \hbar \left| \langle \psi_{\mathrm{I}}(k) | \partial_t \psi_{\mathrm{II}}(k) \rangle / (E_{\mathrm{I}}(k) - E_{\mathrm{II}}(k)) \right| \right]$
in the $(\phi,\gamma_0)$ plane for (a) $\gamma=2$ and (b) $\gamma=0.5$, 
shown on a logarithmic color scale.
Here $A_{\max}$ is obtained by maximizing the adiabatic criterion over the entire Brillouin zone, thus capturing the worst-case nonadiabatic coupling at each $(\phi,\gamma_0)$.
In both cases, strong nonadiabatic hotspots are sharply localized near 
$\phi=\pi/2$ and $3\pi/2$, where the instantaneous gap between the doublon bands becomes minimal.
The overall structure is nearly identical between the two panels, indicating that the dominant adiabatic bottlenecks are governed primarily by transient gap narrowing, rather than by the underlying single-particle topology.
Small asymmetries in the peak profiles along the $\gamma_0$ direction can be resolved in the logarithmic scale, reflecting residual $\gamma_0$-dependent modifications of the doublon band structure.
These gap constrictions lead to strong nonadiabatic coupling independent of the
topological phase, motivating the gap-adapted nonlinear driving protocol
employed in the main text and described here.
\label{fig:adiabaticcriteria}}
\end{figure}

In the main text, we adopt a gap-resolved nonlinear driving protocol,
$\omega=\eta(\Delta_{\min})^{2}$ in Eq.~(6), to ensure high-fidelity
adiabatic pumping of the doublon band.
The purpose of this section is to quantitatively justify this choice.
By evaluating the adiabatic criterion between the two bound-state bands and mapping
its maximal value over the parameter space $(\phi,\gamma_0)$
(Fig.~\ref{fig:adiabaticcriteria}), we demonstrate that pronounced peaks of
non-adiabatic coupling universally emerge near $\phi=\pi/2$ and $3\pi/2$
due to transient gap narrowing, independent of the topological character of the system.
A constant-rate drive would therefore induce significant Landau--Zener transitions at these
critical phases.
In contrast, the gap-adapted protocol automatically slows down the evolution where the
gap becomes small, thereby suppressing interband excitations and ensuring the quantized
pumping dynamics reported in Fig.4 in the main text.

When considering two bound bands $E_{\mathrm{I}}(k)$ and $E_{\mathrm{II}}(k)$, their adiabatic criteria are evaluated using the formula
\begin{equation}
A = \hbar \left| \frac{\langle \psi_{\mathrm{I}}(k) \mid \partial_t \psi_{\mathrm{II}}(k) \rangle}{E_{\mathrm{I}}(k) - E_{\mathrm{II}}(k)} \right| \ll 1
\end{equation}
where $|\psi_{\mathrm{I}}(k)\rangle$ and $|\psi_{\mathrm{II}}(k)\rangle$ are the instantaneous eigenstates corresponding to these two energy bands~\cite{1981QuantumMechanics,2015PhysRevA92043406}.
The maximal adiabatic criterion is shown in Fig.~\ref{fig:adiabaticcriteria} for two representative values of $\gamma$. 
At each $(\phi,\gamma_0)$, we evaluate the adiabatic criterion over the full Brillouin zone and take the maximum value $A_{\max} = \max_k A(k)$, which characterizes the worst-case nonadiabatic coupling between the two doublon bands.
All other parameters are fixed at $J=1$, $\Delta_0=20$, $\delta_0=0.5$ and $U=100$. 
As seen from Figs.~\ref{fig:adiabaticcriteria}(a) and (b), the dominant nonadiabatic hotspots are sharply localized near $\phi=\pi/2$ and $3\pi/2$ across the entire $\gamma_0$ range. 
Remarkably, the overall structure is nearly identical for both $\gamma=2$ and $\gamma=0.5$, 
demonstrating that the primary adiabatic bottlenecks are governed by transient gap narrowing in the doublon spectrum, 
rather than by the underlying single-particle topology.
At the same time, the logarithmic color scale reveals subtle $\gamma_0$-dependent asymmetries in the peak profiles, 
which reflect residual modifications of the doublon band structure induced by the dynamical gauge field. 
These effects do not shift the location of the nonadiabatic hotspots, but slightly reshape their intensity distribution.

To suppress non-adiabatic transitions at these vulnerable points, we implement a nonlinear driving protocol in which the effective driving rate is adapted to the local bandgap.  
Specifically, we set $\omega = \eta (\Delta_{\min})^2$, where $\Delta_{\min}$ denotes the minimum energy gap between the two bound-state bands after scanning the full Brillouin zone in $k$.
The dimensionless constant is chosen as $\eta = 0.01$ in Fig.4 in the main text.
This gap-resolved scheme effectively slows down the time evolution near constrictions in the spectrum, thereby minimizing Landau-Zener tunneling and ensuring high-fidelity adiabatic pumping.

\section{Extended Dynamical Results for Occupation-Selective Pumping}

\begin{figure*}[htp]
\center
\includegraphics[width=1\textwidth]{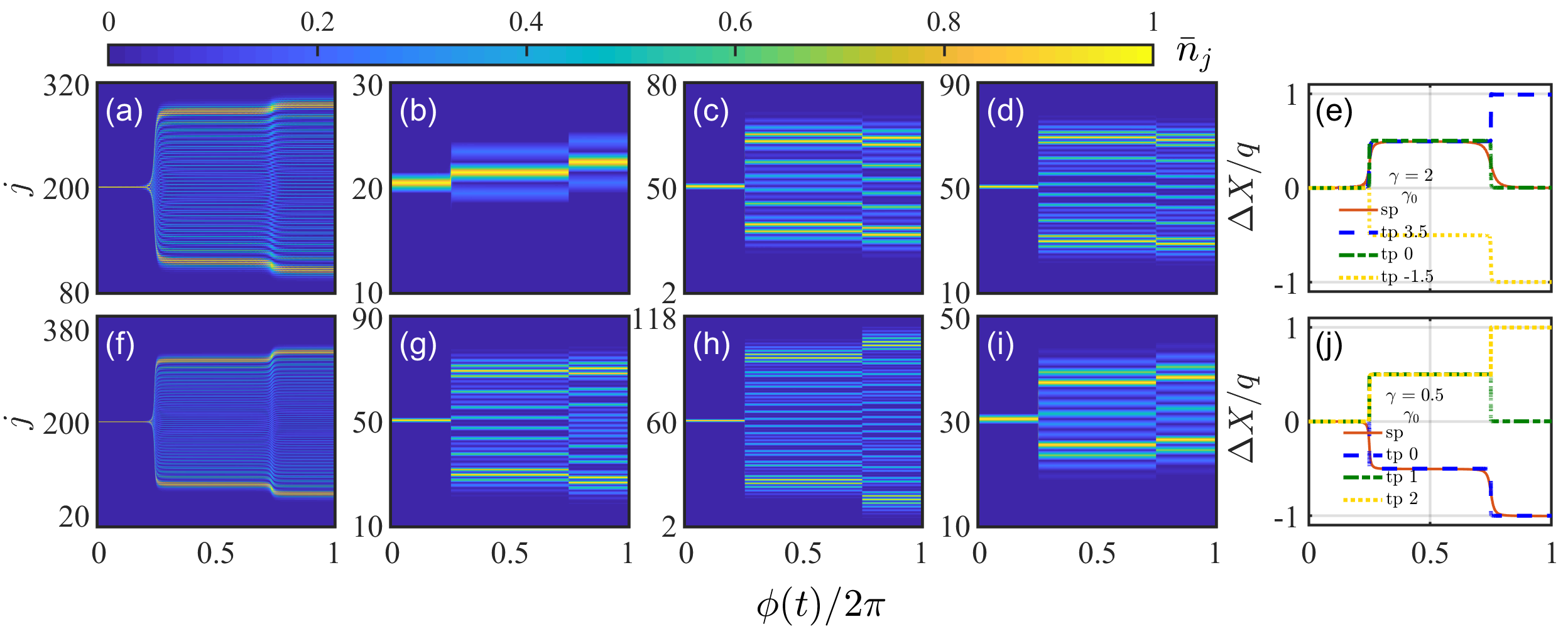}
\caption{
\textbf{Extended dynamical results for occupation-selective topological pumping.}
Topological quantum pumping of single- and two-particle states under density-dependent dynamical gauge fields for multiple parameter choices.
(a)--(d) and (f)--(i): Spatiotemporal evolution of the normalized density distribution $\bar{n}_j = n_j / n_j^{\max}$ over one pumping cycle.
The first row ($\gamma = 2$) corresponds to a topologically trivial single-particle pump:
(a) single-particle dynamics;
(b)--(d) doublon dynamics for $\gamma_0 = 3.5$, $0$, and $-1.5$, respectively.
The second row ($\gamma = 0.5$) corresponds to a topologically nontrivial single-particle pump:
(f) single-particle dynamics;
(g)--(i) doublon dynamics for $\gamma_0 = 0$, $1$, and $2$, respectively.
(e) and (j): Time evolution of the center-of-mass displacement $\Delta X(t)/q$.
The results extend those shown in the main text to a broader set of parameters and illustrate the persistence of quantized pumping in the doublon sector.
\label{fig:DensityEvolutionsm}}
\end{figure*}

In the main text, we highlighted two representative regimes that capture the essential physics of occupation-selective topological pumping: (i) topology-enabled doublon transport in a single-particle trivial pump, and (ii) counter-propagating pumping between single- and double-occupancy sectors. Here we present extended dynamical results over a broader set of parameters. Figure~\ref{fig:DensityEvolutionsm} complements the main-text results by showing the density evolution for multiple values of the density-dependent hopping amplitude $\gamma_0$ for different choices of the static intra-cell hopping $\gamma$. These results demonstrate that the pumping behaviors highlighted in the main text persist over a wider parameter range.

Fig.~\ref{fig:DensityEvolutionsm} presents a systematic exploration of topological quantum pumping across multiple parameter regimes, extending the representative cases shown in the main text. The dynamics are organized into two blocks corresponding to distinct single-particle topologies: $\gamma = 2$ (first row), where the single-particle system is topologically trivial, and $\gamma = 0.5$ (second row), where it exhibits a nontrivial Chern number.
All other parameters are fixed at $J=1$, $\Delta_0=20$, $\delta_0=0.5$ and $U=100$. 
For $\gamma = 2$, panel (a) shows the single-particle density evolution, which returns to its initial position after one cycle—consistent with zero net displacement ($\Delta X/q = 0$). Panels (b)–(d) display the two-particle pumping for $\gamma_0 = 3.5$, $0$, and $-1.5$, respectively. Their center-of-mass displacements, summarized in panel (e), reveal a quantized response that directly maps onto the many-body topological phase diagram: 
the doublon shifts rightward by one unit cell ($\Delta X/q = +1$) for $\gamma_0 = 3.5$ ($C_d = +1$), remains stationary ($\Delta X/q = 0$) for $\gamma_0 = 0$ ($C_d = 0$), and shifts leftward ($\Delta X/q = -1$) for $\gamma_0 = -1.5$ ($C_d = -1$).
For $\gamma = 0.5$, the single-particle pump becomes nontrivial, as seen in panel (f) and the red solid line in panel (j) ($\Delta X/q = -1$). Remarkably, the two-particle transport—shown in panels (g)–(i) for $\gamma_0 = 0$, $1$, and $2$—again follows the same $\gamma_0$-dependent pattern: 
leftward shift ($\Delta X/q = -1$) at $\gamma_0 = 0$, no transport ($\Delta X/q = 0$) at $\gamma_0 = 1$, and rightward shift ($\Delta X/q = +1$) at $\gamma_0 = 2$. 

These results show that the many-body pumping response of the doublon is consistently controlled by the density-dependent hopping amplitude $\gamma_0$, irrespective of the underlying single-particle behavior. The quantized center-of-mass displacements observed across different parameter choices further illustrate the robustness of occupation-selective pumping over an extended parameter range.

\section{Density-Dependent Gauge Field from Floquet driving}
\label{app:floquet}

In this section, we explicitly show how the dynamical density-dependent gauge field can be obtained through Floquet engineering.
Starting from a time-dependent Hamiltonian that combines (i) a slow Thouless-pump modulation with phase $\phi(t)=\omega t$ and (ii) fast periodic drives of both the intra-cell tunneling and the on-site interaction at frequency $\Omega$ with $\omega \ll \Omega$, we derive the effective stroboscopic Hamiltonian in the high-frequency limit. 
This procedure establishes a direct correspondence between the experimentally controllable parameters $(\alpha,\beta,U_0,\Omega)$ and the dynamical gauge-field parameters $(\gamma,\gamma_0)$ appearing in the main model [Eq.(1) in the main text], thereby demonstrating the experimental feasibility of realizing occupation-dependent tunneling via Floquet engineering.

We consider a periodically driven bosonic system described by the time-dependent Hamiltonian
\begin{equation}
\begin{aligned}
& \hat{\mathcal{H}}(t)=\sum_j^N\left[J+\delta_0 \sin (\pi j+\phi(t))\right] \hat{a}_j^{\dagger} \hat{a}_{j+1}+\text {H.c.} \\
& +(\alpha+i \beta \sin \Omega t)\sum_j^{N / 2} \hat{a}_{2 j-1}^{\dagger} \hat{a}_{2 j}+\text {H.c.}+\sum_j^N \Delta_0 \cos (\pi j+\phi(t)) \hat{n}_j \\
& +\sum_j^N \frac{U+(-1)^j U_0 \cos (\Omega t)}{2} \hat{n}_j\left(\hat{n}_j-1\right)
\end{aligned}
\label{eq:driven_H_app}
\end{equation}
where $\phi(t) = \omega t$ with $\omega \ll \Omega$, indicating that $\phi(t)$ varies slowly compared to the fast driving frequency $\Omega$. This separation of timescales allows us to apply the high-frequency Floquet formalism to derive an effective static Hamiltonian.
The Hamiltonian~\eqref{eq:driven_H_app} consists of several experimentally controllable components:
- The terms $\left[J+\delta_0 \sin (\pi j+\phi(t))\right]$ and $\Delta_0 \cos (\pi j+\phi(t))$ represent slow modulations of the nearest-neighbor hopping and on-site potential, respectively, with frequency $\omega$. These are responsible for the Thouless pumping dynamics in the effective model.
- The intra-cell hopping $(\alpha + i\beta \sin\Omega t)$ is rapidly oscillating at frequency $\Omega$, which can be implemented via Raman-assisted tunneling to generate complex tunneling amplitudes~\cite{PhysRevLett.114.125301,PhysRevLett.109.145301}.
- The on-site interaction is modulated as $U_j(t) = [U + (-1)^j U_0 \cos(\Omega t)]/2$, achievable via time-dependent Feshbach resonance~\cite{PhysRevLett.90.230401,WinklerK2006,RevModPhys.82.1225,ClarkLoganW2017}. 

In the high-frequency limit ($\Omega \gg J, \delta_0, \Delta_0, U, \alpha, \beta, U_0$), the effective stroboscopic dynamics are governed by the first-order expansion~\cite{PhysRevX.4.031027,2015Bukov04032015,2016MHJPB,RevModPhys.89.011004,2023PhysRevB.108.075435}: 
\begin{equation}
\hat{\mathcal{H}}_{\mathrm{eff}} = \hat{\mathcal{H}}_0 + \frac{\left[\hat{\mathcal{H}}_{+1}, \hat{\mathcal{H}}_{-1}\right]}{\Omega}
\label{eq:heff_app}
\end{equation}
where $\hat{\mathcal{H}}_0$ is the time average of $\hat{\mathcal{H}}(t)$ over one fast period $T = 2\pi/\Omega$, and $\hat{\mathcal{H}}_{\pm1}$ are the Fourier components of the time-periodic part at frequencies $\pm \Omega$.
We identify the fast-oscillating terms as:
\begin{equation}
\begin{aligned}
& H_{-1}=\frac{U_0}{4} \sum_j^N(-1)^j \hat{n}_j\left(\hat{n}_j-1\right)-\frac{\beta}{2} \sum_j^{N / 2}\left(\hat{a}_{2 j-1}^{\dagger} \hat{a}_{2 j}-\text {H.c.}\right), \\
& H_{+1}=\frac{U_0}{4} \sum_j^N(-1)^j \hat{n}_j\left(\hat{n}_j-1\right)+\frac{\beta}{2} \sum_j^{N / 2}\left(\hat{a}_{2 j-1}^{\dagger} \hat{a}_{2 j}-\text {H.c.}\right) .
\end{aligned}
\end{equation}
The commutator $\left[\hat{\mathcal{H}}_{+1}, \hat{\mathcal{H}}_{-1}\right]$ generates an effective density-dependent tunneling term. Using the bosonic commutation relations and noting that $\left[\hat{n}_j, \hat{a}_j\right] = -\hat{a}_j$, we compute:
\begin{equation}
\frac{\left[\hat{\mathcal{H}}_{+1}, \hat{\mathcal{H}}_{-1}\right]}{\Omega} = \sum_j^{N / 2} \hat{a}_{2 j-1}^{\dagger}\left(\frac{\beta U_0}{2 \Omega} \hat{n}_{2 j-1}+\frac{\beta U_0}{2 \Omega} \hat{n}_{2 j}\right) \hat{a}_{2 j}+\text { H.c. }.
\end{equation}
Combining this with the time-averaged part $\hat{\mathcal{H}}_0$, which retains the slow modulations and the static interaction $U$, we obtain the effective Hamiltonian:
\begin{equation}
\begin{aligned}
& \hat{\mathcal{H}}_{\mathrm{eff}}(t)=\sum_j^N\left[J+\delta_0 \sin (\pi j+\phi(t))\right] \hat{a}_j^{\dagger} \hat{a}_{j+1}+\text { H.c. } \\
& +\sum_j^N \Delta_0 \cos (\pi j+\phi(t)) \hat{n}_j+\frac{U}{2} \sum_j^N \hat{n}_j\left(\hat{n}_j-1\right) \\
& +\sum_j^{N / 2} \hat{a}_{2 j-1}^{\dagger}\left(\alpha+\frac{\beta U_0}{2 \Omega} \hat{n}_{2 j-1}+\frac{\beta U_0}{2 \Omega} \hat{n}_{2 j}\right) \hat{a}_{2 j}+\text { H.c. }.
\end{aligned}
\label{eq:heff_result_app}
\end{equation}

Comparing Eq.~\eqref{eq:heff_result_app} with the main model in Eq.(1) of the main text, we identify that the Floquet driving generates a density-dependent tunneling of the form
\begin{equation}
\gamma + \frac{\beta U_0}{2\Omega}(\hat n_{2j-1}+\hat n_{2j})
\end{equation}
with $\gamma=\alpha$. 
Here, the amplitude $\beta U_0/(2\Omega)$ is a tunable parameter controlled by the strength of Raman-assisted tunneling ($\beta$), interaction modulation ($U_0$), and driving frequency ($\Omega$).

To reproduce the time-dependent structure $\gamma_0 \sin\phi(t)$ in Eq.(1) of the main text, this Floquet-induced amplitude can be further modulated on the slow timescale of the pump. 
In particular, by varying either $\beta$ or $U_0$ at the same frequency $\omega$ as the pumping phase $\phi(t)=\omega t$, one obtains
\begin{equation}
\frac{\beta U_0}{2\Omega} \;\rightarrow\; \gamma_0 \sin\phi(t),
\end{equation}
thereby realizing an occupation-dependent tunneling that is synchronized with the pumping cycle.

This separation of timescales—fast Floquet driving to generate density-dependent hopping and slow modulation to imprint the pump phase—establishes a direct correspondence between the experimentally controllable parameters $(\beta, U_0, \Omega)$ and the dynamical gauge-field term in Eq.(1) of the main text. 
It also clarifies that the gauge field in our model is not static, but coevolves with the pumping parameter, which is essential for the occupation-selective topological effects discussed in the main text.

\section{Extension to Three-Particle Bound States: Effective Triolon Hamiltonian and Topology}
\label{sec:triolon}

To further demonstrate that the occupation-selective topological mechanism identified in the main text is generic rather than specific to two-particle bound states, we extend the analysis to the three-particle sector in the strong-interaction regime 
$U \gg (J,\delta_0,\gamma,\gamma_0,\Delta_0)$.
Following the terminology of doublons for two-particle on-site bound states, we refer to a three-particle on-site bound state as a \emph{triolon}. 
As we show below, the dynamical gauge field continues to reshape the effective tunneling experienced by the bound state, leading to triolon topological phases that are likewise distinct from the single-particle sector.

As in the doublon case, we separate the Hamiltonian [Eq.~(1) in the main text] into the dominant interaction part [Eq.~\eqref{Dominantpart}]
and the perturbation part [Eq.~\eqref{Perturpart}].
For a three-particle system, the dominant Hamiltonian $\hat{H}_{\mathrm d}$ divides the Hilbert space into three degenerate subspaces $\mathcal{U}$, $\mathcal{V}$ and $\mathcal{W}$. 
The subspace $\mathcal{U}\equiv\left\{|3\rangle_{j}\right\}$ contains triolon bound states in which three particles occupy the same site, and its degenerate energy is $E_{j}=3U$. 
The subspace $\mathcal{V}\equiv\left\{|2\rangle_{j}|1\rangle_{k}\right\}$ contains states in which two particles occupy the same site and the third particle occupies a different site, with degenerate energy $E_{jk}=U$ for $j\neq k$. 
The subspace $\mathcal{W}\equiv\left\{|1\rangle_{j}|1\rangle_{k}|1\rangle_{m}\right\}$ contains states in which all three particles occupy different sites, with degenerate energy $E_{jkm}=0$ for $j\neq k\neq m$.
The projection operator onto the triolon subspace is
\begin{equation}
\hat{P}=\sum_{j}|3\rangle_{j}\langle 3|_{j},
\end{equation}
and the resolvent operator entering degenerate perturbation theory is
\begin{equation}
\hat{S}=
\sum_{j \neq k} \frac{1}{E_{j}-E_{jk}}|2\rangle_{j}|1\rangle_{k}\langle 1|_{k}\langle 2|_{j}
+\sum_{j \neq k \neq m} \frac{1}{E_{j}-E_{jkm}}|1\rangle_{j}|1\rangle_{k}|1\rangle_{m}\langle 1|_{m}\langle 1|_{k}\langle 1|_{j}.
\end{equation}

Using degenerate perturbation theory~\cite{1977JPC101289,2011AOP3262793}, the effective triolon Hamiltonian projected onto $\mathcal U$ is obtained up to third order as
\begin{equation}
\begin{aligned}
\hat{H}_{\mathrm{eff}}
=&\,
\hat{h}_{0}+\hat{h}_{1}+\hat{h}_{2}+\hat{h}_{3} \\
=&\,
E_{j} \hat{P}
+\hat{P} \hat{H}_{\mathrm{p}} \hat{P}
+\hat{P} \hat{H}_{\mathrm{p}} \hat{S} \hat{H}_{\mathrm{p}} \hat{P}
+\hat{P}\hat{H}_{\mathrm{p}}\hat{S}\hat{H}_{\mathrm{p}}\hat{S}\hat{H}_{\mathrm{p}}\hat{P} \\
&-\frac{1}{2}\hat{P}\hat{H}_{\mathrm{p}}\hat{S}^2\hat{H}_{\mathrm{p}}\hat{P}\hat{H}_{\mathrm{p}}\hat{P}
-\frac{1}{2}\hat{P}\hat{H}_{\mathrm{p}}\hat{P}\hat{H}_{\mathrm{p}}\hat{S}^2\hat{H}_{\mathrm{p}}\hat{P}.
\end{aligned}
\end{equation}

At zeroth order, one has
\begin{equation}\label{ZeroTermThreeP}
\hat{h}_{0}=3U\sum_{j}|3\rangle_{j}\langle 3|_{j}.
\end{equation}

At first order,
\begin{equation}\label{FirstTermThreeP}
\hat{h}_1=\sum_{j=1}^N 3 \Delta_0 \cos (\pi j+\phi(t))|3\rangle_j\langle 3|_j.
\end{equation}

The second-order contribution yields an on-site energy renormalization,
\begin{equation}\label{SecondTermThreeP}
\begin{aligned}
\hat{h}_2
=
\left[
\frac{\left(J+\delta_0\sin\phi(t)\right)^{2}}{2U}
+\frac{\left(J-\delta_0\sin\phi(t)+\gamma+2\gamma_0\sin\phi(t)\right)^{2}}{2U}
\right]
\sum_{j=1}^{N}|3\rangle_{j}\langle3|_{j}.
\end{aligned}
\end{equation}

The leading triolon hopping appears at third order,
\begin{equation}\label{ThirdTermThreeP}
\begin{aligned}
\hat{h}_{3}
&= \hat{P}\hat{H}_{\mathrm p}\hat{S}\hat{H}_{\mathrm p}\hat{S}\hat{H}_{\mathrm p}\hat{P}
- \frac{1}{2}\hat{P}\hat{H}_{\mathrm p}\hat{S}^{2}\hat{H}_{\mathrm p}\hat{P}\hat{H}_{\mathrm p}\hat{P}
- \frac{1}{2}\hat{P}\hat{H}_{\mathrm p}\hat{P}\hat{H}_{\mathrm p}\hat{S}^{2}\hat{H}_{\mathrm p}\hat{P} \\
&= \sum_{j=1}^{N} \left\{ \mathcal{J}'_{1} + \mathcal{J}'_{2} \frac{1 - \cos(\pi j)}{2} \right\} |3\rangle_{j} \langle3|_{j+1} +\text { H.c. } \\
&\quad - \sum_{j=1}^{N} 3\Delta_{0} \cos(\pi j + \phi(t))
\left[
\frac{\left(J + \delta_{0} \sin\phi(t)\right)^{2}}{(2U)^{2}}
+ \frac{\left(J - \delta_{0} \sin\phi(t) + \gamma + 2\gamma_{0} \sin\phi(t)\right)^{2}}{(2U)^{2}}
\right]
|3\rangle_{j} \langle3|_{j}.
\end{aligned}
\end{equation}
Here
\begin{equation}
\mathcal{J}'_{1} = \frac{\left[J + \delta_{0} \sin(\pi j + \phi(t))\right]^{3}}{(2U)^{2}},
\end{equation}
and
\begin{equation}
\mathcal{J}'_{2}
=
\frac{
3\left(J - \delta_{0} \sin\phi(t)\right)^{2}\left(\gamma + 2\gamma_{0} \sin\phi(t)\right)
+ 3\left(J - \delta_{0} \sin\phi(t)\right)\left(\gamma + 2\gamma_{0} \sin\phi(t)\right)^{2}
+ \left(\gamma + 2\gamma_{0} \sin\phi(t)\right)^{3}
}{(2U)^{2}}.
\end{equation}

\begin{figure}[htp]
\center
\includegraphics[width=0.6\textwidth]{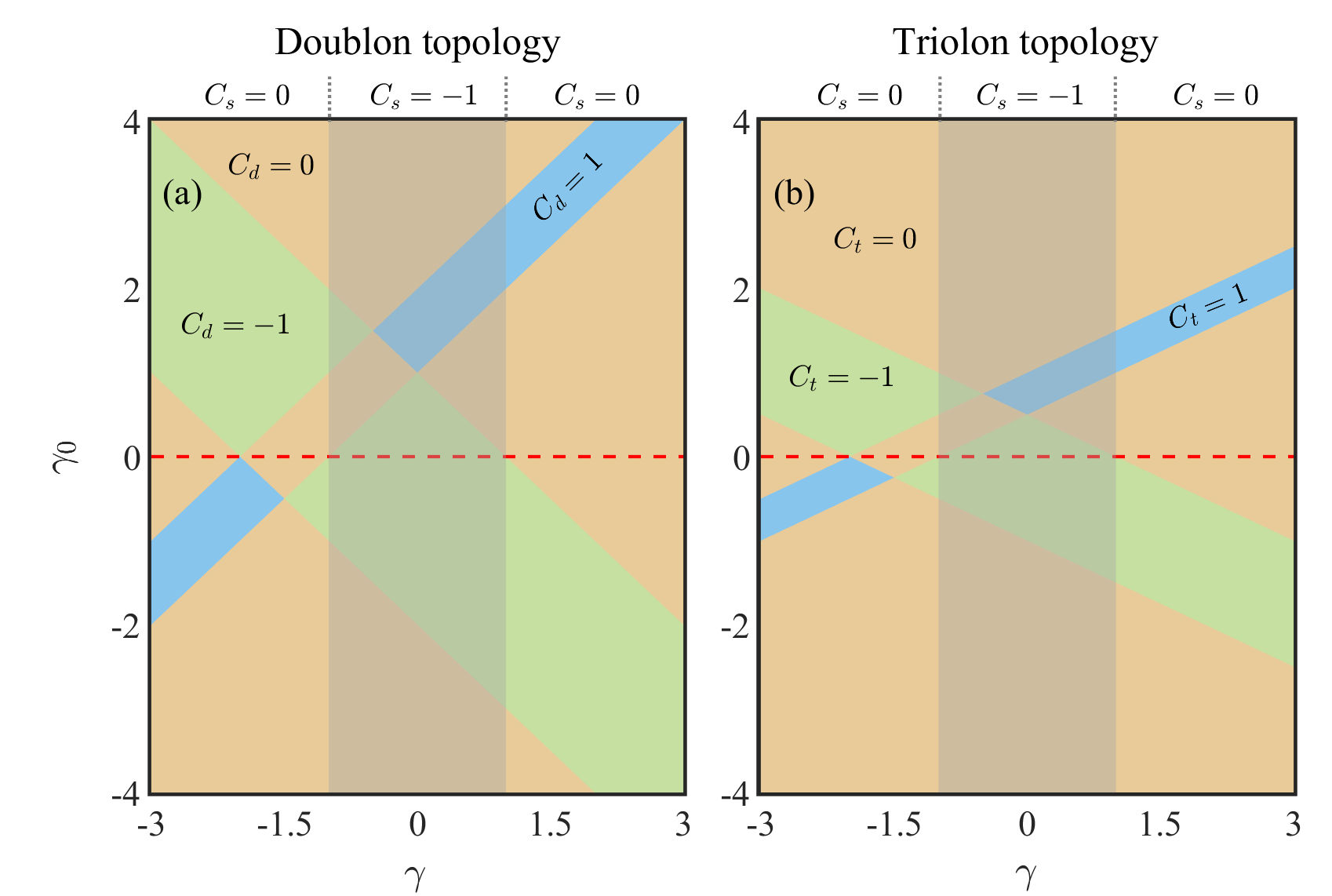}
\caption{\textbf{Extension of occupation-selective topology to three-particle bound states.}
(a) Chern number $C_d$ of the lower doublon band in the $(\gamma,\gamma_0)$ plane, reproduced for comparison. 
(b) Chern number $C_t$ of the lower triolon band obtained from the effective three-particle Hamiltonian. 
The gray shaded regions indicate the single-particle topological phase with $C_s=-1$. 
Along the density-independent line $\gamma_0=0$ (red dashed line), the doublon and triolon sectors both follow the single-particle topology. 
For $\gamma_0\neq0$, the dynamical gauge field generates topological regions unique to the higher-occupancy bound states, including $C_t=\pm1$ even when $C_s=0$. 
The shifted phase boundaries between the doublon and triolon diagrams further show that different occupation sectors experience distinct effective pumping structures under the same driving cycle.}
\label{fig:TriolonPhasediagram}
\end{figure}

Summing Eqs.~\eqref{ZeroTermThreeP}--\eqref{ThirdTermThreeP}, we obtain the effective single-particle Hamiltonian for the triolon manifold. 
In direct analogy with the doublon case, the triolon behaves as an emergent composite particle moving in a driven superlattice, but now with an effective tunneling amplitude that depends more strongly on the density-dependent gauge field. 
Importantly, the dynamical gauge-field contribution enters through the combination $\gamma+2\gamma_0\sin\phi(t)$, reflecting the higher occupation of the bound state. 
This leads to topological phase boundaries that are shifted relative to those of the doublon sector, showing that different bound states experience distinct effective pumping structures under the same adiabatic cycle.

Topological transitions occur when the triolon gap closes. 
Keeping the same parameters as in Fig.~3 of the main text, the phase boundaries of the triolon sector are found to be
\begin{equation}
\gamma_0=\frac{\pm\gamma+1}{2},
\qquad
\gamma_0=\frac{\pm2\pm\gamma}{2}.
\end{equation}
Away from these critical lines, the Chern number $C_t$ of the lower triolon band is computed numerically, yielding the phase diagram shown in Fig.~\ref{fig:TriolonPhasediagram}(b). 
For comparison, the doublon phase diagram $C_d$ is shown in Fig.~\ref{fig:TriolonPhasediagram}(a), while $C_s$ denotes the single-particle topology.
Along the density-independent limit $\gamma_0=0$, the doublon and triolon topologies both follow the single-particle topology $C_s$, consistent with the absence of a dynamical gauge-field correction. 
For $\gamma_0\neq0$, however, the dynamical gauge field independently controls the doublon and triolon sectors, producing parameter regions with $C_d=\pm1$ and $C_t=\pm1$ even when $C_s=0$. 
Moreover, the shifted phase boundaries between the doublon and triolon diagrams indicate that higher-occupancy bound states respond differently to the same gauge-field modulation. 
This demonstrates that the occupation-selective topological mechanism discussed in the main text naturally extends beyond two particles and persists for higher-order bound states.


\begin{thebibliography}{110}%
\makeatletter
\providecommand \@ifxundefined [1]{%
 \@ifx{#1\undefined}
}%
\providecommand \@ifnum [1]{%
 \ifnum #1\expandafter \@firstoftwo
 \else \expandafter \@secondoftwo
 \fi
}%
\providecommand \@ifx [1]{%
 \ifx #1\expandafter \@firstoftwo
 \else \expandafter \@secondoftwo
 \fi
}%
\providecommand \natexlab [1]{#1}%
\providecommand \enquote  [1]{``#1''}%
\providecommand \bibnamefont  [1]{#1}%
\providecommand \bibfnamefont [1]{#1}%
\providecommand \citenamefont [1]{#1}%
\providecommand \href@noop [0]{\@secondoftwo}%
\providecommand \href [0]{\begingroup \@sanitize@url \@href}%
\providecommand \@href[1]{\@@startlink{#1}\@@href}%
\providecommand \@@href[1]{\endgroup#1\@@endlink}%
\providecommand \@sanitize@url [0]{\catcode `\\12\catcode `\$12\catcode `\&12\catcode `\#12\catcode `\^12\catcode `\_12\catcode `\%12\relax}%
\providecommand \@@startlink[1]{}%
\providecommand \@@endlink[0]{}%
\providecommand \url  [0]{\begingroup\@sanitize@url \@url }%
\providecommand \@url [1]{\endgroup\@href {#1}{\urlprefix }}%
\providecommand \urlprefix  [0]{URL }%
\providecommand \Eprint [0]{\href }%
\providecommand \doibase [0]{https://doi.org/}%
\providecommand \selectlanguage [0]{\@gobble}%
\providecommand \bibinfo  [0]{\@secondoftwo}%
\providecommand \bibfield  [0]{\@secondoftwo}%
\providecommand \translation [1]{[#1]}%
\providecommand \BibitemOpen [0]{}%
\providecommand \bibitemStop [0]{}%
\providecommand \bibitemNoStop [0]{.\EOS\space}%
\providecommand \EOS [0]{\spacefactor3000\relax}%
\providecommand \BibitemShut  [1]{\csname bibitem#1\endcsname}%
\let\auto@bib@innerbib\@empty
\bibitem [{\citenamefont {Hasan}\ and\ \citenamefont {Kane}(2010)}]{HasanKane2010}%
  \BibitemOpen
  \bibfield  {author} {\bibinfo {author} {\bibfnamefont {M.~Z.}\ \bibnamefont {Hasan}}\ and\ \bibinfo {author} {\bibfnamefont {C.~L.}\ \bibnamefont {Kane}},\ }\bibfield  {title} {\bibinfo {title} {Colloquium: Topological insulators},\ }\href {https://doi.org/10.1103/RevModPhys.82.3045} {\bibfield  {journal} {\bibinfo  {journal} {Rev. Mod. Phys.}\ }\textbf {\bibinfo {volume} {82}},\ \bibinfo {pages} {3045} (\bibinfo {year} {2010})}\BibitemShut {NoStop}%
\bibitem [{\citenamefont {Qi}\ and\ \citenamefont {Zhang}(2011)}]{QiZhang2011}%
  \BibitemOpen
  \bibfield  {author} {\bibinfo {author} {\bibfnamefont {X.-L.}\ \bibnamefont {Qi}}\ and\ \bibinfo {author} {\bibfnamefont {S.-C.}\ \bibnamefont {Zhang}},\ }\bibfield  {title} {\bibinfo {title} {Topological insulators and superconductors},\ }\href {https://doi.org/10.1103/RevModPhys.83.1057} {\bibfield  {journal} {\bibinfo  {journal} {Rev. Mod. Phys.}\ }\textbf {\bibinfo {volume} {83}},\ \bibinfo {pages} {1057} (\bibinfo {year} {2011})}\BibitemShut {NoStop}%
\bibitem [{\citenamefont {Cooper}\ \emph {et~al.}(2019)\citenamefont {Cooper}, \citenamefont {Dalibard},\ and\ \citenamefont {Spielman}}]{Cooper2019}%
  \BibitemOpen
  \bibfield  {author} {\bibinfo {author} {\bibfnamefont {N.~R.}\ \bibnamefont {Cooper}}, \bibinfo {author} {\bibfnamefont {J.}~\bibnamefont {Dalibard}},\ and\ \bibinfo {author} {\bibfnamefont {I.~B.}\ \bibnamefont {Spielman}},\ }\bibfield  {title} {\bibinfo {title} {Topological quantum matter},\ }\href {https://doi.org/10.1103/RevModPhys.91.015005} {\bibfield  {journal} {\bibinfo  {journal} {Rev. Mod. Phys.}\ }\textbf {\bibinfo {volume} {91}},\ \bibinfo {pages} {015005} (\bibinfo {year} {2019})}\BibitemShut {NoStop}%
\bibitem [{\citenamefont {Thouless}(1983)}]{Thouless1983}%
  \BibitemOpen
  \bibfield  {author} {\bibinfo {author} {\bibfnamefont {D.~J.}\ \bibnamefont {Thouless}},\ }\bibfield  {title} {\bibinfo {title} {Quantization of particle transport},\ }\href {https://doi.org/10.1103/PhysRevB.27.6083} {\bibfield  {journal} {\bibinfo  {journal} {Phys. Rev. B}\ }\textbf {\bibinfo {volume} {27}},\ \bibinfo {pages} {6083} (\bibinfo {year} {1983})}\BibitemShut {NoStop}%
\bibitem [{\citenamefont {Citro}\ and\ \citenamefont {Aidelsburger}(2023)}]{2023TPandTopo}%
  \BibitemOpen
  \bibfield  {author} {\bibinfo {author} {\bibfnamefont {R.}~\bibnamefont {Citro}}\ and\ \bibinfo {author} {\bibfnamefont {M.}~\bibnamefont {Aidelsburger}},\ }\bibfield  {title} {\bibinfo {title} {Thouless pumping and topology},\ }\href {https://doi.org/10.1038/s42254-022-00545-0} {\bibfield  {journal} {\bibinfo  {journal} {Nat. Rev. Phys.}\ }\textbf {\bibinfo {volume} {5}},\ \bibinfo {pages} {87} (\bibinfo {year} {2023})}\BibitemShut {NoStop}%
\bibitem [{\citenamefont {Chen}\ \emph {et~al.}(2025)\citenamefont {Chen}, \citenamefont {Fu}, \citenamefont {Xu},\ and\ \citenamefont {Ye}}]{LaserPhotonReview2025}%
  \BibitemOpen
  \bibfield  {author} {\bibinfo {author} {\bibfnamefont {Y.}~\bibnamefont {Chen}}, \bibinfo {author} {\bibfnamefont {Q.}~\bibnamefont {Fu}}, \bibinfo {author} {\bibfnamefont {Z.}~\bibnamefont {Xu}},\ and\ \bibinfo {author} {\bibfnamefont {F.}~\bibnamefont {Ye}},\ }\bibfield  {title} {\bibinfo {title} {Progress on new-type optical thouless pumping},\ }\href {https://doi.org/https://doi.org/10.1002/lpor.202501528} {\bibfield  {journal} {\bibinfo  {journal} {Laser \& Photonics Rev.}\ ,\ \bibinfo {pages} {e01528}} (\bibinfo {year} {2025})}\BibitemShut {NoStop}%
\bibitem [{\citenamefont {Nakajima}\ \emph {et~al.}(2016)\citenamefont {Nakajima}, \citenamefont {Tomita}, \citenamefont {Taie}, \citenamefont {Ichinose}, \citenamefont {Ozawa}, \citenamefont {Wang}, \citenamefont {Troyer},\ and\ \citenamefont {Takahashi}}]{Nakajima2016}%
  \BibitemOpen
  \bibfield  {author} {\bibinfo {author} {\bibfnamefont {S.}~\bibnamefont {Nakajima}}, \bibinfo {author} {\bibfnamefont {T.}~\bibnamefont {Tomita}}, \bibinfo {author} {\bibfnamefont {S.}~\bibnamefont {Taie}}, \bibinfo {author} {\bibfnamefont {T.}~\bibnamefont {Ichinose}}, \bibinfo {author} {\bibfnamefont {H.}~\bibnamefont {Ozawa}}, \bibinfo {author} {\bibfnamefont {L.}~\bibnamefont {Wang}}, \bibinfo {author} {\bibfnamefont {M.}~\bibnamefont {Troyer}},\ and\ \bibinfo {author} {\bibfnamefont {Y.}~\bibnamefont {Takahashi}},\ }\bibfield  {title} {\bibinfo {title} {Topological thouless pumping of ultracold fermions},\ }\href {https://doi.org/10.1038/nphys3622} {\bibfield  {journal} {\bibinfo  {journal} {Nat. Phys.}\ }\textbf {\bibinfo {volume} {12}},\ \bibinfo {pages} {296} (\bibinfo {year} {2016})}\BibitemShut {NoStop}%
\bibitem [{\citenamefont {Lohse}\ \emph {et~al.}(2016)\citenamefont {Lohse}, \citenamefont {Schweizer}, \citenamefont {Price}, \citenamefont {Goldman},\ and\ \citenamefont {Bloch}}]{Lohse2016}%
  \BibitemOpen
  \bibfield  {author} {\bibinfo {author} {\bibfnamefont {M.}~\bibnamefont {Lohse}}, \bibinfo {author} {\bibfnamefont {C.}~\bibnamefont {Schweizer}}, \bibinfo {author} {\bibfnamefont {H.~M.}\ \bibnamefont {Price}}, \bibinfo {author} {\bibfnamefont {N.}~\bibnamefont {Goldman}},\ and\ \bibinfo {author} {\bibfnamefont {I.}~\bibnamefont {Bloch}},\ }\bibfield  {title} {\bibinfo {title} {A thouless quantum pump with ultracold bosonic atoms in optical superlattices},\ }\href {https://doi.org/10.1038/nature18318} {\bibfield  {journal} {\bibinfo  {journal} {Nature}\ }\textbf {\bibinfo {volume} {534}},\ \bibinfo {pages} {354} (\bibinfo {year} {2016})}\BibitemShut {NoStop}%
\bibitem [{\citenamefont {Walter}\ \emph {et~al.}(2023)\citenamefont {Walter}, \citenamefont {Zhu}, \citenamefont {G{\"a}chter} \emph {et~al.}}]{Walter2023HubbardPump}%
  \BibitemOpen
  \bibfield  {author} {\bibinfo {author} {\bibfnamefont {A.-S.}\ \bibnamefont {Walter}}, \bibinfo {author} {\bibfnamefont {Z.}~\bibnamefont {Zhu}}, \bibinfo {author} {\bibfnamefont {M.}~\bibnamefont {G{\"a}chter}}, \emph {et~al.},\ }\bibfield  {title} {\bibinfo {title} {Quantization and its breakdown in a hubbard--thouless pump},\ }\href {https://doi.org/10.1038/s41567-023-02145-w} {\bibfield  {journal} {\bibinfo  {journal} {Nat. Phys.}\ }\textbf {\bibinfo {volume} {19}},\ \bibinfo {pages} {1471} (\bibinfo {year} {2023})}\BibitemShut {NoStop}%
\bibitem [{\citenamefont {Viebahn}\ \emph {et~al.}(2024)\citenamefont {Viebahn}, \citenamefont {Walter}, \citenamefont {Bertok}, \citenamefont {Zhu}, \citenamefont {G\"achter}, \citenamefont {Aligia}, \citenamefont {Heidrich-Meisner},\ and\ \citenamefont {Esslinger}}]{Viebahn2024InteractionsPump}%
  \BibitemOpen
  \bibfield  {author} {\bibinfo {author} {\bibfnamefont {K.}~\bibnamefont {Viebahn}}, \bibinfo {author} {\bibfnamefont {A.-S.}\ \bibnamefont {Walter}}, \bibinfo {author} {\bibfnamefont {E.}~\bibnamefont {Bertok}}, \bibinfo {author} {\bibfnamefont {Z.}~\bibnamefont {Zhu}}, \bibinfo {author} {\bibfnamefont {M.}~\bibnamefont {G\"achter}}, \bibinfo {author} {\bibfnamefont {A.~A.}\ \bibnamefont {Aligia}}, \bibinfo {author} {\bibfnamefont {F.}~\bibnamefont {Heidrich-Meisner}},\ and\ \bibinfo {author} {\bibfnamefont {T.}~\bibnamefont {Esslinger}},\ }\bibfield  {title} {\bibinfo {title} {Interactions enable thouless pumping in a nonsliding lattice},\ }\href {https://doi.org/10.1103/PhysRevX.14.021049} {\bibfield  {journal} {\bibinfo  {journal} {Phys. Rev. X}\ }\textbf {\bibinfo {volume} {14}},\ \bibinfo {pages} {021049} (\bibinfo {year} {2024})}\BibitemShut {NoStop}%
\bibitem [{\citenamefont {Arg{\"{u}}ello-Luengo}\ \emph {et~al.}(2024)\citenamefont {Arg{\"{u}}ello-Luengo}, \citenamefont {Mark}, \citenamefont {Ferlaino}, \citenamefont {Lewenstein}, \citenamefont {Barbiero},\ and\ \citenamefont {Juli{\`{a}}-Farr{\'{e}}}}]{ArguelloLuengo2024HubbardPump}%
  \BibitemOpen
  \bibfield  {author} {\bibinfo {author} {\bibfnamefont {J.}~\bibnamefont {Arg{\"{u}}ello-Luengo}}, \bibinfo {author} {\bibfnamefont {M.~J.}\ \bibnamefont {Mark}}, \bibinfo {author} {\bibfnamefont {F.}~\bibnamefont {Ferlaino}}, \bibinfo {author} {\bibfnamefont {M.}~\bibnamefont {Lewenstein}}, \bibinfo {author} {\bibfnamefont {L.}~\bibnamefont {Barbiero}},\ and\ \bibinfo {author} {\bibfnamefont {S.}~\bibnamefont {Juli{\`{a}}-Farr{\'{e}}}},\ }\bibfield  {title} {\bibinfo {title} {Stabilization of {H}ubbard-{T}houless pumps through nonlocal fermionic repulsion},\ }\href {https://doi.org/10.22331/q-2024-03-14-1285} {\bibfield  {journal} {\bibinfo  {journal} {Quantum}\ }\textbf {\bibinfo {volume} {8}},\ \bibinfo {pages} {1285} (\bibinfo {year} {2024})}\BibitemShut {NoStop}%
\bibitem [{\citenamefont {Koh}\ \emph {et~al.}(2022)\citenamefont {Koh}, \citenamefont {Tai},\ and\ \citenamefont {Lee}}]{2022PhysRevLett.129.140502}%
  \BibitemOpen
  \bibfield  {author} {\bibinfo {author} {\bibfnamefont {J.~M.}\ \bibnamefont {Koh}}, \bibinfo {author} {\bibfnamefont {T.}~\bibnamefont {Tai}},\ and\ \bibinfo {author} {\bibfnamefont {C.~H.}\ \bibnamefont {Lee}},\ }\bibfield  {title} {\bibinfo {title} {Simulation of interaction-induced chiral topological dynamics on a digital quantum computer},\ }\href {https://doi.org/10.1103/PhysRevLett.129.140502} {\bibfield  {journal} {\bibinfo  {journal} {Phys. Rev. Lett.}\ }\textbf {\bibinfo {volume} {129}},\ \bibinfo {pages} {140502} (\bibinfo {year} {2022})}\BibitemShut {NoStop}%
\bibitem [{\citenamefont {Sridhar}\ \emph {et~al.}(2024)\citenamefont {Sridhar}, \citenamefont {Ghosh}, \citenamefont {Srinivasan}, \citenamefont {Miller},\ and\ \citenamefont {Dutt}}]{Sridhar2024SyntheticPump}%
  \BibitemOpen
  \bibfield  {author} {\bibinfo {author} {\bibfnamefont {S.~K.}\ \bibnamefont {Sridhar}}, \bibinfo {author} {\bibfnamefont {S.}~\bibnamefont {Ghosh}}, \bibinfo {author} {\bibfnamefont {D.}~\bibnamefont {Srinivasan}}, \bibinfo {author} {\bibfnamefont {A.~R.}\ \bibnamefont {Miller}},\ and\ \bibinfo {author} {\bibfnamefont {A.}~\bibnamefont {Dutt}},\ }\bibfield  {title} {\bibinfo {title} {Quantized topological pumping in floquet synthetic dimensions with a driven dissipative photonic molecule},\ }\href {https://doi.org/10.1038/s41567-024-02413-3} {\bibfield  {journal} {\bibinfo  {journal} {Nat. Phys.}\ }\textbf {\bibinfo {volume} {20}},\ \bibinfo {pages} {843} (\bibinfo {year} {2024})}\BibitemShut {NoStop}%
\bibitem [{\citenamefont {Liu}\ \emph {et~al.}(2025{\natexlab{a}})\citenamefont {Liu}, \citenamefont {Zhang}, \citenamefont {Shi} \emph {et~al.}}]{Liu2025SCProcessorPump}%
  \BibitemOpen
  \bibfield  {author} {\bibinfo {author} {\bibfnamefont {Y.}~\bibnamefont {Liu}}, \bibinfo {author} {\bibfnamefont {Y.-R.}\ \bibnamefont {Zhang}}, \bibinfo {author} {\bibfnamefont {Y.-H.}\ \bibnamefont {Shi}}, \emph {et~al.},\ }\bibfield  {title} {\bibinfo {title} {Interplay between disorder and topology in thouless pumping on a superconducting quantum processor},\ }\href {https://doi.org/10.1038/s41467-024-55343-2} {\bibfield  {journal} {\bibinfo  {journal} {Nat. Commun.}\ }\textbf {\bibinfo {volume} {16}},\ \bibinfo {pages} {108} (\bibinfo {year} {2025}{\natexlab{a}})}\BibitemShut {NoStop}%
\bibitem [{\citenamefont {Zhu}\ \emph {et~al.}(2025)\citenamefont {Zhu}, \citenamefont {Kiefer}, \citenamefont {Jele}, \citenamefont {G\"achter}, \citenamefont {Bisson}, \citenamefont {Viebahn},\ and\ \citenamefont {Esslinger}}]{xh3v-tky4}%
  \BibitemOpen
  \bibfield  {author} {\bibinfo {author} {\bibfnamefont {Z.}~\bibnamefont {Zhu}}, \bibinfo {author} {\bibfnamefont {Y.}~\bibnamefont {Kiefer}}, \bibinfo {author} {\bibfnamefont {S.}~\bibnamefont {Jele}}, \bibinfo {author} {\bibfnamefont {M.}~\bibnamefont {G\"achter}}, \bibinfo {author} {\bibfnamefont {G.}~\bibnamefont {Bisson}}, \bibinfo {author} {\bibfnamefont {K.}~\bibnamefont {Viebahn}},\ and\ \bibinfo {author} {\bibfnamefont {T.}~\bibnamefont {Esslinger}},\ }\bibfield  {title} {\bibinfo {title} {Splitting and connecting singlets in atomic quantum circuits},\ }\href {https://doi.org/10.1103/xh3v-tky4} {\bibfield  {journal} {\bibinfo  {journal} {Phys. Rev. X}\ }\textbf {\bibinfo {volume} {15}},\ \bibinfo {pages} {041032} (\bibinfo {year} {2025})}\BibitemShut {NoStop}%
\bibitem [{\citenamefont {Tuloup}\ \emph {et~al.}(2023)\citenamefont {Tuloup}, \citenamefont {Bomantara},\ and\ \citenamefont {Gong}}]{Tuloup2023Breakdown}%
  \BibitemOpen
  \bibfield  {author} {\bibinfo {author} {\bibfnamefont {T.}~\bibnamefont {Tuloup}}, \bibinfo {author} {\bibfnamefont {R.~W.}\ \bibnamefont {Bomantara}},\ and\ \bibinfo {author} {\bibfnamefont {J.}~\bibnamefont {Gong}},\ }\bibfield  {title} {\bibinfo {title} {{Breakdown of quantization in nonlinear Thouless pumping}},\ }\href {https://doi.org/10.1088/1367-2630/acef4d} {\bibfield  {journal} {\bibinfo  {journal} {New J. Phys.}\ }\textbf {\bibinfo {volume} {25}},\ \bibinfo {pages} {083048} (\bibinfo {year} {2023})}\BibitemShut {NoStop}%
\bibitem [{\citenamefont {Ke}\ \emph {et~al.}(2017)\citenamefont {Ke}, \citenamefont {Qin}, \citenamefont {Kivshar},\ and\ \citenamefont {Lee}}]{2017PhysRevA.95.063630}%
  \BibitemOpen
  \bibfield  {author} {\bibinfo {author} {\bibfnamefont {Y.}~\bibnamefont {Ke}}, \bibinfo {author} {\bibfnamefont {X.}~\bibnamefont {Qin}}, \bibinfo {author} {\bibfnamefont {Y.~S.}\ \bibnamefont {Kivshar}},\ and\ \bibinfo {author} {\bibfnamefont {C.}~\bibnamefont {Lee}},\ }\bibfield  {title} {\bibinfo {title} {Multiparticle wannier states and thouless pumping of interacting bosons},\ }\href {https://doi.org/10.1103/PhysRevA.95.063630} {\bibfield  {journal} {\bibinfo  {journal} {Phys. Rev. A}\ }\textbf {\bibinfo {volume} {95}},\ \bibinfo {pages} {063630} (\bibinfo {year} {2017})}\BibitemShut {NoStop}%
\bibitem [{\citenamefont {Lin}\ \emph {et~al.}(2020)\citenamefont {Lin}, \citenamefont {Ke},\ and\ \citenamefont {Lee}}]{2020PhysRevA101023620}%
  \BibitemOpen
  \bibfield  {author} {\bibinfo {author} {\bibfnamefont {L.}~\bibnamefont {Lin}}, \bibinfo {author} {\bibfnamefont {Y.}~\bibnamefont {Ke}},\ and\ \bibinfo {author} {\bibfnamefont {C.}~\bibnamefont {Lee}},\ }\bibfield  {title} {\bibinfo {title} {Interaction-induced topological bound states and thouless pumping in a one-dimensional optical lattice},\ }\href {https://doi.org/10.1103/PhysRevA.101.023620} {\bibfield  {journal} {\bibinfo  {journal} {Phys. Rev. A}\ }\textbf {\bibinfo {volume} {101}},\ \bibinfo {pages} {023620} (\bibinfo {year} {2020})}\BibitemShut {NoStop}%
\bibitem [{\citenamefont {Liu}\ \emph {et~al.}(2023)\citenamefont {Liu}, \citenamefont {Hu}, \citenamefont {Zhang}, \citenamefont {Ke},\ and\ \citenamefont {Lee}}]{2023PhysRevResearch.5.013020}%
  \BibitemOpen
  \bibfield  {author} {\bibinfo {author} {\bibfnamefont {W.}~\bibnamefont {Liu}}, \bibinfo {author} {\bibfnamefont {S.}~\bibnamefont {Hu}}, \bibinfo {author} {\bibfnamefont {L.}~\bibnamefont {Zhang}}, \bibinfo {author} {\bibfnamefont {Y.}~\bibnamefont {Ke}},\ and\ \bibinfo {author} {\bibfnamefont {C.}~\bibnamefont {Lee}},\ }\bibfield  {title} {\bibinfo {title} {Correlated topological pumping of interacting bosons assisted by bloch oscillations},\ }\href {https://doi.org/10.1103/PhysRevResearch.5.013020} {\bibfield  {journal} {\bibinfo  {journal} {Phys. Rev. Res.}\ }\textbf {\bibinfo {volume} {5}},\ \bibinfo {pages} {013020} (\bibinfo {year} {2023})}\BibitemShut {NoStop}%
\bibitem [{\citenamefont {Huang}\ \emph {et~al.}(2024)\citenamefont {Huang}, \citenamefont {Ke}, \citenamefont {Zhong}, \citenamefont {Kivshar},\ and\ \citenamefont {Lee}}]{2024PhysRevLett133140202}%
  \BibitemOpen
  \bibfield  {author} {\bibinfo {author} {\bibfnamefont {B.}~\bibnamefont {Huang}}, \bibinfo {author} {\bibfnamefont {Y.}~\bibnamefont {Ke}}, \bibinfo {author} {\bibfnamefont {H.}~\bibnamefont {Zhong}}, \bibinfo {author} {\bibfnamefont {Y.~S.}\ \bibnamefont {Kivshar}},\ and\ \bibinfo {author} {\bibfnamefont {C.}~\bibnamefont {Lee}},\ }\bibfield  {title} {\bibinfo {title} {Interaction-induced multiparticle bound states in the continuum},\ }\href {https://doi.org/10.1103/PhysRevLett.133.140202} {\bibfield  {journal} {\bibinfo  {journal} {Phys. Rev. Lett.}\ }\textbf {\bibinfo {volume} {133}},\ \bibinfo {pages} {140202} (\bibinfo {year} {2024})}\BibitemShut {NoStop}%
\bibitem [{\citenamefont {Hu}\ \emph {et~al.}(2020)\citenamefont {Hu}, \citenamefont {Ke},\ and\ \citenamefont {Lee}}]{PhysRevA.101.052323}%
  \BibitemOpen
  \bibfield  {author} {\bibinfo {author} {\bibfnamefont {S.}~\bibnamefont {Hu}}, \bibinfo {author} {\bibfnamefont {Y.}~\bibnamefont {Ke}},\ and\ \bibinfo {author} {\bibfnamefont {C.}~\bibnamefont {Lee}},\ }\bibfield  {title} {\bibinfo {title} {Topological quantum transport and spatial entanglement distribution via a disordered bulk channel},\ }\href {https://doi.org/10.1103/PhysRevA.101.052323} {\bibfield  {journal} {\bibinfo  {journal} {Phys. Rev. A}\ }\textbf {\bibinfo {volume} {101}},\ \bibinfo {pages} {052323} (\bibinfo {year} {2020})}\BibitemShut {NoStop}%
\bibitem [{\citenamefont {Cerjan}\ \emph {et~al.}(2020)\citenamefont {Cerjan}, \citenamefont {Wang}, \citenamefont {Huang}, \citenamefont {Chen},\ and\ \citenamefont {Rechtsman}}]{Cerjan2020DisorderedPhotonic}%
  \BibitemOpen
  \bibfield  {author} {\bibinfo {author} {\bibfnamefont {A.}~\bibnamefont {Cerjan}}, \bibinfo {author} {\bibfnamefont {M.}~\bibnamefont {Wang}}, \bibinfo {author} {\bibfnamefont {S.}~\bibnamefont {Huang}}, \bibinfo {author} {\bibfnamefont {K.~P.}\ \bibnamefont {Chen}},\ and\ \bibinfo {author} {\bibfnamefont {M.~C.}\ \bibnamefont {Rechtsman}},\ }\bibfield  {title} {\bibinfo {title} {Thouless pumping in disordered photonic systems},\ }\href {https://doi.org/10.1038/s41377-020-00408-2} {\bibfield  {journal} {\bibinfo  {journal} {Light: Sci. Appl.}\ }\textbf {\bibinfo {volume} {9}},\ \bibinfo {pages} {178} (\bibinfo {year} {2020})}\BibitemShut {NoStop}%
\bibitem [{\citenamefont {Grusdt}\ and\ \citenamefont {H\"oning}(2014)}]{2014PhysRevA90053623}%
  \BibitemOpen
  \bibfield  {author} {\bibinfo {author} {\bibfnamefont {F.}~\bibnamefont {Grusdt}}\ and\ \bibinfo {author} {\bibfnamefont {M.}~\bibnamefont {H\"oning}},\ }\bibfield  {title} {\bibinfo {title} {Realization of fractional chern insulators in the thin-torus limit with ultracold bosons},\ }\href {https://doi.org/10.1103/PhysRevA.90.053623} {\bibfield  {journal} {\bibinfo  {journal} {Phys. Rev. A}\ }\textbf {\bibinfo {volume} {90}},\ \bibinfo {pages} {053623} (\bibinfo {year} {2014})}\BibitemShut {NoStop}%
\bibitem [{\citenamefont {Zeng}\ \emph {et~al.}(2016)\citenamefont {Zeng}, \citenamefont {Zhu},\ and\ \citenamefont {Sheng}}]{2016PhysRevB94235139}%
  \BibitemOpen
  \bibfield  {author} {\bibinfo {author} {\bibfnamefont {T.-S.}\ \bibnamefont {Zeng}}, \bibinfo {author} {\bibfnamefont {W.}~\bibnamefont {Zhu}},\ and\ \bibinfo {author} {\bibfnamefont {D.~N.}\ \bibnamefont {Sheng}},\ }\bibfield  {title} {\bibinfo {title} {Fractional charge pumping of interacting bosons in one-dimensional superlattice},\ }\href {https://doi.org/10.1103/PhysRevB.94.235139} {\bibfield  {journal} {\bibinfo  {journal} {Phys. Rev. B}\ }\textbf {\bibinfo {volume} {94}},\ \bibinfo {pages} {235139} (\bibinfo {year} {2016})}\BibitemShut {NoStop}%
\bibitem [{\citenamefont {Taddia}\ \emph {et~al.}(2017)\citenamefont {Taddia}, \citenamefont {Cornfeld}, \citenamefont {Rossini}, \citenamefont {Mazza}, \citenamefont {Sela},\ and\ \citenamefont {Fazio}}]{2017PhysRevLett118230402}%
  \BibitemOpen
  \bibfield  {author} {\bibinfo {author} {\bibfnamefont {L.}~\bibnamefont {Taddia}}, \bibinfo {author} {\bibfnamefont {E.}~\bibnamefont {Cornfeld}}, \bibinfo {author} {\bibfnamefont {D.}~\bibnamefont {Rossini}}, \bibinfo {author} {\bibfnamefont {L.}~\bibnamefont {Mazza}}, \bibinfo {author} {\bibfnamefont {E.}~\bibnamefont {Sela}},\ and\ \bibinfo {author} {\bibfnamefont {R.}~\bibnamefont {Fazio}},\ }\bibfield  {title} {\bibinfo {title} {Topological fractional pumping with alkaline-earth-like atoms in synthetic lattices},\ }\href {https://doi.org/10.1103/PhysRevLett.118.230402} {\bibfield  {journal} {\bibinfo  {journal} {Phys. Rev. Lett.}\ }\textbf {\bibinfo {volume} {118}},\ \bibinfo {pages} {230402} (\bibinfo {year} {2017})}\BibitemShut {NoStop}%
\bibitem [{\citenamefont {Lee}\ \emph {et~al.}(2018{\natexlab{a}})\citenamefont {Lee}, \citenamefont {Wang}, \citenamefont {Chen},\ and\ \citenamefont {Zhang}}]{2018PhysRevB.98.094434}%
  \BibitemOpen
  \bibfield  {author} {\bibinfo {author} {\bibfnamefont {C.~H.}\ \bibnamefont {Lee}}, \bibinfo {author} {\bibfnamefont {Y.}~\bibnamefont {Wang}}, \bibinfo {author} {\bibfnamefont {Y.}~\bibnamefont {Chen}},\ and\ \bibinfo {author} {\bibfnamefont {X.}~\bibnamefont {Zhang}},\ }\bibfield  {title} {\bibinfo {title} {Electromagnetic response of quantum hall systems in dimensions five and six and beyond},\ }\href {https://doi.org/10.1103/PhysRevB.98.094434} {\bibfield  {journal} {\bibinfo  {journal} {Phys. Rev. B}\ }\textbf {\bibinfo {volume} {98}},\ \bibinfo {pages} {094434} (\bibinfo {year} {2018}{\natexlab{a}})}\BibitemShut {NoStop}%
\bibitem [{\citenamefont {Qin}\ \emph {et~al.}(2024)\citenamefont {Qin}, \citenamefont {Shen}, \citenamefont {Li},\ and\ \citenamefont {Lee}}]{2024PhysRevA.109.053311}%
  \BibitemOpen
  \bibfield  {author} {\bibinfo {author} {\bibfnamefont {F.}~\bibnamefont {Qin}}, \bibinfo {author} {\bibfnamefont {R.}~\bibnamefont {Shen}}, \bibinfo {author} {\bibfnamefont {L.}~\bibnamefont {Li}},\ and\ \bibinfo {author} {\bibfnamefont {C.~H.}\ \bibnamefont {Lee}},\ }\bibfield  {title} {\bibinfo {title} {Kinked linear response from non-hermitian cold-atom pumping},\ }\href {https://doi.org/10.1103/PhysRevA.109.053311} {\bibfield  {journal} {\bibinfo  {journal} {Phys. Rev. A}\ }\textbf {\bibinfo {volume} {109}},\ \bibinfo {pages} {053311} (\bibinfo {year} {2024})}\BibitemShut {NoStop}%
\bibitem [{\citenamefont {J\"urgensen}\ \emph {et~al.}(2025)\citenamefont {J\"urgensen}, \citenamefont {Steiner}, \citenamefont {Refael},\ and\ \citenamefont {Rechtsman}}]{Juergensen2025FractionalPump}%
  \BibitemOpen
  \bibfield  {author} {\bibinfo {author} {\bibfnamefont {M.}~\bibnamefont {J\"urgensen}}, \bibinfo {author} {\bibfnamefont {J.}~\bibnamefont {Steiner}}, \bibinfo {author} {\bibfnamefont {G.}~\bibnamefont {Refael}},\ and\ \bibinfo {author} {\bibfnamefont {M.~C.}\ \bibnamefont {Rechtsman}},\ }\bibfield  {title} {\bibinfo {title} {Multiband fractional thouless pumps},\ }\href {https://doi.org/10.1103/4d5s-n4gn} {\bibfield  {journal} {\bibinfo  {journal} {Phys. Rev. Lett.}\ }\textbf {\bibinfo {volume} {135}},\ \bibinfo {pages} {166601} (\bibinfo {year} {2025})}\BibitemShut {NoStop}%
\bibitem [{\citenamefont {J{\"u}rgensen}\ \emph {et~al.}(2021)\citenamefont {J{\"u}rgensen}, \citenamefont {Mukherjee},\ and\ \citenamefont {Rechtsman}}]{Jurgensen2021QuantizedNonlinear}%
  \BibitemOpen
  \bibfield  {author} {\bibinfo {author} {\bibfnamefont {M.}~\bibnamefont {J{\"u}rgensen}}, \bibinfo {author} {\bibfnamefont {S.}~\bibnamefont {Mukherjee}},\ and\ \bibinfo {author} {\bibfnamefont {M.~C.}\ \bibnamefont {Rechtsman}},\ }\bibfield  {title} {\bibinfo {title} {Quantized nonlinear thouless pumping},\ }\href {https://doi.org/10.1038/s41586-021-03688-9} {\bibfield  {journal} {\bibinfo  {journal} {Nature}\ }\textbf {\bibinfo {volume} {596}},\ \bibinfo {pages} {63} (\bibinfo {year} {2021})}\BibitemShut {NoStop}%
\bibitem [{\citenamefont {J{\"u}rgensen}\ and\ \citenamefont {Rechtsman}(2022)}]{Jurgensen2022ChernNumber}%
  \BibitemOpen
  \bibfield  {author} {\bibinfo {author} {\bibfnamefont {M.}~\bibnamefont {J{\"u}rgensen}}\ and\ \bibinfo {author} {\bibfnamefont {M.~C.}\ \bibnamefont {Rechtsman}},\ }\bibfield  {title} {\bibinfo {title} {Chern number governs soliton motion in nonlinear thouless pumps},\ }\href {https://doi.org/10.1103/PhysRevLett.128.113901} {\bibfield  {journal} {\bibinfo  {journal} {Phys. Rev. Lett.}\ }\textbf {\bibinfo {volume} {128}},\ \bibinfo {pages} {113901} (\bibinfo {year} {2022})}\BibitemShut {NoStop}%
\bibitem [{\citenamefont {Fu}\ \emph {et~al.}(2022)\citenamefont {Fu}, \citenamefont {Wang}, \citenamefont {Kartashov}, \citenamefont {Konotop},\ and\ \citenamefont {Ye}}]{Fu2022NonlinearThouless}%
  \BibitemOpen
  \bibfield  {author} {\bibinfo {author} {\bibfnamefont {Q.}~\bibnamefont {Fu}}, \bibinfo {author} {\bibfnamefont {P.}~\bibnamefont {Wang}}, \bibinfo {author} {\bibfnamefont {Y.~V.}\ \bibnamefont {Kartashov}}, \bibinfo {author} {\bibfnamefont {V.~V.}\ \bibnamefont {Konotop}},\ and\ \bibinfo {author} {\bibfnamefont {F.}~\bibnamefont {Ye}},\ }\bibfield  {title} {\bibinfo {title} {Nonlinear thouless pumping: Solitons and transport breakdown},\ }\href {https://doi.org/10.1103/PhysRevLett.128.154101} {\bibfield  {journal} {\bibinfo  {journal} {Phys. Rev. Lett.}\ }\textbf {\bibinfo {volume} {128}},\ \bibinfo {pages} {154101} (\bibinfo {year} {2022})}\BibitemShut {NoStop}%
\bibitem [{\citenamefont {Mostaan}\ \emph {et~al.}(2022)\citenamefont {Mostaan}, \citenamefont {Grusdt},\ and\ \citenamefont {Goldman}}]{Mostaan2022QuantizedTopological}%
  \BibitemOpen
  \bibfield  {author} {\bibinfo {author} {\bibfnamefont {N.}~\bibnamefont {Mostaan}}, \bibinfo {author} {\bibfnamefont {F.}~\bibnamefont {Grusdt}},\ and\ \bibinfo {author} {\bibfnamefont {N.}~\bibnamefont {Goldman}},\ }\bibfield  {title} {\bibinfo {title} {Quantized topological pumping of solitons in nonlinear photonics and ultracold atomic mixtures},\ }\href {https://doi.org/10.1038/s41467-022-33478-4} {\bibfield  {journal} {\bibinfo  {journal} {Nat. Commun.}\ }\textbf {\bibinfo {volume} {13}},\ \bibinfo {pages} {5997} (\bibinfo {year} {2022})}\BibitemShut {NoStop}%
\bibitem [{\citenamefont {J{\"u}rgensen}\ \emph {et~al.}(2023)\citenamefont {J{\"u}rgensen}, \citenamefont {Mukherjee}, \citenamefont {J{\"o}rg},\ and\ \citenamefont {Rechtsman}}]{Jurgensen2023QuantizedFractional}%
  \BibitemOpen
  \bibfield  {author} {\bibinfo {author} {\bibfnamefont {M.}~\bibnamefont {J{\"u}rgensen}}, \bibinfo {author} {\bibfnamefont {S.}~\bibnamefont {Mukherjee}}, \bibinfo {author} {\bibfnamefont {C.}~\bibnamefont {J{\"o}rg}},\ and\ \bibinfo {author} {\bibfnamefont {M.~C.}\ \bibnamefont {Rechtsman}},\ }\bibfield  {title} {\bibinfo {title} {Quantized fractional thouless pumping of solitons},\ }\href {https://doi.org/10.1038/s41567-022-01871-x} {\bibfield  {journal} {\bibinfo  {journal} {Nat. Phys.}\ }\textbf {\bibinfo {volume} {19}},\ \bibinfo {pages} {420} (\bibinfo {year} {2023})}\BibitemShut {NoStop}%
\bibitem [{\citenamefont {Xiao}\ \emph {et~al.}(2025)\citenamefont {Xiao}, \citenamefont {You}, \citenamefont {Ke},\ and\ \citenamefont {Lee}}]{Xiao2025Nonlinear}%
  \BibitemOpen
  \bibfield  {author} {\bibinfo {author} {\bibfnamefont {L.}~\bibnamefont {Xiao}}, \bibinfo {author} {\bibfnamefont {X.}~\bibnamefont {You}}, \bibinfo {author} {\bibfnamefont {Y.}~\bibnamefont {Ke}},\ and\ \bibinfo {author} {\bibfnamefont {C.}~\bibnamefont {Lee}},\ }\bibfield  {title} {\bibinfo {title} {{Nonlinear topological pumping of solitons with time-dependent interactions}},\ }\href {https://doi.org/10.15302/frontphys.2025.062202} {\bibfield  {journal} {\bibinfo  {journal} {Front. Phys.}\ }\textbf {\bibinfo {volume} {20}},\ \bibinfo {pages} {062202} (\bibinfo {year} {2025})}\BibitemShut {NoStop}%
\bibitem [{\citenamefont {You}\ \emph {et~al.}(2025)\citenamefont {You}, \citenamefont {Xiao}, \citenamefont {Huang}, \citenamefont {Ke},\ and\ \citenamefont {Lee}}]{2025PhysRevA111033306}%
  \BibitemOpen
  \bibfield  {author} {\bibinfo {author} {\bibfnamefont {X.}~\bibnamefont {You}}, \bibinfo {author} {\bibfnamefont {L.}~\bibnamefont {Xiao}}, \bibinfo {author} {\bibfnamefont {B.}~\bibnamefont {Huang}}, \bibinfo {author} {\bibfnamefont {Y.}~\bibnamefont {Ke}},\ and\ \bibinfo {author} {\bibfnamefont {C.}~\bibnamefont {Lee}},\ }\bibfield  {title} {\bibinfo {title} {Nonlinear topological pumping of edge solitons},\ }\href {https://doi.org/10.1103/PhysRevA.111.033306} {\bibfield  {journal} {\bibinfo  {journal} {Phys. Rev. A}\ }\textbf {\bibinfo {volume} {111}},\ \bibinfo {pages} {033306} (\bibinfo {year} {2025})}\BibitemShut {NoStop}%
\bibitem [{\citenamefont {Ravets}\ \emph {et~al.}(2025)\citenamefont {Ravets}, \citenamefont {Pernet}, \citenamefont {Mostaan}, \citenamefont {Goldman},\ and\ \citenamefont {Bloch}}]{Ravets2025KerrPump}%
  \BibitemOpen
  \bibfield  {author} {\bibinfo {author} {\bibfnamefont {S.}~\bibnamefont {Ravets}}, \bibinfo {author} {\bibfnamefont {N.}~\bibnamefont {Pernet}}, \bibinfo {author} {\bibfnamefont {N.}~\bibnamefont {Mostaan}}, \bibinfo {author} {\bibfnamefont {N.}~\bibnamefont {Goldman}},\ and\ \bibinfo {author} {\bibfnamefont {J.}~\bibnamefont {Bloch}},\ }\bibfield  {title} {\bibinfo {title} {Thouless pumping in a driven-dissipative kerr resonator array},\ }\href {https://doi.org/10.1103/PhysRevLett.134.093801} {\bibfield  {journal} {\bibinfo  {journal} {Phys. Rev. Lett.}\ }\textbf {\bibinfo {volume} {134}},\ \bibinfo {pages} {093801} (\bibinfo {year} {2025})}\BibitemShut {NoStop}%
\bibitem [{\citenamefont {Tao}\ \emph {et~al.}(2025)\citenamefont {Tao}, \citenamefont {Zhang},\ and\ \citenamefont {Xu}}]{202596f5-qszj}%
  \BibitemOpen
  \bibfield  {author} {\bibinfo {author} {\bibfnamefont {Y.-L.}\ \bibnamefont {Tao}}, \bibinfo {author} {\bibfnamefont {Y.}~\bibnamefont {Zhang}},\ and\ \bibinfo {author} {\bibfnamefont {Y.}~\bibnamefont {Xu}},\ }\bibfield  {title} {\bibinfo {title} {Nonlinearity-induced fractional thouless pumping of solitons},\ }\href {https://doi.org/10.1103/96f5-qszj} {\bibfield  {journal} {\bibinfo  {journal} {Phys. Rev. Lett.}\ }\textbf {\bibinfo {volume} {135}},\ \bibinfo {pages} {097202} (\bibinfo {year} {2025})}\BibitemShut {NoStop}%
\bibitem [{\citenamefont {Gong}\ \emph {et~al.}(2009{\natexlab{a}})\citenamefont {Gong}, \citenamefont {Morales-Molina},\ and\ \citenamefont {H\"anggi}}]{2009PhysRevLett103133002}%
  \BibitemOpen
  \bibfield  {author} {\bibinfo {author} {\bibfnamefont {J.}~\bibnamefont {Gong}}, \bibinfo {author} {\bibfnamefont {L.}~\bibnamefont {Morales-Molina}},\ and\ \bibinfo {author} {\bibfnamefont {P.}~\bibnamefont {H\"anggi}},\ }\bibfield  {title} {\bibinfo {title} {Many-body coherent destruction of tunneling},\ }\href {https://doi.org/10.1103/PhysRevLett.103.133002} {\bibfield  {journal} {\bibinfo  {journal} {Phys. Rev. Lett.}\ }\textbf {\bibinfo {volume} {103}},\ \bibinfo {pages} {133002} (\bibinfo {year} {2009}{\natexlab{a}})}\BibitemShut {NoStop}%
\bibitem [{\citenamefont {Keilmann}\ \emph {et~al.}(2011)\citenamefont {Keilmann}, \citenamefont {Lanzmich}, \citenamefont {McCulloch},\ and\ \citenamefont {Roncaglia}}]{Keilmann2011DDPeierls}%
  \BibitemOpen
  \bibfield  {author} {\bibinfo {author} {\bibfnamefont {T.}~\bibnamefont {Keilmann}}, \bibinfo {author} {\bibfnamefont {S.}~\bibnamefont {Lanzmich}}, \bibinfo {author} {\bibfnamefont {I.}~\bibnamefont {McCulloch}},\ and\ \bibinfo {author} {\bibfnamefont {M.}~\bibnamefont {Roncaglia}},\ }\bibfield  {title} {\bibinfo {title} {Statistically induced phase transitions and anyons in 1d optical lattices},\ }\href {https://doi.org/10.1103/PhysRevLett.107.255304} {\bibfield  {journal} {\bibinfo  {journal} {Phys. Rev. Lett.}\ }\textbf {\bibinfo {volume} {107}},\ \bibinfo {pages} {255304} (\bibinfo {year} {2011})}\BibitemShut {NoStop}%
\bibitem [{\citenamefont {Greschner}\ and\ \citenamefont {Santos}(2014)}]{Greschner2014DDGauge}%
  \BibitemOpen
  \bibfield  {author} {\bibinfo {author} {\bibfnamefont {S.}~\bibnamefont {Greschner}}\ and\ \bibinfo {author} {\bibfnamefont {L.}~\bibnamefont {Santos}},\ }\bibfield  {title} {\bibinfo {title} {Density-dependent synthetic magnetism for ultracold bosons in optical lattices},\ }\href {https://doi.org/10.1103/PhysRevLett.113.215303} {\bibfield  {journal} {\bibinfo  {journal} {Phys. Rev. Lett.}\ }\textbf {\bibinfo {volume} {113}},\ \bibinfo {pages} {215303} (\bibinfo {year} {2014})}\BibitemShut {NoStop}%
\bibitem [{\citenamefont {Meinert}\ \emph {et~al.}(2016{\natexlab{a}})\citenamefont {Meinert}, \citenamefont {Mark}, \citenamefont {Lauber}, \citenamefont {Daley},\ and\ \citenamefont {N\"agerl}}]{2016PhysRevLett116205301}%
  \BibitemOpen
  \bibfield  {author} {\bibinfo {author} {\bibfnamefont {F.}~\bibnamefont {Meinert}}, \bibinfo {author} {\bibfnamefont {M.~J.}\ \bibnamefont {Mark}}, \bibinfo {author} {\bibfnamefont {K.}~\bibnamefont {Lauber}}, \bibinfo {author} {\bibfnamefont {A.~J.}\ \bibnamefont {Daley}},\ and\ \bibinfo {author} {\bibfnamefont {H.-C.}\ \bibnamefont {N\"agerl}},\ }\bibfield  {title} {\bibinfo {title} {Floquet engineering of correlated tunneling in the bose-hubbard model with ultracold atoms},\ }\href {https://doi.org/10.1103/PhysRevLett.116.205301} {\bibfield  {journal} {\bibinfo  {journal} {Phys. Rev. Lett.}\ }\textbf {\bibinfo {volume} {116}},\ \bibinfo {pages} {205301} (\bibinfo {year} {2016}{\natexlab{a}})}\BibitemShut {NoStop}%
\bibitem [{\citenamefont {Yang}\ \emph {et~al.}(2017)\citenamefont {Yang}, \citenamefont {Hu}, \citenamefont {Lee},\ and\ \citenamefont {Papi\ifmmode~\acute{c}\else \'{c}\fi{}}}]{2017PhysRevLett.118.146403}%
  \BibitemOpen
  \bibfield  {author} {\bibinfo {author} {\bibfnamefont {B.}~\bibnamefont {Yang}}, \bibinfo {author} {\bibfnamefont {Z.-X.}\ \bibnamefont {Hu}}, \bibinfo {author} {\bibfnamefont {C.~H.}\ \bibnamefont {Lee}},\ and\ \bibinfo {author} {\bibfnamefont {Z.}~\bibnamefont {Papi\ifmmode~\acute{c}\else \'{c}\fi{}}},\ }\bibfield  {title} {\bibinfo {title} {Generalized pseudopotentials for the anisotropic fractional quantum hall effect},\ }\href {https://doi.org/10.1103/PhysRevLett.118.146403} {\bibfield  {journal} {\bibinfo  {journal} {Phys. Rev. Lett.}\ }\textbf {\bibinfo {volume} {118}},\ \bibinfo {pages} {146403} (\bibinfo {year} {2017})}\BibitemShut {NoStop}%
\bibitem [{\citenamefont {Lee}\ \emph {et~al.}(2018{\natexlab{b}})\citenamefont {Lee}, \citenamefont {Ho}, \citenamefont {Yang}, \citenamefont {Gong},\ and\ \citenamefont {Papi\ifmmode~\acute{c}\else \'{c}\fi{}}}]{2018PhysRevLett.121.237401}%
  \BibitemOpen
  \bibfield  {author} {\bibinfo {author} {\bibfnamefont {C.~H.}\ \bibnamefont {Lee}}, \bibinfo {author} {\bibfnamefont {W.~W.}\ \bibnamefont {Ho}}, \bibinfo {author} {\bibfnamefont {B.}~\bibnamefont {Yang}}, \bibinfo {author} {\bibfnamefont {J.}~\bibnamefont {Gong}},\ and\ \bibinfo {author} {\bibfnamefont {Z.}~\bibnamefont {Papi\ifmmode~\acute{c}\else \'{c}\fi{}}},\ }\bibfield  {title} {\bibinfo {title} {Floquet mechanism for non-abelian fractional quantum hall states},\ }\href {https://doi.org/10.1103/PhysRevLett.121.237401} {\bibfield  {journal} {\bibinfo  {journal} {Phys. Rev. Lett.}\ }\textbf {\bibinfo {volume} {121}},\ \bibinfo {pages} {237401} (\bibinfo {year} {2018}{\natexlab{b}})}\BibitemShut {NoStop}%
\bibitem [{\citenamefont {Shen}\ and\ \citenamefont {Lee}(2022)}]{2022CommunPhys5238}%
  \BibitemOpen
  \bibfield  {author} {\bibinfo {author} {\bibfnamefont {R.}~\bibnamefont {Shen}}\ and\ \bibinfo {author} {\bibfnamefont {C.~H.}\ \bibnamefont {Lee}},\ }\bibfield  {title} {\bibinfo {title} {Non-hermitian skin clusters from strong interactions},\ }\href {https://doi.org/10.1038/s42005-022-01015-w} {\bibfield  {journal} {\bibinfo  {journal} {Commun. Phys.}\ }\textbf {\bibinfo {volume} {5}},\ \bibinfo {pages} {238} (\bibinfo {year} {2022})}\BibitemShut {NoStop}%
\bibitem [{\citenamefont {Faugno}\ and\ \citenamefont {Ozawa}(2022)}]{2022PhysRevLett.129.180401}%
  \BibitemOpen
  \bibfield  {author} {\bibinfo {author} {\bibfnamefont {W.~N.}\ \bibnamefont {Faugno}}\ and\ \bibinfo {author} {\bibfnamefont {T.}~\bibnamefont {Ozawa}},\ }\bibfield  {title} {\bibinfo {title} {Interaction-induced non-hermitian topological phases from a dynamical gauge field},\ }\href {https://doi.org/10.1103/PhysRevLett.129.180401} {\bibfield  {journal} {\bibinfo  {journal} {Phys. Rev. Lett.}\ }\textbf {\bibinfo {volume} {129}},\ \bibinfo {pages} {180401} (\bibinfo {year} {2022})}\BibitemShut {NoStop}%
\bibitem [{\citenamefont {Faugno}\ \emph {et~al.}(2024)\citenamefont {Faugno}, \citenamefont {Salerno},\ and\ \citenamefont {Ozawa}}]{2024PhysRevLett132023401}%
  \BibitemOpen
  \bibfield  {author} {\bibinfo {author} {\bibfnamefont {W.~N.}\ \bibnamefont {Faugno}}, \bibinfo {author} {\bibfnamefont {M.}~\bibnamefont {Salerno}},\ and\ \bibinfo {author} {\bibfnamefont {T.}~\bibnamefont {Ozawa}},\ }\bibfield  {title} {\bibinfo {title} {Density dependent gauge field inducing emergent su-schrieffer-heeger physics, solitons, and condensates in a discrete nonlinear schr\"odinger equation},\ }\href {https://doi.org/10.1103/PhysRevLett.132.023401} {\bibfield  {journal} {\bibinfo  {journal} {Phys. Rev. Lett.}\ }\textbf {\bibinfo {volume} {132}},\ \bibinfo {pages} {023401} (\bibinfo {year} {2024})}\BibitemShut {NoStop}%
\bibitem [{\citenamefont {Koh}\ \emph {et~al.}(2024)\citenamefont {Koh}, \citenamefont {Tai},\ and\ \citenamefont {Lee}}]{2024NatCommun155807}%
  \BibitemOpen
  \bibfield  {author} {\bibinfo {author} {\bibfnamefont {J.}~\bibnamefont {Koh}}, \bibinfo {author} {\bibfnamefont {T.}~\bibnamefont {Tai}},\ and\ \bibinfo {author} {\bibfnamefont {C.}~\bibnamefont {Lee}},\ }\bibfield  {title} {\bibinfo {title} {Realization of higher-order topological lattices on a quantum computer},\ }\href {https://doi.org/10.1038/s41467-024-49648-5} {\bibfield  {journal} {\bibinfo  {journal} {Nat. Commun.}\ }\textbf {\bibinfo {volume} {15}},\ \bibinfo {pages} {5807} (\bibinfo {year} {2024})}\BibitemShut {NoStop}%
\bibitem [{\citenamefont {Padhan}\ \emph {et~al.}(2025)\citenamefont {Padhan}, \citenamefont {Barbiero},\ and\ \citenamefont {Mishra}}]{2025lw8k-7h6p}%
  \BibitemOpen
  \bibfield  {author} {\bibinfo {author} {\bibfnamefont {A.}~\bibnamefont {Padhan}}, \bibinfo {author} {\bibfnamefont {L.}~\bibnamefont {Barbiero}},\ and\ \bibinfo {author} {\bibfnamefont {T.}~\bibnamefont {Mishra}},\ }\bibfield  {title} {\bibinfo {title} {Correlated-hopping-induced topological order in an atomic mixture},\ }\href {https://doi.org/10.1103/lw8k-7h6p} {\bibfield  {journal} {\bibinfo  {journal} {Phys. Rev. A}\ }\textbf {\bibinfo {volume} {112}},\ \bibinfo {pages} {L011305} (\bibinfo {year} {2025})}\BibitemShut {NoStop}%
\bibitem [{\citenamefont {Lei}\ and\ \citenamefont {Li}(2026)}]{arXivLei2025}%
  \BibitemOpen
  \bibfield  {author} {\bibinfo {author} {\bibfnamefont {Z.}~\bibnamefont {Lei}}\ and\ \bibinfo {author} {\bibfnamefont {L.}~\bibnamefont {Li}},\ }\bibfield  {title} {\bibinfo {title} {Inter-species topological phases via a dynamical gauge field},\ }\href {https://doi.org/10.1007/s11433-025-2882-5} {\bibfield  {journal} {\bibinfo  {journal} {Sci. China Phys. Mech. Astron.}\ }\textbf {\bibinfo {volume} {69}},\ \bibinfo {pages} {257811} (\bibinfo {year} {2026})}\BibitemShut {NoStop}%
\bibitem [{\citenamefont {Kogut}(1979)}]{Kogut1979}%
  \BibitemOpen
  \bibfield  {author} {\bibinfo {author} {\bibfnamefont {J.~B.}\ \bibnamefont {Kogut}},\ }\bibfield  {title} {\bibinfo {title} {An introduction to lattice gauge theory and spin systems},\ }\href {https://doi.org/10.1103/RevModPhys.51.659} {\bibfield  {journal} {\bibinfo  {journal} {Rev. Mod. Phys.}\ }\textbf {\bibinfo {volume} {51}},\ \bibinfo {pages} {659} (\bibinfo {year} {1979})}\BibitemShut {NoStop}%
\bibitem [{\citenamefont {Kogut}(1983)}]{1983RevModPhys55775}%
  \BibitemOpen
  \bibfield  {author} {\bibinfo {author} {\bibfnamefont {J.~B.}\ \bibnamefont {Kogut}},\ }\bibfield  {title} {\bibinfo {title} {The lattice gauge theory approach to quantum chromodynamics},\ }\href {https://doi.org/10.1103/RevModPhys.55.775} {\bibfield  {journal} {\bibinfo  {journal} {Rev. Mod. Phys.}\ }\textbf {\bibinfo {volume} {55}},\ \bibinfo {pages} {775} (\bibinfo {year} {1983})}\BibitemShut {NoStop}%
\bibitem [{\citenamefont {Wiese}(2013)}]{Wiese2013}%
  \BibitemOpen
  \bibfield  {author} {\bibinfo {author} {\bibfnamefont {U.-J.}\ \bibnamefont {Wiese}},\ }\bibfield  {title} {\bibinfo {title} {Ultracold quantum gases and lattice systems: quantum simulation of lattice gauge theories},\ }\href {https://doi.org/10.1002/andp.201300104} {\bibfield  {journal} {\bibinfo  {journal} {Annalen der Physik}\ }\textbf {\bibinfo {volume} {525}},\ \bibinfo {pages} {777} (\bibinfo {year} {2013})}\BibitemShut {NoStop}%
\bibitem [{\citenamefont {Schweizer}\ \emph {et~al.}(2019{\natexlab{a}})\citenamefont {Schweizer} \emph {et~al.}}]{Schweizer2019Z2}%
  \BibitemOpen
  \bibfield  {author} {\bibinfo {author} {\bibfnamefont {C.}~\bibnamefont {Schweizer}} \emph {et~al.},\ }\bibfield  {title} {\bibinfo {title} {Floquet approach to {Z}$_2$ lattice gauge theories with ultracold atoms in optical lattices},\ }\href {https://doi.org/10.1038/s41567-019-0649-7} {\bibfield  {journal} {\bibinfo  {journal} {Nat. Phys.}\ }\textbf {\bibinfo {volume} {15}},\ \bibinfo {pages} {1168} (\bibinfo {year} {2019}{\natexlab{a}})}\BibitemShut {NoStop}%
\bibitem [{\citenamefont {Ba{\~n}uls}\ \emph {et~al.}(2020)\citenamefont {Ba{\~n}uls}, \citenamefont {Blatt}, \citenamefont {Catani}, \citenamefont {Celi}, \citenamefont {Cirac}, \citenamefont {Dalmonte}, \citenamefont {Fallani}, \citenamefont {Jansen}, \citenamefont {Lewenstein}, \citenamefont {Montangero} \emph {et~al.}}]{Banuls2020QT}%
  \BibitemOpen
  \bibfield  {author} {\bibinfo {author} {\bibfnamefont {M.~C.}\ \bibnamefont {Ba{\~n}uls}}, \bibinfo {author} {\bibfnamefont {R.}~\bibnamefont {Blatt}}, \bibinfo {author} {\bibfnamefont {J.}~\bibnamefont {Catani}}, \bibinfo {author} {\bibfnamefont {A.}~\bibnamefont {Celi}}, \bibinfo {author} {\bibfnamefont {J.~I.}\ \bibnamefont {Cirac}}, \bibinfo {author} {\bibfnamefont {M.}~\bibnamefont {Dalmonte}}, \bibinfo {author} {\bibfnamefont {L.}~\bibnamefont {Fallani}}, \bibinfo {author} {\bibfnamefont {K.}~\bibnamefont {Jansen}}, \bibinfo {author} {\bibfnamefont {M.}~\bibnamefont {Lewenstein}}, \bibinfo {author} {\bibfnamefont {S.}~\bibnamefont {Montangero}}, \emph {et~al.},\ }\bibfield  {title} {\bibinfo {title} {Simulating lattice gauge theories within quantum technologies},\ }\href {https://doi.org/10.1140/epjd/e2020-100571-8} {\bibfield  {journal} {\bibinfo  {journal} {Eur. Phys. J. D}\ }\textbf {\bibinfo {volume} {74}},\ \bibinfo {pages} {165} (\bibinfo {year} {2020})}\BibitemShut {NoStop}%
\bibitem [{\citenamefont {Cheng}\ and\ \citenamefont {Zhai}(2024)}]{Cheng2024EmergentU1}%
  \BibitemOpen
  \bibfield  {author} {\bibinfo {author} {\bibfnamefont {Y.}~\bibnamefont {Cheng}}\ and\ \bibinfo {author} {\bibfnamefont {H.}~\bibnamefont {Zhai}},\ }\bibfield  {title} {\bibinfo {title} {Emergent u(1) lattice gauge theory in rydberg atom arrays},\ }\href {https://doi.org/10.1038/s42254-024-00749-6} {\bibfield  {journal} {\bibinfo  {journal} {Nat. Rev. Phys.}\ }\textbf {\bibinfo {volume} {6}},\ \bibinfo {pages} {566} (\bibinfo {year} {2024})}\BibitemShut {NoStop}%
\bibitem [{\citenamefont {Halimeh}\ \emph {et~al.}(2022{\natexlab{a}})\citenamefont {Halimeh}, \citenamefont {McCulloch}, \citenamefont {Yang},\ and\ \citenamefont {Hauke}}]{2022PRXQuantum3040316}%
  \BibitemOpen
  \bibfield  {author} {\bibinfo {author} {\bibfnamefont {J.~C.}\ \bibnamefont {Halimeh}}, \bibinfo {author} {\bibfnamefont {I.~P.}\ \bibnamefont {McCulloch}}, \bibinfo {author} {\bibfnamefont {B.}~\bibnamefont {Yang}},\ and\ \bibinfo {author} {\bibfnamefont {P.}~\bibnamefont {Hauke}},\ }\bibfield  {title} {\bibinfo {title} {Tuning the topological $\ensuremath{\theta}$-angle in cold-atom quantum simulators of gauge theories},\ }\href {https://doi.org/10.1103/PRXQuantum.3.040316} {\bibfield  {journal} {\bibinfo  {journal} {PRX Quantum}\ }\textbf {\bibinfo {volume} {3}},\ \bibinfo {pages} {040316} (\bibinfo {year} {2022}{\natexlab{a}})}\BibitemShut {NoStop}%
\bibitem [{\citenamefont {Cheng}\ \emph {et~al.}(2022)\citenamefont {Cheng}, \citenamefont {Liu}, \citenamefont {Zheng}, \citenamefont {Zhang},\ and\ \citenamefont {Zhai}}]{2022PRXQuantum3040317}%
  \BibitemOpen
  \bibfield  {author} {\bibinfo {author} {\bibfnamefont {Y.}~\bibnamefont {Cheng}}, \bibinfo {author} {\bibfnamefont {S.}~\bibnamefont {Liu}}, \bibinfo {author} {\bibfnamefont {W.}~\bibnamefont {Zheng}}, \bibinfo {author} {\bibfnamefont {P.}~\bibnamefont {Zhang}},\ and\ \bibinfo {author} {\bibfnamefont {H.}~\bibnamefont {Zhai}},\ }\bibfield  {title} {\bibinfo {title} {Tunable confinement-deconfinement transition in an ultracold-atom quantum simulator},\ }\href {https://doi.org/10.1103/PRXQuantum.3.040317} {\bibfield  {journal} {\bibinfo  {journal} {PRX Quantum}\ }\textbf {\bibinfo {volume} {3}},\ \bibinfo {pages} {040317} (\bibinfo {year} {2022})}\BibitemShut {NoStop}%
\bibitem [{\citenamefont {Zhang}\ \emph {et~al.}(2025)\citenamefont {Zhang}, \citenamefont {Liu}, \citenamefont {Cheng} \emph {et~al.}}]{Zhang2025MicroConfinement}%
  \BibitemOpen
  \bibfield  {author} {\bibinfo {author} {\bibfnamefont {W.-Y.}\ \bibnamefont {Zhang}}, \bibinfo {author} {\bibfnamefont {Y.}~\bibnamefont {Liu}}, \bibinfo {author} {\bibfnamefont {Y.}~\bibnamefont {Cheng}}, \emph {et~al.},\ }\bibfield  {title} {\bibinfo {title} {Observation of microscopic confinement dynamics by a tunable topological $\theta$-angle},\ }\href {https://doi.org/10.1038/s41567-024-02702-x} {\bibfield  {journal} {\bibinfo  {journal} {Nat. Phys.}\ }\textbf {\bibinfo {volume} {21}},\ \bibinfo {pages} {155} (\bibinfo {year} {2025})}\BibitemShut {NoStop}%
\bibitem [{\citenamefont {Banerjee}\ \emph {et~al.}(2012)\citenamefont {Banerjee}, \citenamefont {Dalmonte}, \citenamefont {M{\"u}ller}, \citenamefont {Rico}, \citenamefont {Stebler}, \citenamefont {Wiese},\ and\ \citenamefont {Zoller}}]{PhysRevLett.109.175302}%
  \BibitemOpen
  \bibfield  {author} {\bibinfo {author} {\bibfnamefont {D.}~\bibnamefont {Banerjee}}, \bibinfo {author} {\bibfnamefont {M.}~\bibnamefont {Dalmonte}}, \bibinfo {author} {\bibfnamefont {M.}~\bibnamefont {M{\"u}ller}}, \bibinfo {author} {\bibfnamefont {E.}~\bibnamefont {Rico}}, \bibinfo {author} {\bibfnamefont {P.}~\bibnamefont {Stebler}}, \bibinfo {author} {\bibfnamefont {U.-J.}\ \bibnamefont {Wiese}},\ and\ \bibinfo {author} {\bibfnamefont {P.}~\bibnamefont {Zoller}},\ }\bibfield  {title} {\bibinfo {title} {Atomic quantum simulation of dynamical gauge fields coupled to fermionic matter: From string breaking to evolution after a quench},\ }\href {https://doi.org/10.1103/PhysRevLett.109.175302} {\bibfield  {journal} {\bibinfo  {journal} {Phys. Rev. Lett.}\ }\textbf {\bibinfo {volume} {109}},\ \bibinfo {pages} {175302} (\bibinfo {year} {2012})}\BibitemShut {NoStop}%
\bibitem [{\citenamefont {Marcos}\ \emph {et~al.}(2013)\citenamefont {Marcos}, \citenamefont {Rabl}, \citenamefont {Rico},\ and\ \citenamefont {Zoller}}]{2013PhysRevLett111110504}%
  \BibitemOpen
  \bibfield  {author} {\bibinfo {author} {\bibfnamefont {D.}~\bibnamefont {Marcos}}, \bibinfo {author} {\bibfnamefont {P.}~\bibnamefont {Rabl}}, \bibinfo {author} {\bibfnamefont {E.}~\bibnamefont {Rico}},\ and\ \bibinfo {author} {\bibfnamefont {P.}~\bibnamefont {Zoller}},\ }\bibfield  {title} {\bibinfo {title} {Superconducting circuits for quantum simulation of dynamical gauge fields},\ }\href {https://doi.org/10.1103/PhysRevLett.111.110504} {\bibfield  {journal} {\bibinfo  {journal} {Phys. Rev. Lett.}\ }\textbf {\bibinfo {volume} {111}},\ \bibinfo {pages} {110504} (\bibinfo {year} {2013})}\BibitemShut {NoStop}%
\bibitem [{\citenamefont {Pichler}\ \emph {et~al.}(2016)\citenamefont {Pichler}, \citenamefont {Dalmonte}, \citenamefont {Rico}, \citenamefont {Zoller},\ and\ \citenamefont {Montangero}}]{2016PhysRevX6011023}%
  \BibitemOpen
  \bibfield  {author} {\bibinfo {author} {\bibfnamefont {T.}~\bibnamefont {Pichler}}, \bibinfo {author} {\bibfnamefont {M.}~\bibnamefont {Dalmonte}}, \bibinfo {author} {\bibfnamefont {E.}~\bibnamefont {Rico}}, \bibinfo {author} {\bibfnamefont {P.}~\bibnamefont {Zoller}},\ and\ \bibinfo {author} {\bibfnamefont {S.}~\bibnamefont {Montangero}},\ }\bibfield  {title} {\bibinfo {title} {Real-time dynamics in u(1) lattice gauge theories with tensor networks},\ }\href {https://doi.org/10.1103/PhysRevX.6.011023} {\bibfield  {journal} {\bibinfo  {journal} {Phys. Rev. X}\ }\textbf {\bibinfo {volume} {6}},\ \bibinfo {pages} {011023} (\bibinfo {year} {2016})}\BibitemShut {NoStop}%
\bibitem [{\citenamefont {Verdel}\ \emph {et~al.}(2020)\citenamefont {Verdel}, \citenamefont {Liu}, \citenamefont {Whitsitt}, \citenamefont {Gorshkov},\ and\ \citenamefont {Heyl}}]{2020PhysRevB102014308}%
  \BibitemOpen
  \bibfield  {author} {\bibinfo {author} {\bibfnamefont {R.}~\bibnamefont {Verdel}}, \bibinfo {author} {\bibfnamefont {F.}~\bibnamefont {Liu}}, \bibinfo {author} {\bibfnamefont {S.}~\bibnamefont {Whitsitt}}, \bibinfo {author} {\bibfnamefont {A.~V.}\ \bibnamefont {Gorshkov}},\ and\ \bibinfo {author} {\bibfnamefont {M.}~\bibnamefont {Heyl}},\ }\bibfield  {title} {\bibinfo {title} {Real-time dynamics of string breaking in quantum spin chains},\ }\href {https://doi.org/10.1103/PhysRevB.102.014308} {\bibfield  {journal} {\bibinfo  {journal} {Phys. Rev. B}\ }\textbf {\bibinfo {volume} {102}},\ \bibinfo {pages} {014308} (\bibinfo {year} {2020})}\BibitemShut {NoStop}%
\bibitem [{\citenamefont {Surace}\ \emph {et~al.}(2020)\citenamefont {Surace}, \citenamefont {Mazza}, \citenamefont {Giudici}, \citenamefont {Lerose}, \citenamefont {Gambassi},\ and\ \citenamefont {Dalmonte}}]{2020PhysRevX10021041}%
  \BibitemOpen
  \bibfield  {author} {\bibinfo {author} {\bibfnamefont {F.~M.}\ \bibnamefont {Surace}}, \bibinfo {author} {\bibfnamefont {P.~P.}\ \bibnamefont {Mazza}}, \bibinfo {author} {\bibfnamefont {G.}~\bibnamefont {Giudici}}, \bibinfo {author} {\bibfnamefont {A.}~\bibnamefont {Lerose}}, \bibinfo {author} {\bibfnamefont {A.}~\bibnamefont {Gambassi}},\ and\ \bibinfo {author} {\bibfnamefont {M.}~\bibnamefont {Dalmonte}},\ }\bibfield  {title} {\bibinfo {title} {Lattice gauge theories and string dynamics in rydberg atom quantum simulators},\ }\href {https://doi.org/10.1103/PhysRevX.10.021041} {\bibfield  {journal} {\bibinfo  {journal} {Phys. Rev. X}\ }\textbf {\bibinfo {volume} {10}},\ \bibinfo {pages} {021041} (\bibinfo {year} {2020})}\BibitemShut {NoStop}%
\bibitem [{\citenamefont {De}\ \emph {et~al.}(2024)\citenamefont {De}, \citenamefont {Lerose}, \citenamefont {Luo}, \citenamefont {Surace}, \citenamefont {Schuckert}, \citenamefont {Bennewitz}, \citenamefont {Ware}, \citenamefont {Morong}, \citenamefont {Collins}, \citenamefont {Davoudi}, \citenamefont {Gorshkov}, \citenamefont {Katz},\ and\ \citenamefont {Monroe}}]{De2024StringBreakingDynamics}%
  \BibitemOpen
  \bibfield  {author} {\bibinfo {author} {\bibfnamefont {A.}~\bibnamefont {De}}, \bibinfo {author} {\bibfnamefont {A.}~\bibnamefont {Lerose}}, \bibinfo {author} {\bibfnamefont {D.}~\bibnamefont {Luo}}, \bibinfo {author} {\bibfnamefont {F.~M.}\ \bibnamefont {Surace}}, \bibinfo {author} {\bibfnamefont {A.}~\bibnamefont {Schuckert}}, \bibinfo {author} {\bibfnamefont {E.~R.}\ \bibnamefont {Bennewitz}}, \bibinfo {author} {\bibfnamefont {B.}~\bibnamefont {Ware}}, \bibinfo {author} {\bibfnamefont {W.}~\bibnamefont {Morong}}, \bibinfo {author} {\bibfnamefont {K.~S.}\ \bibnamefont {Collins}}, \bibinfo {author} {\bibfnamefont {Z.}~\bibnamefont {Davoudi}}, \bibinfo {author} {\bibfnamefont {A.~V.}\ \bibnamefont {Gorshkov}}, \bibinfo {author} {\bibfnamefont {O.}~\bibnamefont {Katz}},\ and\ \bibinfo {author} {\bibfnamefont {C.}~\bibnamefont {Monroe}},\ }\bibfield  {title} {\bibinfo {title} {Observation of string-breaking dynamics in a quantum simulator},\ }\bibfield  {journal} {\bibinfo  {journal} {arXiv:2410.13815
  [quant-ph]}\ }\href {https://doi.org/10.48550/arXiv.2410.13815} {10.48550/arXiv.2410.13815} (\bibinfo {year} {2024})\BibitemShut {NoStop}%
\bibitem [{\citenamefont {Gonz{\'a}lez-Cuadra}\ \emph {et~al.}(2025)\citenamefont {Gonz{\'a}lez-Cuadra}, \citenamefont {Hamdan}, \citenamefont {Zache}, \citenamefont {Braverman}, \citenamefont {Kornja{\v c}a}, \citenamefont {Lukin}, \citenamefont {Cant{\'u}}, \citenamefont {Liu}, \citenamefont {Wang}, \citenamefont {Keesling}, \citenamefont {Lukin}, \citenamefont {Zoller},\ and\ \citenamefont {Bylinskii}}]{GonzalezCuadra2025StringBreakingRydberg}%
  \BibitemOpen
  \bibfield  {author} {\bibinfo {author} {\bibfnamefont {D.}~\bibnamefont {Gonz{\'a}lez-Cuadra}}, \bibinfo {author} {\bibfnamefont {M.}~\bibnamefont {Hamdan}}, \bibinfo {author} {\bibfnamefont {T.~V.}\ \bibnamefont {Zache}}, \bibinfo {author} {\bibfnamefont {B.}~\bibnamefont {Braverman}}, \bibinfo {author} {\bibfnamefont {M.}~\bibnamefont {Kornja{\v c}a}}, \bibinfo {author} {\bibfnamefont {A.}~\bibnamefont {Lukin}}, \bibinfo {author} {\bibfnamefont {S.~H.}\ \bibnamefont {Cant{\'u}}}, \bibinfo {author} {\bibfnamefont {F.}~\bibnamefont {Liu}}, \bibinfo {author} {\bibfnamefont {S.-T.}\ \bibnamefont {Wang}}, \bibinfo {author} {\bibfnamefont {A.}~\bibnamefont {Keesling}}, \bibinfo {author} {\bibfnamefont {M.~D.}\ \bibnamefont {Lukin}}, \bibinfo {author} {\bibfnamefont {P.}~\bibnamefont {Zoller}},\ and\ \bibinfo {author} {\bibfnamefont {A.}~\bibnamefont {Bylinskii}},\ }\bibfield  {title} {\bibinfo {title} {Observation of string breaking on a (2 + 1)d rydberg quantum simulator},\ }\href
  {https://doi.org/10.1038/s41586-025-09051-6} {\bibfield  {journal} {\bibinfo  {journal} {Nature}\ }\textbf {\bibinfo {volume} {642}},\ \bibinfo {pages} {321} (\bibinfo {year} {2025})}\BibitemShut {NoStop}%
\bibitem [{\citenamefont {Cochran}\ \emph {et~al.}(2025)\citenamefont {Cochran}, \citenamefont {Jobst}, \citenamefont {Rosenberg}, \citenamefont {Lensky}, \citenamefont {Gyawali}, \citenamefont {Eassa}, \citenamefont {Will}, \citenamefont {Szasz}, \citenamefont {Abanin}, \citenamefont {Acharya}, \citenamefont {Aghababaie~Beni}, \citenamefont {Andersen}, \citenamefont {Ansmann}, \citenamefont {Arute}, \citenamefont {Arya}, \citenamefont {Asfaw}, \citenamefont {Atalaya}, \citenamefont {Babbush}, \citenamefont {Ballard} \emph {et~al.}}]{Cochran2025VisualizingStrings}%
  \BibitemOpen
  \bibfield  {author} {\bibinfo {author} {\bibfnamefont {T.~A.}\ \bibnamefont {Cochran}}, \bibinfo {author} {\bibfnamefont {B.}~\bibnamefont {Jobst}}, \bibinfo {author} {\bibfnamefont {E.}~\bibnamefont {Rosenberg}}, \bibinfo {author} {\bibfnamefont {Y.~D.}\ \bibnamefont {Lensky}}, \bibinfo {author} {\bibfnamefont {G.}~\bibnamefont {Gyawali}}, \bibinfo {author} {\bibfnamefont {N.}~\bibnamefont {Eassa}}, \bibinfo {author} {\bibfnamefont {M.}~\bibnamefont {Will}}, \bibinfo {author} {\bibfnamefont {A.}~\bibnamefont {Szasz}}, \bibinfo {author} {\bibfnamefont {D.}~\bibnamefont {Abanin}}, \bibinfo {author} {\bibfnamefont {R.}~\bibnamefont {Acharya}}, \bibinfo {author} {\bibfnamefont {L.}~\bibnamefont {Aghababaie~Beni}}, \bibinfo {author} {\bibfnamefont {T.~I.}\ \bibnamefont {Andersen}}, \bibinfo {author} {\bibfnamefont {M.}~\bibnamefont {Ansmann}}, \bibinfo {author} {\bibfnamefont {F.}~\bibnamefont {Arute}}, \bibinfo {author} {\bibfnamefont {K.}~\bibnamefont {Arya}}, \bibinfo {author} {\bibfnamefont
  {A.}~\bibnamefont {Asfaw}}, \bibinfo {author} {\bibfnamefont {J.}~\bibnamefont {Atalaya}}, \bibinfo {author} {\bibfnamefont {R.}~\bibnamefont {Babbush}}, \bibinfo {author} {\bibfnamefont {B.}~\bibnamefont {Ballard}}, \emph {et~al.},\ }\bibfield  {title} {\bibinfo {title} {Visualizing dynamics of charges and strings in (2 + 1)d lattice gauge theories},\ }\href {https://doi.org/10.1038/s41586-025-08999-9} {\bibfield  {journal} {\bibinfo  {journal} {Nature}\ }\textbf {\bibinfo {volume} {642}},\ \bibinfo {pages} {315} (\bibinfo {year} {2025})}\BibitemShut {NoStop}%
\bibitem [{\citenamefont {Liu}\ \emph {et~al.}(2025{\natexlab{b}})\citenamefont {Liu}, \citenamefont {Zhang}, \citenamefont {Zhu}, \citenamefont {He}, \citenamefont {Yuan},\ and\ \citenamefont {Pan}}]{2025mwy1-v9hk}%
  \BibitemOpen
  \bibfield  {author} {\bibinfo {author} {\bibfnamefont {Y.}~\bibnamefont {Liu}}, \bibinfo {author} {\bibfnamefont {W.-Y.}\ \bibnamefont {Zhang}}, \bibinfo {author} {\bibfnamefont {Z.-H.}\ \bibnamefont {Zhu}}, \bibinfo {author} {\bibfnamefont {M.-G.}\ \bibnamefont {He}}, \bibinfo {author} {\bibfnamefont {Z.-S.}\ \bibnamefont {Yuan}},\ and\ \bibinfo {author} {\bibfnamefont {J.-W.}\ \bibnamefont {Pan}},\ }\bibfield  {title} {\bibinfo {title} {String-breaking mechanism in a lattice schwinger model simulator},\ }\href {https://doi.org/10.1103/mwy1-v9hk} {\bibfield  {journal} {\bibinfo  {journal} {Phys. Rev. Lett.}\ }\textbf {\bibinfo {volume} {135}},\ \bibinfo {pages} {101902} (\bibinfo {year} {2025}{\natexlab{b}})}\BibitemShut {NoStop}%
\bibitem [{\citenamefont {Smith}\ \emph {et~al.}(2017)\citenamefont {Smith}, \citenamefont {Knolle}, \citenamefont {Kovrizhin},\ and\ \citenamefont {Moessner}}]{2017PhysRevLett118266601}%
  \BibitemOpen
  \bibfield  {author} {\bibinfo {author} {\bibfnamefont {A.}~\bibnamefont {Smith}}, \bibinfo {author} {\bibfnamefont {J.}~\bibnamefont {Knolle}}, \bibinfo {author} {\bibfnamefont {D.~L.}\ \bibnamefont {Kovrizhin}},\ and\ \bibinfo {author} {\bibfnamefont {R.}~\bibnamefont {Moessner}},\ }\bibfield  {title} {\bibinfo {title} {Disorder-free localization},\ }\href {https://doi.org/10.1103/PhysRevLett.118.266601} {\bibfield  {journal} {\bibinfo  {journal} {Phys. Rev. Lett.}\ }\textbf {\bibinfo {volume} {118}},\ \bibinfo {pages} {266601} (\bibinfo {year} {2017})}\BibitemShut {NoStop}%
\bibitem [{\citenamefont {Karpov}\ \emph {et~al.}(2021)\citenamefont {Karpov}, \citenamefont {Verdel}, \citenamefont {Huang}, \citenamefont {Schmitt},\ and\ \citenamefont {Heyl}}]{2021PhysRevLett126130401}%
  \BibitemOpen
  \bibfield  {author} {\bibinfo {author} {\bibfnamefont {P.}~\bibnamefont {Karpov}}, \bibinfo {author} {\bibfnamefont {R.}~\bibnamefont {Verdel}}, \bibinfo {author} {\bibfnamefont {Y.-P.}\ \bibnamefont {Huang}}, \bibinfo {author} {\bibfnamefont {M.}~\bibnamefont {Schmitt}},\ and\ \bibinfo {author} {\bibfnamefont {M.}~\bibnamefont {Heyl}},\ }\bibfield  {title} {\bibinfo {title} {Disorder-free localization in an interacting 2d lattice gauge theory},\ }\href {https://doi.org/10.1103/PhysRevLett.126.130401} {\bibfield  {journal} {\bibinfo  {journal} {Phys. Rev. Lett.}\ }\textbf {\bibinfo {volume} {126}},\ \bibinfo {pages} {130401} (\bibinfo {year} {2021})}\BibitemShut {NoStop}%
\bibitem [{\citenamefont {Halimeh}\ \emph {et~al.}(2022{\natexlab{b}})\citenamefont {Halimeh}, \citenamefont {Homeier}, \citenamefont {Zhao}, \citenamefont {Bohrdt}, \citenamefont {Grusdt}, \citenamefont {Hauke},\ and\ \citenamefont {Knolle}}]{2022PRXQuantum3020345}%
  \BibitemOpen
  \bibfield  {author} {\bibinfo {author} {\bibfnamefont {J.~C.}\ \bibnamefont {Halimeh}}, \bibinfo {author} {\bibfnamefont {L.}~\bibnamefont {Homeier}}, \bibinfo {author} {\bibfnamefont {H.}~\bibnamefont {Zhao}}, \bibinfo {author} {\bibfnamefont {A.}~\bibnamefont {Bohrdt}}, \bibinfo {author} {\bibfnamefont {F.}~\bibnamefont {Grusdt}}, \bibinfo {author} {\bibfnamefont {P.}~\bibnamefont {Hauke}},\ and\ \bibinfo {author} {\bibfnamefont {J.}~\bibnamefont {Knolle}},\ }\bibfield  {title} {\bibinfo {title} {Enhancing disorder-free localization through dynamically emergent local symmetries},\ }\href {https://doi.org/10.1103/PRXQuantum.3.020345} {\bibfield  {journal} {\bibinfo  {journal} {PRX Quantum}\ }\textbf {\bibinfo {volume} {3}},\ \bibinfo {pages} {020345} (\bibinfo {year} {2022}{\natexlab{b}})}\BibitemShut {NoStop}%
\bibitem [{\citenamefont {Chakraborty}\ \emph {et~al.}(2023)\citenamefont {Chakraborty}, \citenamefont {Heyl}, \citenamefont {Karpov},\ and\ \citenamefont {Moessner}}]{2023PhysRevLett131220402}%
  \BibitemOpen
  \bibfield  {author} {\bibinfo {author} {\bibfnamefont {N.}~\bibnamefont {Chakraborty}}, \bibinfo {author} {\bibfnamefont {M.}~\bibnamefont {Heyl}}, \bibinfo {author} {\bibfnamefont {P.}~\bibnamefont {Karpov}},\ and\ \bibinfo {author} {\bibfnamefont {R.}~\bibnamefont {Moessner}},\ }\bibfield  {title} {\bibinfo {title} {Spectral response of disorder-free localized lattice gauge theories},\ }\href {https://doi.org/10.1103/PhysRevLett.131.220402} {\bibfield  {journal} {\bibinfo  {journal} {Phys. Rev. Lett.}\ }\textbf {\bibinfo {volume} {131}},\ \bibinfo {pages} {220402} (\bibinfo {year} {2023})}\BibitemShut {NoStop}%
\bibitem [{\citenamefont {Rico}\ \emph {et~al.}(2014)\citenamefont {Rico}, \citenamefont {Pichler}, \citenamefont {Dalmonte}, \citenamefont {Zoller},\ and\ \citenamefont {Montangero}}]{2014PhysRevLett112201601}%
  \BibitemOpen
  \bibfield  {author} {\bibinfo {author} {\bibfnamefont {E.}~\bibnamefont {Rico}}, \bibinfo {author} {\bibfnamefont {T.}~\bibnamefont {Pichler}}, \bibinfo {author} {\bibfnamefont {M.}~\bibnamefont {Dalmonte}}, \bibinfo {author} {\bibfnamefont {P.}~\bibnamefont {Zoller}},\ and\ \bibinfo {author} {\bibfnamefont {S.}~\bibnamefont {Montangero}},\ }\bibfield  {title} {\bibinfo {title} {Tensor networks for lattice gauge theories and atomic quantum simulation},\ }\href {https://doi.org/10.1103/PhysRevLett.112.201601} {\bibfield  {journal} {\bibinfo  {journal} {Phys. Rev. Lett.}\ }\textbf {\bibinfo {volume} {112}},\ \bibinfo {pages} {201601} (\bibinfo {year} {2014})}\BibitemShut {NoStop}%
\bibitem [{\citenamefont {Huang}\ \emph {et~al.}(2019)\citenamefont {Huang}, \citenamefont {Banerjee},\ and\ \citenamefont {Heyl}}]{2019PhysRevLett122250401}%
  \BibitemOpen
  \bibfield  {author} {\bibinfo {author} {\bibfnamefont {Y.-P.}\ \bibnamefont {Huang}}, \bibinfo {author} {\bibfnamefont {D.}~\bibnamefont {Banerjee}},\ and\ \bibinfo {author} {\bibfnamefont {M.}~\bibnamefont {Heyl}},\ }\bibfield  {title} {\bibinfo {title} {Dynamical quantum phase transitions in u(1) quantum link models},\ }\href {https://doi.org/10.1103/PhysRevLett.122.250401} {\bibfield  {journal} {\bibinfo  {journal} {Phys. Rev. Lett.}\ }\textbf {\bibinfo {volume} {122}},\ \bibinfo {pages} {250401} (\bibinfo {year} {2019})}\BibitemShut {NoStop}%
\bibitem [{\citenamefont {Yao}\ \emph {et~al.}(2022)\citenamefont {Yao}, \citenamefont {Pan}, \citenamefont {Liu},\ and\ \citenamefont {Zhai}}]{2022PhysRevB105125123}%
  \BibitemOpen
  \bibfield  {author} {\bibinfo {author} {\bibfnamefont {Z.}~\bibnamefont {Yao}}, \bibinfo {author} {\bibfnamefont {L.}~\bibnamefont {Pan}}, \bibinfo {author} {\bibfnamefont {S.}~\bibnamefont {Liu}},\ and\ \bibinfo {author} {\bibfnamefont {H.}~\bibnamefont {Zhai}},\ }\bibfield  {title} {\bibinfo {title} {Quantum many-body scars and quantum criticality},\ }\href {https://doi.org/10.1103/PhysRevB.105.125123} {\bibfield  {journal} {\bibinfo  {journal} {Phys. Rev. B}\ }\textbf {\bibinfo {volume} {105}},\ \bibinfo {pages} {125123} (\bibinfo {year} {2022})}\BibitemShut {NoStop}%
\bibitem [{\citenamefont {Wang}\ \emph {et~al.}(2023)\citenamefont {Wang}, \citenamefont {Zhang}, \citenamefont {Yao}, \citenamefont {Liu}, \citenamefont {Zhu}, \citenamefont {Zheng}, \citenamefont {Wang}, \citenamefont {Zhai}, \citenamefont {Yuan},\ and\ \citenamefont {Pan}}]{2023PhysRevLett131050401}%
  \BibitemOpen
  \bibfield  {author} {\bibinfo {author} {\bibfnamefont {H.-Y.}\ \bibnamefont {Wang}}, \bibinfo {author} {\bibfnamefont {W.-Y.}\ \bibnamefont {Zhang}}, \bibinfo {author} {\bibfnamefont {Z.}~\bibnamefont {Yao}}, \bibinfo {author} {\bibfnamefont {Y.}~\bibnamefont {Liu}}, \bibinfo {author} {\bibfnamefont {Z.-H.}\ \bibnamefont {Zhu}}, \bibinfo {author} {\bibfnamefont {Y.-G.}\ \bibnamefont {Zheng}}, \bibinfo {author} {\bibfnamefont {X.-K.}\ \bibnamefont {Wang}}, \bibinfo {author} {\bibfnamefont {H.}~\bibnamefont {Zhai}}, \bibinfo {author} {\bibfnamefont {Z.-S.}\ \bibnamefont {Yuan}},\ and\ \bibinfo {author} {\bibfnamefont {J.-W.}\ \bibnamefont {Pan}},\ }\bibfield  {title} {\bibinfo {title} {Interrelated thermalization and quantum criticality in a lattice gauge simulator},\ }\href {https://doi.org/10.1103/PhysRevLett.131.050401} {\bibfield  {journal} {\bibinfo  {journal} {Phys. Rev. Lett.}\ }\textbf {\bibinfo {volume} {131}},\ \bibinfo {pages} {050401} (\bibinfo {year} {2023})}\BibitemShut {NoStop}%
\bibitem [{\citenamefont {Shen}\ \emph {et~al.}(2024)\citenamefont {Shen}, \citenamefont {Qin}, \citenamefont {Desaules}, \citenamefont {Papi\ifmmode~\acute{c}\else \'{c}\fi{}},\ and\ \citenamefont {Lee}}]{2024PhysRevLett133216601}%
  \BibitemOpen
  \bibfield  {author} {\bibinfo {author} {\bibfnamefont {R.}~\bibnamefont {Shen}}, \bibinfo {author} {\bibfnamefont {F.}~\bibnamefont {Qin}}, \bibinfo {author} {\bibfnamefont {J.-Y.}\ \bibnamefont {Desaules}}, \bibinfo {author} {\bibfnamefont {Z.}~\bibnamefont {Papi\ifmmode~\acute{c}\else \'{c}\fi{}}},\ and\ \bibinfo {author} {\bibfnamefont {C.~H.}\ \bibnamefont {Lee}},\ }\bibfield  {title} {\bibinfo {title} {Enhanced many-body quantum scars from the non-hermitian fock skin effect},\ }\href {https://doi.org/10.1103/PhysRevLett.133.216601} {\bibfield  {journal} {\bibinfo  {journal} {Phys. Rev. Lett.}\ }\textbf {\bibinfo {volume} {133}},\ \bibinfo {pages} {216601} (\bibinfo {year} {2024})}\BibitemShut {NoStop}%
\bibitem [{\citenamefont {Yang}\ \emph {et~al.}(2025)\citenamefont {Yang}, \citenamefont {Yuan},\ and\ \citenamefont {Lee}}]{2025CommunPhys}%
  \BibitemOpen
  \bibfield  {author} {\bibinfo {author} {\bibfnamefont {M.}~\bibnamefont {Yang}}, \bibinfo {author} {\bibfnamefont {L.}~\bibnamefont {Yuan}},\ and\ \bibinfo {author} {\bibfnamefont {C.~H.}\ \bibnamefont {Lee}},\ }\bibfield  {title} {\bibinfo {title} {Non-hermitian strong bosonic clustering through interaction-induced caging},\ }\href {https://doi.org/10.1038/s42005-025-02274-z} {\bibfield  {journal} {\bibinfo  {journal} {Commun. Phys.}\ }\textbf {\bibinfo {volume} {8}},\ \bibinfo {pages} {388} (\bibinfo {year} {2025})}\BibitemShut {NoStop}%
\bibitem [{\citenamefont {Zohar}\ \emph {et~al.}(2016)\citenamefont {Zohar}, \citenamefont {Cirac},\ and\ \citenamefont {Reznik}}]{Zohar_2016}%
  \BibitemOpen
  \bibfield  {author} {\bibinfo {author} {\bibfnamefont {E.}~\bibnamefont {Zohar}}, \bibinfo {author} {\bibfnamefont {J.~I.}\ \bibnamefont {Cirac}},\ and\ \bibinfo {author} {\bibfnamefont {B.}~\bibnamefont {Reznik}},\ }\bibfield  {title} {\bibinfo {title} {Quantum simulations of lattice gauge theories using ultracold atoms in optical lattices},\ }\href {https://doi.org/10.1088/0034-4885/79/1/014401} {\bibfield  {journal} {\bibinfo  {journal} {Rep. Prog. Phys.}\ }\textbf {\bibinfo {volume} {79}},\ \bibinfo {pages} {014401} (\bibinfo {year} {2016})}\BibitemShut {NoStop}%
\bibitem [{\citenamefont {Schweizer}\ \emph {et~al.}(2019{\natexlab{b}})\citenamefont {Schweizer}, \citenamefont {Grusdt}, \citenamefont {Berngruber}, \citenamefont {Barbiero}, \citenamefont {Demler}, \citenamefont {Goldman}, \citenamefont {Bloch},\ and\ \citenamefont {Aidelsburger}}]{2019NP151168}%
  \BibitemOpen
  \bibfield  {author} {\bibinfo {author} {\bibfnamefont {C.}~\bibnamefont {Schweizer}}, \bibinfo {author} {\bibfnamefont {F.}~\bibnamefont {Grusdt}}, \bibinfo {author} {\bibfnamefont {M.}~\bibnamefont {Berngruber}}, \bibinfo {author} {\bibfnamefont {L.}~\bibnamefont {Barbiero}}, \bibinfo {author} {\bibfnamefont {E.}~\bibnamefont {Demler}}, \bibinfo {author} {\bibfnamefont {N.}~\bibnamefont {Goldman}}, \bibinfo {author} {\bibfnamefont {I.}~\bibnamefont {Bloch}},\ and\ \bibinfo {author} {\bibfnamefont {M.}~\bibnamefont {Aidelsburger}},\ }\bibfield  {title} {\bibinfo {title} {Floquet approach to $\mathbb{Z}_{2}$ lattice gauge theories with ultracold atoms in optical lattices},\ }\href {https://doi.org/10.1038/s41567-019-0649-7} {\bibfield  {journal} {\bibinfo  {journal} {Nat. Phys.}\ }\textbf {\bibinfo {volume} {15}},\ \bibinfo {pages} {1168} (\bibinfo {year} {2019}{\natexlab{b}})}\BibitemShut {NoStop}%
\bibitem [{\citenamefont {Görg}\ \emph {et~al.}(2019)\citenamefont {Görg}, \citenamefont {Sandholzer}, \citenamefont {Minguzzi}, \citenamefont {Desbuquois}, \citenamefont {Messer},\ and\ \citenamefont {Esslinger}}]{2019NP151161}%
  \BibitemOpen
  \bibfield  {author} {\bibinfo {author} {\bibfnamefont {F.}~\bibnamefont {Görg}}, \bibinfo {author} {\bibfnamefont {K.}~\bibnamefont {Sandholzer}}, \bibinfo {author} {\bibfnamefont {J.}~\bibnamefont {Minguzzi}}, \bibinfo {author} {\bibfnamefont {R.}~\bibnamefont {Desbuquois}}, \bibinfo {author} {\bibfnamefont {M.}~\bibnamefont {Messer}},\ and\ \bibinfo {author} {\bibfnamefont {T.}~\bibnamefont {Esslinger}},\ }\bibfield  {title} {\bibinfo {title} {Realization of density-dependent peierls phases to engineer quantized gauge fields coupled to ultracold matter},\ }\href {https://doi.org/10.1038/s41567-019-0615-4} {\bibfield  {journal} {\bibinfo  {journal} {Nat. Phys.}\ }\textbf {\bibinfo {volume} {15}},\ \bibinfo {pages} {1161} (\bibinfo {year} {2019})}\BibitemShut {NoStop}%
\bibitem [{\citenamefont {Shen}\ \emph {et~al.}(2023)\citenamefont {Shen}, \citenamefont {Chen}, \citenamefont {Aliyu}, \citenamefont {Qin}, \citenamefont {Zhong}, \citenamefont {Loh},\ and\ \citenamefont {Lee}}]{2023PhysRevLett.131.080403}%
  \BibitemOpen
  \bibfield  {author} {\bibinfo {author} {\bibfnamefont {R.}~\bibnamefont {Shen}}, \bibinfo {author} {\bibfnamefont {T.}~\bibnamefont {Chen}}, \bibinfo {author} {\bibfnamefont {M.~M.}\ \bibnamefont {Aliyu}}, \bibinfo {author} {\bibfnamefont {F.}~\bibnamefont {Qin}}, \bibinfo {author} {\bibfnamefont {Y.}~\bibnamefont {Zhong}}, \bibinfo {author} {\bibfnamefont {H.}~\bibnamefont {Loh}},\ and\ \bibinfo {author} {\bibfnamefont {C.~H.}\ \bibnamefont {Lee}},\ }\bibfield  {title} {\bibinfo {title} {Proposal for observing yang-lee criticality in rydberg atomic arrays},\ }\href {https://doi.org/10.1103/PhysRevLett.131.080403} {\bibfield  {journal} {\bibinfo  {journal} {Phys. Rev. Lett.}\ }\textbf {\bibinfo {volume} {131}},\ \bibinfo {pages} {080403} (\bibinfo {year} {2023})}\BibitemShut {NoStop}%
\bibitem [{\citenamefont {Cheng}\ \emph {et~al.}(2025)\citenamefont {Cheng}, \citenamefont {Wang}, \citenamefont {Zhang}, \citenamefont {Chen},\ and\ \citenamefont {Nie}}]{2025PhysRevA111013319}%
  \BibitemOpen
  \bibfield  {author} {\bibinfo {author} {\bibfnamefont {X.-C.}\ \bibnamefont {Cheng}}, \bibinfo {author} {\bibfnamefont {Z.-Y.}\ \bibnamefont {Wang}}, \bibinfo {author} {\bibfnamefont {J.}~\bibnamefont {Zhang}}, \bibinfo {author} {\bibfnamefont {S.}~\bibnamefont {Chen}},\ and\ \bibinfo {author} {\bibfnamefont {X.}~\bibnamefont {Nie}},\ }\bibfield  {title} {\bibinfo {title} {Density-dependent gauge field with raman lattices},\ }\href {https://doi.org/10.1103/PhysRevA.111.013319} {\bibfield  {journal} {\bibinfo  {journal} {Phys. Rev. A}\ }\textbf {\bibinfo {volume} {111}},\ \bibinfo {pages} {013319} (\bibinfo {year} {2025})}\BibitemShut {NoStop}%
\bibitem [{\citenamefont {Hu}\ \emph {et~al.}(2025)\citenamefont {Hu}, \citenamefont {Wang}, \citenamefont {Lian},\ and\ \citenamefont {Wang}}]{2025wztw-l8wg}%
  \BibitemOpen
  \bibfield  {author} {\bibinfo {author} {\bibfnamefont {Y.-M.}\ \bibnamefont {Hu}}, \bibinfo {author} {\bibfnamefont {Z.}~\bibnamefont {Wang}}, \bibinfo {author} {\bibfnamefont {B.}~\bibnamefont {Lian}},\ and\ \bibinfo {author} {\bibfnamefont {Z.}~\bibnamefont {Wang}},\ }\bibfield  {title} {\bibinfo {title} {Many-body non-hermitian skin effect with exact steady states in the dissipative quantum link model},\ }\href {https://doi.org/10.1103/wztw-l8wg} {\bibfield  {journal} {\bibinfo  {journal} {Phys. Rev. Lett.}\ }\textbf {\bibinfo {volume} {135}},\ \bibinfo {pages} {260401} (\bibinfo {year} {2025})}\BibitemShut {NoStop}%
\bibitem [{202()}]{2026supplemental}%
  \BibitemOpen
  \href@noop {} {}\bibinfo {note} {See Supplemental Materials for details of (S1) two-body spectrum and band isolation for well-defined topology; (S2) derivation and validation of the effective doublon Hamiltonian; (S3) adiabaticity and gap-adapted diving protocol; (S4) extended dynamical results for occupation-selective pumping; (S5) density-dependent gauge field from Floquet driving; (S6) extension to three-particle bound states: effective triolon Hamiltonian and topology}\BibitemShut {NoStop}%
\bibitem [{\citenamefont {Su}\ \emph {et~al.}(1980)\citenamefont {Su}, \citenamefont {Schrieffer},\ and\ \citenamefont {Heeger}}]{1980PhysRevB222099}%
  \BibitemOpen
  \bibfield  {author} {\bibinfo {author} {\bibfnamefont {W.~P.}\ \bibnamefont {Su}}, \bibinfo {author} {\bibfnamefont {J.~R.}\ \bibnamefont {Schrieffer}},\ and\ \bibinfo {author} {\bibfnamefont {A.~J.}\ \bibnamefont {Heeger}},\ }\bibfield  {title} {\bibinfo {title} {Soliton excitations in polyacetylene},\ }\href {https://doi.org/10.1103/PhysRevB.22.2099} {\bibfield  {journal} {\bibinfo  {journal} {Phys. Rev. B}\ }\textbf {\bibinfo {volume} {22}},\ \bibinfo {pages} {2099} (\bibinfo {year} {1980})}\BibitemShut {NoStop}%
\bibitem [{\citenamefont {Heeger}\ \emph {et~al.}(1988)\citenamefont {Heeger}, \citenamefont {Kivelson}, \citenamefont {Schrieffer},\ and\ \citenamefont {Su}}]{1988RevModPhys60781}%
  \BibitemOpen
  \bibfield  {author} {\bibinfo {author} {\bibfnamefont {A.~J.}\ \bibnamefont {Heeger}}, \bibinfo {author} {\bibfnamefont {S.}~\bibnamefont {Kivelson}}, \bibinfo {author} {\bibfnamefont {J.~R.}\ \bibnamefont {Schrieffer}},\ and\ \bibinfo {author} {\bibfnamefont {W.~P.}\ \bibnamefont {Su}},\ }\bibfield  {title} {\bibinfo {title} {Solitons in conducting polymers},\ }\href {https://doi.org/10.1103/RevModPhys.60.781} {\bibfield  {journal} {\bibinfo  {journal} {Rev. Mod. Phys.}\ }\textbf {\bibinfo {volume} {60}},\ \bibinfo {pages} {781} (\bibinfo {year} {1988})}\BibitemShut {NoStop}%
\bibitem [{\citenamefont {Jim\'enez-Garc\'{\i}a}\ \emph {et~al.}(2015)\citenamefont {Jim\'enez-Garc\'{\i}a}, \citenamefont {LeBlanc}, \citenamefont {Williams}, \citenamefont {Beeler}, \citenamefont {Qu}, \citenamefont {Gong}, \citenamefont {Zhang},\ and\ \citenamefont {Spielman}}]{PhysRevLett.114.125301}%
  \BibitemOpen
  \bibfield  {author} {\bibinfo {author} {\bibfnamefont {K.}~\bibnamefont {Jim\'enez-Garc\'{\i}a}}, \bibinfo {author} {\bibfnamefont {L.~J.}\ \bibnamefont {LeBlanc}}, \bibinfo {author} {\bibfnamefont {R.~A.}\ \bibnamefont {Williams}}, \bibinfo {author} {\bibfnamefont {M.~C.}\ \bibnamefont {Beeler}}, \bibinfo {author} {\bibfnamefont {C.}~\bibnamefont {Qu}}, \bibinfo {author} {\bibfnamefont {M.}~\bibnamefont {Gong}}, \bibinfo {author} {\bibfnamefont {C.}~\bibnamefont {Zhang}},\ and\ \bibinfo {author} {\bibfnamefont {I.~B.}\ \bibnamefont {Spielman}},\ }\bibfield  {title} {\bibinfo {title} {Tunable spin-orbit coupling via strong driving in ultracold-atom systems},\ }\href {https://doi.org/10.1103/PhysRevLett.114.125301} {\bibfield  {journal} {\bibinfo  {journal} {Phys. Rev. Lett.}\ }\textbf {\bibinfo {volume} {114}},\ \bibinfo {pages} {125301} (\bibinfo {year} {2015})}\BibitemShut {NoStop}%
\bibitem [{\citenamefont {Hauke}\ \emph {et~al.}(2012)\citenamefont {Hauke}, \citenamefont {Tieleman}, \citenamefont {Celi}, \citenamefont {\"Olschl\"ager}, \citenamefont {Simonet}, \citenamefont {Struck}, \citenamefont {Weinberg}, \citenamefont {Windpassinger}, \citenamefont {Sengstock}, \citenamefont {Lewenstein},\ and\ \citenamefont {Eckardt}}]{PhysRevLett.109.145301}%
  \BibitemOpen
  \bibfield  {author} {\bibinfo {author} {\bibfnamefont {P.}~\bibnamefont {Hauke}}, \bibinfo {author} {\bibfnamefont {O.}~\bibnamefont {Tieleman}}, \bibinfo {author} {\bibfnamefont {A.}~\bibnamefont {Celi}}, \bibinfo {author} {\bibfnamefont {C.}~\bibnamefont {\"Olschl\"ager}}, \bibinfo {author} {\bibfnamefont {J.}~\bibnamefont {Simonet}}, \bibinfo {author} {\bibfnamefont {J.}~\bibnamefont {Struck}}, \bibinfo {author} {\bibfnamefont {M.}~\bibnamefont {Weinberg}}, \bibinfo {author} {\bibfnamefont {P.}~\bibnamefont {Windpassinger}}, \bibinfo {author} {\bibfnamefont {K.}~\bibnamefont {Sengstock}}, \bibinfo {author} {\bibfnamefont {M.}~\bibnamefont {Lewenstein}},\ and\ \bibinfo {author} {\bibfnamefont {A.}~\bibnamefont {Eckardt}},\ }\bibfield  {title} {\bibinfo {title} {Non-abelian gauge fields and topological insulators in shaken optical lattices},\ }\href {https://doi.org/10.1103/PhysRevLett.109.145301} {\bibfield  {journal} {\bibinfo  {journal} {Phys. Rev. Lett.}\ }\textbf {\bibinfo {volume} {109}},\ \bibinfo
  {pages} {145301} (\bibinfo {year} {2012})}\BibitemShut {NoStop}%
\bibitem [{\citenamefont {Kevrekidis}\ \emph {et~al.}(2003)\citenamefont {Kevrekidis}, \citenamefont {Theocharis}, \citenamefont {Frantzeskakis},\ and\ \citenamefont {Malomed}}]{PhysRevLett.90.230401}%
  \BibitemOpen
  \bibfield  {author} {\bibinfo {author} {\bibfnamefont {P.~G.}\ \bibnamefont {Kevrekidis}}, \bibinfo {author} {\bibfnamefont {G.}~\bibnamefont {Theocharis}}, \bibinfo {author} {\bibfnamefont {D.~J.}\ \bibnamefont {Frantzeskakis}},\ and\ \bibinfo {author} {\bibfnamefont {B.~A.}\ \bibnamefont {Malomed}},\ }\bibfield  {title} {\bibinfo {title} {Feshbach resonance management for bose-einstein condensates},\ }\href {https://doi.org/10.1103/PhysRevLett.90.230401} {\bibfield  {journal} {\bibinfo  {journal} {Phys. Rev. Lett.}\ }\textbf {\bibinfo {volume} {90}},\ \bibinfo {pages} {230401} (\bibinfo {year} {2003})}\BibitemShut {NoStop}%
\bibitem [{\citenamefont {Winkler}\ \emph {et~al.}(2006)\citenamefont {Winkler}, \citenamefont {Thalhammer}, \citenamefont {Lang}, \citenamefont {Grimm}, \citenamefont {Denschlag}, \citenamefont {Daley}, \citenamefont {Kantian}, \citenamefont {B\"{u}chler},\ and\ \citenamefont {Zoller}}]{WinklerK2006}%
  \BibitemOpen
  \bibfield  {author} {\bibinfo {author} {\bibfnamefont {K.}~\bibnamefont {Winkler}}, \bibinfo {author} {\bibfnamefont {G.}~\bibnamefont {Thalhammer}}, \bibinfo {author} {\bibfnamefont {F.}~\bibnamefont {Lang}}, \bibinfo {author} {\bibfnamefont {R.}~\bibnamefont {Grimm}}, \bibinfo {author} {\bibfnamefont {J.~H.}\ \bibnamefont {Denschlag}}, \bibinfo {author} {\bibfnamefont {A.~J.}\ \bibnamefont {Daley}}, \bibinfo {author} {\bibfnamefont {A.}~\bibnamefont {Kantian}}, \bibinfo {author} {\bibfnamefont {H.~P.}\ \bibnamefont {B\"{u}chler}},\ and\ \bibinfo {author} {\bibfnamefont {P.}~\bibnamefont {Zoller}},\ }\bibfield  {title} {\bibinfo {title} {Repulsively bound atom pairs in an optical lattice},\ }\href {https://doi.org/10.1038/nature04918} {\bibfield  {journal} {\bibinfo  {journal} {Nature}\ }\textbf {\bibinfo {volume} {441}},\ \bibinfo {pages} {853} (\bibinfo {year} {2006})}\BibitemShut {NoStop}%
\bibitem [{\citenamefont {Chin}\ \emph {et~al.}(2010)\citenamefont {Chin}, \citenamefont {Grimm}, \citenamefont {Julienne},\ and\ \citenamefont {Tiesinga}}]{RevModPhys.82.1225}%
  \BibitemOpen
  \bibfield  {author} {\bibinfo {author} {\bibfnamefont {C.}~\bibnamefont {Chin}}, \bibinfo {author} {\bibfnamefont {R.}~\bibnamefont {Grimm}}, \bibinfo {author} {\bibfnamefont {P.}~\bibnamefont {Julienne}},\ and\ \bibinfo {author} {\bibfnamefont {E.}~\bibnamefont {Tiesinga}},\ }\bibfield  {title} {\bibinfo {title} {Feshbach resonances in ultracold gases},\ }\href {https://doi.org/10.1103/RevModPhys.82.1225} {\bibfield  {journal} {\bibinfo  {journal} {Rev. Mod. Phys.}\ }\textbf {\bibinfo {volume} {82}},\ \bibinfo {pages} {1225} (\bibinfo {year} {2010})}\BibitemShut {NoStop}%
\bibitem [{\citenamefont {Clark}\ \emph {et~al.}(2017)\citenamefont {Clark}, \citenamefont {Gaj}, \citenamefont {Feng},\ and\ \citenamefont {Chin}}]{ClarkLoganW2017}%
  \BibitemOpen
  \bibfield  {author} {\bibinfo {author} {\bibfnamefont {L.~W.}\ \bibnamefont {Clark}}, \bibinfo {author} {\bibfnamefont {A.}~\bibnamefont {Gaj}}, \bibinfo {author} {\bibfnamefont {L.}~\bibnamefont {Feng}},\ and\ \bibinfo {author} {\bibfnamefont {C.}~\bibnamefont {Chin}},\ }\bibfield  {title} {\bibinfo {title} {Collective emission of matter-wave jets from driven bose-einstein condensates},\ }\href {https://doi.org/10.1038/nature24272} {\bibfield  {journal} {\bibinfo  {journal} {Nature}\ }\textbf {\bibinfo {volume} {551}},\ \bibinfo {pages} {356} (\bibinfo {year} {2017})}\BibitemShut {NoStop}%
\bibitem [{\citenamefont {Gong}\ \emph {et~al.}(2009{\natexlab{b}})\citenamefont {Gong}, \citenamefont {Morales-Molina},\ and\ \citenamefont {H\"anggi}}]{2009PhysRevLett.103.133002}%
  \BibitemOpen
  \bibfield  {author} {\bibinfo {author} {\bibfnamefont {J.}~\bibnamefont {Gong}}, \bibinfo {author} {\bibfnamefont {L.}~\bibnamefont {Morales-Molina}},\ and\ \bibinfo {author} {\bibfnamefont {P.}~\bibnamefont {H\"anggi}},\ }\bibfield  {title} {\bibinfo {title} {Many-body coherent destruction of tunneling},\ }\href {https://doi.org/10.1103/PhysRevLett.103.133002} {\bibfield  {journal} {\bibinfo  {journal} {Phys. Rev. Lett.}\ }\textbf {\bibinfo {volume} {103}},\ \bibinfo {pages} {133002} (\bibinfo {year} {2009}{\natexlab{b}})}\BibitemShut {NoStop}%
\bibitem [{\citenamefont {Rapp}\ \emph {et~al.}(2012)\citenamefont {Rapp}, \citenamefont {Deng},\ and\ \citenamefont {Santos}}]{2021PhysRevLett.109.203005}%
  \BibitemOpen
  \bibfield  {author} {\bibinfo {author} {\bibfnamefont {A.}~\bibnamefont {Rapp}}, \bibinfo {author} {\bibfnamefont {X.}~\bibnamefont {Deng}},\ and\ \bibinfo {author} {\bibfnamefont {L.}~\bibnamefont {Santos}},\ }\bibfield  {title} {\bibinfo {title} {Ultracold lattice gases with periodically modulated interactions},\ }\href {https://doi.org/10.1103/PhysRevLett.109.203005} {\bibfield  {journal} {\bibinfo  {journal} {Phys. Rev. Lett.}\ }\textbf {\bibinfo {volume} {109}},\ \bibinfo {pages} {203005} (\bibinfo {year} {2012})}\BibitemShut {NoStop}%
\bibitem [{\citenamefont {Liberto}\ \emph {et~al.}(2014)\citenamefont {Liberto}, \citenamefont {Creffield}, \citenamefont {Japaridze},\ and\ \citenamefont {Smith}}]{2014PhysRevA.89.013624}%
  \BibitemOpen
  \bibfield  {author} {\bibinfo {author} {\bibfnamefont {M.~D.}\ \bibnamefont {Liberto}}, \bibinfo {author} {\bibfnamefont {C.~E.}\ \bibnamefont {Creffield}}, \bibinfo {author} {\bibfnamefont {G.~I.}\ \bibnamefont {Japaridze}},\ and\ \bibinfo {author} {\bibfnamefont {C.~M.}\ \bibnamefont {Smith}},\ }\bibfield  {title} {\bibinfo {title} {Quantum simulation of correlated-hopping models with fermions in optical lattices},\ }\href {https://doi.org/10.1103/PhysRevA.89.013624} {\bibfield  {journal} {\bibinfo  {journal} {Phys. Rev. A}\ }\textbf {\bibinfo {volume} {89}},\ \bibinfo {pages} {013624} (\bibinfo {year} {2014})}\BibitemShut {NoStop}%
\bibitem [{\citenamefont {Greschner}\ \emph {et~al.}(2014)\citenamefont {Greschner}, \citenamefont {Sun}, \citenamefont {Poletti},\ and\ \citenamefont {Santos}}]{2014PhysRevLett.113.215303}%
  \BibitemOpen
  \bibfield  {author} {\bibinfo {author} {\bibfnamefont {S.}~\bibnamefont {Greschner}}, \bibinfo {author} {\bibfnamefont {G.}~\bibnamefont {Sun}}, \bibinfo {author} {\bibfnamefont {D.}~\bibnamefont {Poletti}},\ and\ \bibinfo {author} {\bibfnamefont {L.}~\bibnamefont {Santos}},\ }\bibfield  {title} {\bibinfo {title} {Density-dependent synthetic gauge fields using periodically modulated interactions},\ }\href {https://doi.org/10.1103/PhysRevLett.113.215303} {\bibfield  {journal} {\bibinfo  {journal} {Phys. Rev. Lett.}\ }\textbf {\bibinfo {volume} {113}},\ \bibinfo {pages} {215303} (\bibinfo {year} {2014})}\BibitemShut {NoStop}%
\bibitem [{\citenamefont {Meinert}\ \emph {et~al.}(2016{\natexlab{b}})\citenamefont {Meinert}, \citenamefont {Mark}, \citenamefont {Lauber}, \citenamefont {Daley},\ and\ \citenamefont {N\"agerl}}]{2016PhysRevLett.116.205301}%
  \BibitemOpen
  \bibfield  {author} {\bibinfo {author} {\bibfnamefont {F.}~\bibnamefont {Meinert}}, \bibinfo {author} {\bibfnamefont {M.~J.}\ \bibnamefont {Mark}}, \bibinfo {author} {\bibfnamefont {K.}~\bibnamefont {Lauber}}, \bibinfo {author} {\bibfnamefont {A.~J.}\ \bibnamefont {Daley}},\ and\ \bibinfo {author} {\bibfnamefont {H.-C.}\ \bibnamefont {N\"agerl}},\ }\bibfield  {title} {\bibinfo {title} {Floquet engineering of correlated tunneling in the bose-hubbard model with ultracold atoms},\ }\href {https://doi.org/10.1103/PhysRevLett.116.205301} {\bibfield  {journal} {\bibinfo  {journal} {Phys. Rev. Lett.}\ }\textbf {\bibinfo {volume} {116}},\ \bibinfo {pages} {205301} (\bibinfo {year} {2016}{\natexlab{b}})}\BibitemShut {NoStop}%
\bibitem [{\citenamefont {Kolovsky}\ and\ \citenamefont {Buchleitner}(2003)}]{2003PhysRevE.68.056213}%
  \BibitemOpen
  \bibfield  {author} {\bibinfo {author} {\bibfnamefont {A.~R.}\ \bibnamefont {Kolovsky}}\ and\ \bibinfo {author} {\bibfnamefont {A.}~\bibnamefont {Buchleitner}},\ }\bibfield  {title} {\bibinfo {title} {Floquet-bloch operator for the bose-hubbard model with static field},\ }\href {https://doi.org/10.1103/PhysRevE.68.056213} {\bibfield  {journal} {\bibinfo  {journal} {Phys. Rev. E}\ }\textbf {\bibinfo {volume} {68}},\ \bibinfo {pages} {056213} (\bibinfo {year} {2003})}\BibitemShut {NoStop}%
\bibitem [{\citenamefont {Kivelson}(1982)}]{1982PhysRevB.26.4269}%
  \BibitemOpen
  \bibfield  {author} {\bibinfo {author} {\bibfnamefont {S.}~\bibnamefont {Kivelson}},\ }\bibfield  {title} {\bibinfo {title} {Wannier functions in one-dimensional disordered systems: Application to fractionally charged solitons},\ }\href {https://doi.org/10.1103/PhysRevB.26.4269} {\bibfield  {journal} {\bibinfo  {journal} {Phys. Rev. B}\ }\textbf {\bibinfo {volume} {26}},\ \bibinfo {pages} {4269} (\bibinfo {year} {1982})}\BibitemShut {NoStop}%
\bibitem [{\citenamefont {Marzari}\ and\ \citenamefont {Vanderbilt}(1997)}]{1997PhysRevB.56.12847}%
  \BibitemOpen
  \bibfield  {author} {\bibinfo {author} {\bibfnamefont {N.}~\bibnamefont {Marzari}}\ and\ \bibinfo {author} {\bibfnamefont {D.}~\bibnamefont {Vanderbilt}},\ }\bibfield  {title} {\bibinfo {title} {Maximally localized generalized wannier functions for composite energy bands},\ }\href {https://doi.org/10.1103/PhysRevB.56.12847} {\bibfield  {journal} {\bibinfo  {journal} {Phys. Rev. B}\ }\textbf {\bibinfo {volume} {56}},\ \bibinfo {pages} {12847} (\bibinfo {year} {1997})}\BibitemShut {NoStop}%
\bibitem [{\citenamefont {Lee}\ and\ \citenamefont {Ye}(2015)}]{2015PhysRevB.91.085119}%
  \BibitemOpen
  \bibfield  {author} {\bibinfo {author} {\bibfnamefont {C.~H.}\ \bibnamefont {Lee}}\ and\ \bibinfo {author} {\bibfnamefont {P.}~\bibnamefont {Ye}},\ }\bibfield  {title} {\bibinfo {title} {Free-fermion entanglement spectrum through wannier interpolation},\ }\href {https://doi.org/10.1103/PhysRevB.91.085119} {\bibfield  {journal} {\bibinfo  {journal} {Phys. Rev. B}\ }\textbf {\bibinfo {volume} {91}},\ \bibinfo {pages} {085119} (\bibinfo {year} {2015})}\BibitemShut {NoStop}%
\bibitem [{\citenamefont {Takahashi}(1977)}]{1977JPC101289}%
  \BibitemOpen
  \bibfield  {author} {\bibinfo {author} {\bibfnamefont {M.}~\bibnamefont {Takahashi}},\ }\bibfield  {title} {\bibinfo {title} {Half-filled hubbard model at low temperature},\ }\href {https://doi.org/10.1088/0022-3719/10/8/031} {\bibfield  {journal} {\bibinfo  {journal} {J. Phys. C}\ }\textbf {\bibinfo {volume} {10}},\ \bibinfo {pages} {1289} (\bibinfo {year} {1977})}\BibitemShut {NoStop}%
\bibitem [{\citenamefont {Bravyi}\ \emph {et~al.}(2011)\citenamefont {Bravyi}, \citenamefont {DiVincenzo},\ and\ \citenamefont {Loss}}]{2011AOP3262793}%
  \BibitemOpen
  \bibfield  {author} {\bibinfo {author} {\bibfnamefont {S.}~\bibnamefont {Bravyi}}, \bibinfo {author} {\bibfnamefont {D.~P.}\ \bibnamefont {DiVincenzo}},\ and\ \bibinfo {author} {\bibfnamefont {D.}~\bibnamefont {Loss}},\ }\bibfield  {title} {\bibinfo {title} {Schrieffer-wolff transformation for quantum many-body systems},\ }\href {https://doi.org/10.1016/j.aop.2011.06.004} {\bibfield  {journal} {\bibinfo  {journal} {Annals of Physics}\ }\textbf {\bibinfo {volume} {326}},\ \bibinfo {pages} {2793} (\bibinfo {year} {2011})}\BibitemShut {NoStop}%
\bibitem [{\citenamefont {Schiff}(1981)}]{1981QuantumMechanics}%
  \BibitemOpen
  \bibfield  {author} {\bibinfo {author} {\bibfnamefont {L.~I.}\ \bibnamefont {Schiff}},\ }\bibfield  {title} {\bibinfo {title} {Quantum mechanics},\ }\href@noop {} {\bibfield  {journal} {\bibinfo  {journal} {McGraw-Hill, New York}\ } (\bibinfo {year} {1981})}\BibitemShut {NoStop}%
\bibitem [{\citenamefont {Mart\'{\i}nez-Garaot}\ \emph {et~al.}(2015)\citenamefont {Mart\'{\i}nez-Garaot}, \citenamefont {Ruschhaupt}, \citenamefont {Gillet}, \citenamefont {Busch},\ and\ \citenamefont {Muga}}]{2015PhysRevA92043406}%
  \BibitemOpen
  \bibfield  {author} {\bibinfo {author} {\bibfnamefont {S.}~\bibnamefont {Mart\'{\i}nez-Garaot}}, \bibinfo {author} {\bibfnamefont {A.}~\bibnamefont {Ruschhaupt}}, \bibinfo {author} {\bibfnamefont {J.}~\bibnamefont {Gillet}}, \bibinfo {author} {\bibfnamefont {T.}~\bibnamefont {Busch}},\ and\ \bibinfo {author} {\bibfnamefont {J.~G.}\ \bibnamefont {Muga}},\ }\bibfield  {title} {\bibinfo {title} {Fast quasiadiabatic dynamics},\ }\href {https://doi.org/10.1103/PhysRevA.92.043406} {\bibfield  {journal} {\bibinfo  {journal} {Phys. Rev. A}\ }\textbf {\bibinfo {volume} {92}},\ \bibinfo {pages} {043406} (\bibinfo {year} {2015})}\BibitemShut {NoStop}%
\bibitem [{\citenamefont {Goldman}\ and\ \citenamefont {Dalibard}(2014)}]{PhysRevX.4.031027}%
  \BibitemOpen
  \bibfield  {author} {\bibinfo {author} {\bibfnamefont {N.}~\bibnamefont {Goldman}}\ and\ \bibinfo {author} {\bibfnamefont {J.}~\bibnamefont {Dalibard}},\ }\bibfield  {title} {\bibinfo {title} {Periodically driven quantum systems: Effective hamiltonians and engineered gauge fields},\ }\href {https://doi.org/10.1103/PhysRevX.4.031027} {\bibfield  {journal} {\bibinfo  {journal} {Phys. Rev. X}\ }\textbf {\bibinfo {volume} {4}},\ \bibinfo {pages} {031027} (\bibinfo {year} {2014})}\BibitemShut {NoStop}%
\bibitem [{\citenamefont {Bukov}\ \emph {et~al.}(2015)\citenamefont {Bukov}, \citenamefont {D'Alessio},\ and\ \citenamefont {Polkovnikov}}]{2015Bukov04032015}%
  \BibitemOpen
  \bibfield  {author} {\bibinfo {author} {\bibfnamefont {M.}~\bibnamefont {Bukov}}, \bibinfo {author} {\bibfnamefont {L.}~\bibnamefont {D'Alessio}},\ and\ \bibinfo {author} {\bibfnamefont {A.}~\bibnamefont {Polkovnikov}},\ }\bibfield  {title} {\bibinfo {title} {Universal high-frequency behavior of periodically driven systems: from dynamical stabilization to floquet engineering},\ }\href {https://doi.org/10.1080/00018732.2015.1055918} {\bibfield  {journal} {\bibinfo  {journal} {Advances in Physics}\ }\textbf {\bibinfo {volume} {64}},\ \bibinfo {pages} {139} (\bibinfo {year} {2015})}\BibitemShut {NoStop}%
\bibitem [{\citenamefont {Holthaus}(2016)}]{2016MHJPB}%
  \BibitemOpen
  \bibfield  {author} {\bibinfo {author} {\bibfnamefont {M.}~\bibnamefont {Holthaus}},\ }\bibfield  {title} {\bibinfo {title} {Floquet engineering with quasienergy bands of periodically driven optical lattices},\ }\href {https://doi.org/10.1088/0953-4075/49/1/013001} {\bibfield  {journal} {\bibinfo  {journal} {J. Phys. B: At. Mol. Opt. Phys.}\ }\textbf {\bibinfo {volume} {49}},\ \bibinfo {pages} {013001} (\bibinfo {year} {2016})}\BibitemShut {NoStop}%
\bibitem [{\citenamefont {Eckardt}(2017)}]{RevModPhys.89.011004}%
  \BibitemOpen
  \bibfield  {author} {\bibinfo {author} {\bibfnamefont {A.}~\bibnamefont {Eckardt}},\ }\bibfield  {title} {\bibinfo {title} {Colloquium: Atomic quantum gases in periodically driven optical lattices},\ }\href {https://doi.org/10.1103/RevModPhys.89.011004} {\bibfield  {journal} {\bibinfo  {journal} {Rev. Mod. Phys.}\ }\textbf {\bibinfo {volume} {89}},\ \bibinfo {pages} {011004} (\bibinfo {year} {2017})}\BibitemShut {NoStop}%
\bibitem [{\citenamefont {Qin}\ \emph {et~al.}(2023)\citenamefont {Qin}, \citenamefont {Lee},\ and\ \citenamefont {Chen}}]{2023PhysRevB.108.075435}%
  \BibitemOpen
  \bibfield  {author} {\bibinfo {author} {\bibfnamefont {F.}~\bibnamefont {Qin}}, \bibinfo {author} {\bibfnamefont {C.~H.}\ \bibnamefont {Lee}},\ and\ \bibinfo {author} {\bibfnamefont {R.}~\bibnamefont {Chen}},\ }\bibfield  {title} {\bibinfo {title} {Light-induced half-quantized hall effect and axion insulator},\ }\href {https://doi.org/10.1103/PhysRevB.108.075435} {\bibfield  {journal} {\bibinfo  {journal} {Phys. Rev. B}\ }\textbf {\bibinfo {volume} {108}},\ \bibinfo {pages} {075435} (\bibinfo {year} {2023})}\BibitemShut {NoStop}%
\end{thebibliography}
\end{document}